\documentclass[acmsmall]{acmart}
\AtBeginDocument{%
  \providecommand\BibTeX{{%
    \normalfont B\kern-0.5em{\scshape i\kern-0.25em b}\kern-0.8em\TeX}}}

\setcopyright{acmlicensed}
\copyrightyear{2025}
\acmYear{2025}
\acmDOI{XXXXXXX.XXXXXXX}

\usepackage{subcaption} 

\usepackage{multirow}
\usepackage{tabularx}
\usepackage{rotating}
\usepackage{xcolor}
\usepackage{amsthm}

\usepackage{amssymb}
\usepackage{colortbl}
\usepackage{pifont}

\usepackage[normalem]{ulem}

\useunder{\uline}{\ul}{}

\newcommand{\cmark}{\textcolor{green}{\textbf{\ding{51}}}}
\newcommand{\xmark}{\textcolor{black}{\textbf{\ding{55}}}}
\acmJournal{CSUR}
\acmVolume{9}
\acmNumber{9}
\acmArticle{35}
\acmMonth{9}

\begin{document}

\title{Intelligent Forensics in Next-Generation Mobile Networks: Evidence, Methods, and Applications}

\authorsaddresses{Authors' Contact Information: 
Jiacheng Wang, jiacheng.wang@ntu.edu.sg, NTU, Singapore;
Weihong Qin, qinwh25@mails.jlu.edu.cn, Jilin University, Changchun, China;
Jialing He, hejialing@cqu.edu.cn, Chongqing University, Chongqing, China;
Changyuan Zhao, zhao0441@e.ntu.edu.sg, NTU, Singapore;
Dusit Niyato, dniyato@ntu.edu.sg, NTU, Singapore;
Tao Xiang, txiang@cqu.edu.cn, Chongqing University, Chongqing, China.
}


\settopmatter{printacmref=false}
\setcopyright{none}
\renewcommand\footnotetextcopyrightpermission[1]{}


\author{Jiacheng Wang}
\affiliation{%
  \institution{Nanyang Technological University}
  \country{Singapore}
}

\author{Weihong Qin}
\affiliation{%
  \institution{Jilin University}
  \country{China}
}

\author{Jialing He}
\affiliation{%
  \institution{Chongqing University}
  \country{China}
}

\author{Changyuan Zhao}
\affiliation{%
  \institution{Nanyang Technological University}
  \country{Singapore}
}






\author{Dusit Niyato}
\affiliation{%
  \institution{Nanyang Technological University}
  \country{Singapore}
}

\author{Tao Xiang}
\affiliation{%
  \institution{Chongqing University}
  \country{China}
}




\renewcommand{\shortauthors}{J. Wang et al.}

\begin{abstract}
\textcolor{black}{This survey examines intelligent forensics in next-generation mobile networks, arguing that future wireless security must move beyond real-time detection toward accountable post-incident reconstruction. Unlike traditional digital forensics, wireless investigations rely on short-lived, distributed, and heterogeneous evidence, including radio waveforms, channel measurements, device-side artifacts, and network telemetry, affected by calibration, timing uncertainty, privacy constraints, and adversarial manipulation. To address this limitation, this paper develops an evidence-centric framework that treats wireless measurements as first-class forensic artifacts and organizes the field through a unified taxonomy spanning physical-layer, device-layer, network-layer, and cross-layer forensics. We further systematize the forensic workflow into readiness and preservation-by-design, acquisition, correlation and analysis, and reporting and reproducibility, while comparing the complementary roles of traditional methods and artificial intelligence-assisted techniques. Subsequently, we review major application areas, including anomaly discovery, attribution, provenance and localization, authenticity verification, and timeline reconstruction. Finally, we identify key open challenges, including domain shift, resource-aware evidence capture, and the benefits and admissibility risks of generative evidence. Overall, this paper positions wireless forensics as a foundational capability for trustworthy, auditable, and reproducible security in next-generation wireless systems. Readers can understand and streamline wireless forensics processes for specific applications, such as low-altitude wireless networks, vehicular communications, and edge general intelligence.}
\end{abstract}




\maketitle

\section{Introduction}

\subsection{Background and Motivation}
Forensics refers to the disciplined practice of identifying, collecting, examining, and reporting evidence so that conclusions are reproducible, verifiable, and, when necessary, admissible~\cite{akinbi2023digital}. In modern digital infrastructure, incidents rarely stay within a single component; rather, they span software stacks, cloud services, identity systems, and network control functions, where traces may be fragmented or overwritten. Hence, beyond remediation, the key task is forensic reconstruction: determining what happened and why from preserved artifacts rather than assumptions. 
The Capital One incident disclosed in 2019 illustrates this challenge. Public reports indicate that the attacker exploited a web application vulnerability, then obtained cloud credentials and accessed cloud storage services, causing unauthorized exposure of customer related data \cite{nelson2025incident}. Investigators therefore had to correlate heterogeneous evidence, including application and security alerts, identity and access traces, and cloud audit and storage access logs, to reconstruct the intrusion path, establish the timing and scope of data access, and justify the conclusions \cite{wang2022cnn}. More broadly, this case shows that complex systems require accountable reconstruction, not only real time alarms.

This need on intelligent wireless forensics is becoming more significant in wireless communication systems. Future wireless networks are programmable service platforms that connect heterogeneous devices and couple radio access with edge computing under strict latency and reliability constraints. The same infrastructure may simultaneously support consumer traffic, industrial telemetry, vehicular safety messages, and sensing related bit streams, raising the demand for accountable post incident analysis. Forensic capability is especially important because wireless evidence can directly affect cyber physical control, which is distributed across operators, cloud and edge providers, and devices, and is further complicated by softwareized architectures such as cloud native cores, network slicing, and disaggregated radio access networks \cite{khan2019survey}. Hence, wireless forensics is increasingly needed for incident response, operational accountability, safety validation, and policy refinement \cite{rizvi2022application}.
\textcolor{black}{In addition, wireless communication systems are becoming a foundational component of critical digital infrastructure, supporting industrial automation, autonomous transportation, smart cities, and large-scale \textcolor{black}{ internet of things (IoT) }deployments. As these systems evolve toward \textcolor{black}{sixth generation (6G) }architectures, they increasingly integrate programmable networks, edge intelligence, and heterogeneous wireless devices operating across multiple layers of the communication stack. While these advances enable unprecedented connectivity and flexibility, they also introduce new security vulnerabilities and attack surfaces that extend beyond traditional network boundaries. Consequently, securing wireless systems now requires not only detecting attacks and mitigating threats in real time, but also reconstructing incidents with sufficient evidence to understand how attacks unfolded and how defenses should be improved.}
Meanwhile, wireless makes forensic reconstruction harder than many wired settings. Relevant traces are often ephemeral; evidence is inherently multi-layer, spanning radio features, channel dynamics, timing relations, and protocol bit streams; encryption reduces semantic visibility; and telemetry is fragmented across base stations, roadside units, access points, user devices, and edge controllers \cite{stoyanova2020survey}. \textcolor{black}{From a security perspective, incident reconstruction in wireless networks is particularly challenging because decisive evidence frequently resides in transient radio signals, distributed device artifacts, and multi-layer telemetry. Radio waveforms, channel measurements, and device-level state transitions may collectively encode critical clues about malicious activity, yet these signals are often short-lived and difficult to preserve.}

Therefore, wireless forensics workflows must address short-lived observables, environmental uncertainty, distributed evidence and time alignment, scale and resource limits, and adversarial manipulation. Conventional security monitoring and intrusion detection remain indispensable for real-time defense in wireless networks~\cite{yang2026data}, but they do not provide preserved, provenance-aware, and cross-vantage evidence for post incident scrutiny~\cite{nelson2024incident}. Traditional digital forensics is also less effective when decisive wireless traces are short lived, distributed, and constrained by resource and privacy limits~\cite{stoyanova2020survey}. \textcolor{black}{These gaps motivate wireless intelligent forensics, which complements existing defenses through preservation by design, resource aware evidence capture, and auditable, learning assisted analysis that can cope with uncertainty, mobility, and heterogeneous hardware~\cite{alrajeh2017evidence}. By systematically capturing and correlating radio observations with device- and network-layer telemetry, wireless intelligent forensics enables accountable reconstruction of attacks, attribution of malicious transmissions, and validation of security claims under explicit uncertainty assumptions. Guided by this motivation, this survey analyzes wireless evidence sources, forensic workflows, analysis methods, evaluation practices, and case studies across 6G and beyond.}


\subsection{Related Work}
Existing review papers related to this survey can be grouped into digital/network forensics and IoT/mobile/wireless security, both of which provide important foundations.

\emph{\textbf{1) Digital and Network Security:}}
Many surveys establish the general foundations of digital forensics. For example,~\cite{casino2022research} reviews recurring themes such as evidence acquisition, analysis pipelines, tool support, scalability, and standardization. From the network perspective,~\cite{pilli2010network} surveys network forensic frameworks and emphasizes systematic collection, multi-source correlation, and trustworthy reporting, while~\cite{khan2016network} provides a taxonomy and discusses attribution under partial observability and large-scale heterogeneous trace correlation. Related surveys further examine operational limits on preservation and processing:~\cite{quick2014impacts} highlights prioritization and data reduction under growing forensic data volume, and~\cite{van2014digital} motivates Digital Forensics as a Service for scalable processing and shared tooling. In cloud settings,~\cite{pichan2015cloud} analyzes limited physical access, multi-tenancy, and dependence on provider logs and APIs, and~\cite{manral2019systematic} organizes cloud forensic artifacts and challenges across investigation stages. Overall, these works clarify workflows, scalability, and cross-domain evidence, but rarely treat radio and link-layer measurements as first-class evidence or address channel uncertainty and mobility.

\emph{\textbf{2) IoT and Mobile security:}}
IoT security surveys stress that investigations span an ecosystem rather than a single host. The survey in~\cite{stoyanova2020survey} shows that evidence is fragmented across devices, companion mobile apps, gateways, and cloud backends, complicating acquisition, correlation, and chain-of-custody management. Similarly,~\cite{yaqoob2019internet} explains how IoT scale and heterogeneity constrain evidence visibility and attribution, motivating cross-source fusion and structured investigation. The work in~\cite{atlam2020internet} further reviews forensic requirements and tools, emphasizing automation and learning-based assistance for diverse devices and growing evidence volume, while~\cite{servida2019iot} shows that minor artifacts such as application records and device interactions can be decisive when traditional logs are incomplete. Unlike IoT security surveys, wireless-oriented surveys mainly focus on functions. The review in~\cite{danev2012physical} examines physical-layer identification and shows how hardware-imposed radio features support device attribution, while~\cite{xu2015device} extends this to device fingerprinting across protocol layers and discusses robustness issues affecting evidentiary reliability. More recently,~\cite{jagannath2022comprehensive} surveys \textcolor{black}{radio frequency (RF)} fingerprinting with emphasis on deep learning pipelines and datasets, and~\cite{mitchell2014survey,van2018survey} review wireless intrusion and misbehavior detection through semantic and behavioral inconsistencies. In summary, IoT security surveys explain where evidence resides and why cross-domain correlation is necessary, while wireless surveys provide technical blocks for attribution and anomaly sensing. 

\begin{figure*}[t]
    \centering
    \includegraphics[width= 1.0\linewidth]{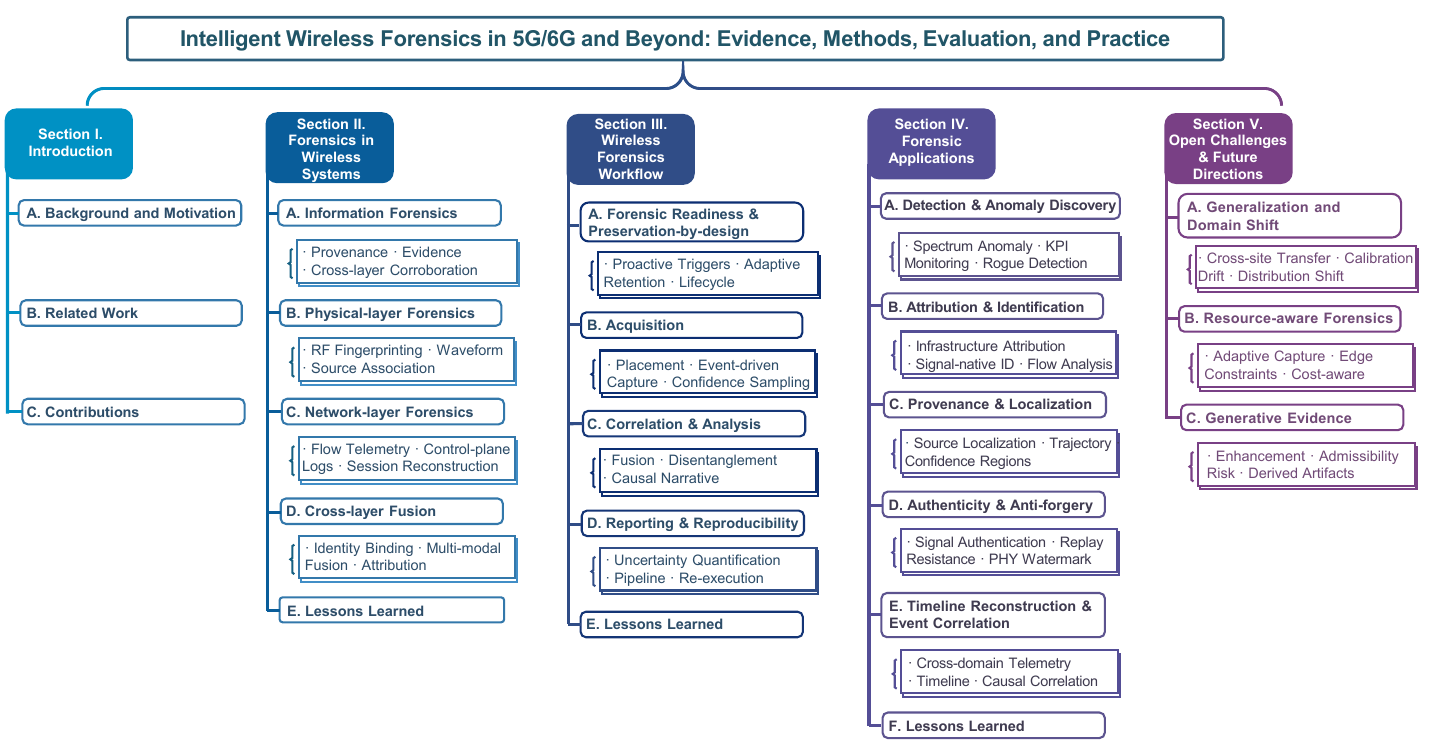}
        \vspace{-0.7cm}
    \caption{Survey organization and taxonomy overview.}
    \label{fig:structure}
    \vspace{-0.8cm}
\end{figure*}

\subsection{Contributions}
While many surveys exist on digital investigations, most works do not study the problem from a wireless evidence perspective. Existing reviews of digital, network, and cloud forensics mainly focus on investigation phases, toolchains, and trace correlation across distributed systems, with emphasis on logs, files, software artifacts, and network telemetry rather than radio observations~\cite{khan2016network,manral2019systematic}. IoT and mobile security surveys are closer to wireless scenarios, as they emphasize heterogeneous endpoints, limited device visibility, and multi-party evidence ownership~\cite{stoyanova2020survey,yaqoob2019internet}. However, they rarely treat radio measurements and time-varying channels as primary evidence requiring calibration and uncertainty modeling. By contrast, wireless security surveys need to address device identification, RF fingerprinting, and anomaly or misbehavior detection~\cite{jagannath2022comprehensive,van2018survey}, focusing on classification robustness and online detection accuracy. Therefore, as summarized in Table I, this contributes to the forensics and wireless fields by treating wireless measurements themselves as forensic evidence and by organizing the field around accountable reconstruction rather than only detection or defense. By connecting traditional wireless measurements to the broader goal of forensic reconstruction, this survey aims to extend wireless communication research toward more accountable and trustworthy digital infrastructures, with its main contributions summarized as follows.

\begin{itemize}
    \item We introduce a unified wireless evidence taxonomy spanning radio, protocol, and architectural traces, with explicit discussion of evidentiary value and uncertainty sources.
    \item We systematize a wireless forensics workflow that distinguishes readiness, acquisition, correlation and analysis, and reporting, and it clarifies where learning assisted methods can be safely inserted.
    \item We summarize analysis methods from signal level identification to cross layer correlation, highlighting design principles needed for defensible conclusions, including provenance and reproducibility considerations.
    \item We present evaluation practices, representative case studies, and open challenges, aiming to connect practical wireless measurements to auditable and reproducible forensic outcomes.
\end{itemize}

\begin{table*}[tpb]
\centering
\caption{Overview of representative related surveys.}
\vspace{-0.3cm}
\label{tab:related_surveys_tableI_style}
\tiny
\renewcommand{\arraystretch}{1}
\begin{tabular}{m{1.5cm}||c|p{4.5cm}|c|c|c}
\hline
\hline
\rowcolor{gray!15}
\textbf{Scope} & \textbf{Ref.} & \textbf{Overview} & \textbf{\shortstack{Wireless\\ evidence}} & \textbf{\shortstack{Uncertainty/\\ calibration}} & \textbf{\shortstack{Forensic\\ reconstruction}} \\

\multirow{7}{*}{\shortstack{Digital and\\ Network\\ Forensics}}
& \cite{casino2022research} & Reviews digital forensics workflows, tools, and open challenges. & \xmark & \xmark & \cmark \\
\cline{2-6}
~ & \cite{pilli2010network} & Surveys network forensic frameworks for collection, correlation, and reporting. & \xmark & \xmark & \cmark \\
\cline{2-6}
~ & \cite{khan2016network} & Taxonomizes network forensics and attribution under partial visibility. & \xmark & \xmark & \cmark \\
\cline{2-6}
~ & \cite{quick2014impacts} & Examines forensic data growth and the need for triage. & \xmark & \xmark & Partially \\
\cline{2-6}
~ & \cite{van2014digital} & Discusses forensic-as-a-service for scalable analysis. & \xmark & \xmark & Partially \\
\cline{2-6}
~ & \cite{pichan2015cloud} & Reviews cloud forensics challenges in access and provider dependence. & \xmark & \xmark & \cmark \\
\cline{2-6}
~ & \cite{manral2019systematic} & Organizes cloud forensic artifacts and challenges by investigation stage. & \xmark & \xmark & \cmark \\
\hline

\multirow{9}{*}{\shortstack{IoT and\\ Mobile\\ Forensics}}
& \cite{stoyanova2020survey} & Surveys IoT forensics across devices, gateways, and clouds. & Partially & \xmark & \cmark \\
\cline{2-6}
~ & \cite{yaqoob2019internet} & Reviews IoT forensic taxonomy, requirements, and scalability issues. & Partially & \xmark & \cmark \\
\cline{2-6}
~ & \cite{atlam2020internet} & Surveys IoT forensic requirements, tools, and automation. & Partially & \xmark & \cmark \\
\cline{2-6}
~ & \cite{servida2019iot} & Highlights trace-centric IoT artifacts and their evidentiary value. & Partially & \xmark & \cmark \\
\cline{2-6}
~ & \cite{danev2012physical} & Reviews physical-layer identification for device attribution. & \cmark & Partially & Partially \\
\cline{2-6}
~ & \cite{xu2015device} & Surveys device fingerprinting and its robustness across layers. & \cmark & \cmark & Partially \\
\cline{2-6}
~ & \cite{jagannath2022comprehensive} & Surveys RF fingerprinting methods, datasets, and challenges. & \cmark & Partially & \xmark \\
\cline{2-6}
~ & \cite{mitchell2014survey} & Reviews wireless intrusion detection threats and methods. & \cmark & \xmark & \xmark \\
\cline{2-6}
~ & \cite{van2018survey} & Surveys misbehavior detection in cooperative ITS. & \cmark & \xmark & \xmark \\
\hline

\textbf{This survey} & -- & Systematizes wireless evidence, workflows, methods, and evaluation from an evidence-centric perspective. & \cmark & \cmark & \cmark \\
\hline
\hline
\end{tabular}
\vspace{-0.5cm}
\end{table*}

The rest of this survey is organized as follows. Section 2 provides a unified taxonomy of forensics in wireless communication systems. Section 3 compares the traditional and \textcolor{black}{artificial intelligence (AI)}-based forensics. After that, Section 4 discusses the application of forensics via several cases, Section 5 gives the open challenges and future detections, and followed by Conclusion in Section 6.

\section{Forensics in Wireless Systems: A Unified Taxonomy}
This section adopts a unified taxonomy for wireless forensics, organized from information forensics to physical layer, device layer, network layer, and cross-layer forensics. For each layer, it summarizes the evidence types, supported claims, limitations and failure modes, adversarial manipulation, and corresponding mitigation and cross-checks.

\subsection{Information Forensics}
\label{subsec:info_forensics}
Wireless investigations treat short-lived over-the-air artifacts as strictly verifiable evidentiary objects. Defensible wireless forensics demands rigorous provenance metadata to guarantee independent reproducibility. This explicit metadata encompasses signal acquisition configurations, temporal synchronization references, receiver calibration statuses, and observation geometries~\cite{Stoyanova2020IoTForensics}. Fulfilling these stringent requirements successfully transforms transient RF phenomena into admissible investigative conclusions~\cite{Lyle2022NISTIR8354}. Wireless evidence naturally distributes across over-the-air observables, endpoint artifacts, and infrastructure-side telemetry. Strong investigative conclusions inherently rely on rigorous cross-layer corroboration~\cite{Stoyanova2020IoTForensics}. Physical-layer forensics extracts re-examinable signal observables to anchor incident manifestation claims and enable probabilistic source association. Device-layer forensics explicitly links these wireless-facing phenomena to endpoint operations, configuration evolution, and security-material usage. Network-layer forensics utilizes control-plane and data-plane telemetry to reconstruct session evolution and infrastructure-side identifier dynamics. Cross-layer forensics systematically fuses these heterogeneous artifacts to strengthen transmitter attribution, validate temporal ordering, and strictly bound alternative explanations under documented uncertainty sources~\cite{Lyle2022NISTIR8354}.

\subsection{Physical-layer Forensics}
\label{subsec:phy_forensics}

Physical-layer forensics extracts over-the-air waveforms to reconstruct wireless incidents. Because receiver-induced distortions can overwhelm intrinsic RF fingerprints, raw signal measurements must be tied to clear capture provenance, including receiver settings and sampling parameters, to support defensible re-examination~\cite{han2023smart}.

\emph{\textbf{1) Evidence Types:}}
Physical-layer wireless evidence primarily falls into three distinct categories. The first category comprises waveform and snapshot artifacts including baseband in-phase and quadrature samples. The second category involves channel and measurement artifacts mapping directly to geometric constraints. The third category encompasses hardware-impairment and transmitter-signature artifacts supporting same-origin assessments~\cite{Jagannath2022RFFSurvey}. Concrete studies operationalize the extraction of these physical-layer artifacts. Highlighting waveform artifacts, the study in~\cite{Hilburn2017SigMF} constructs raw waveform recordings combined with explicit capture descriptors. This standardized metadata allows independent examiners to re-estimate synchronization parameters from the same structured fields. Addressing channel artifacts, the authors in~\cite{Halperin2011CSITool} modify the firmware of an Intel 5300 network interface card. This modification exports per-packet subcarrier-level channel state information to user space. Investigating hardware impairments, the work in~\cite{Brik2008Radiometric} extracts transmitter signatures from standard IEEE 802.11 waveforms. The processing estimates carrier frequency offsets and constellation distortion patterns to perform robust device classification.

\emph{\textbf{2) Supported Forensic Claims:}}
Physical-layer evidence supports specialized forensic claims encompassing incident manifestation determining abnormal radio spectrum activities, source association providing probabilistic transmitter attribution under defined propagation assumptions, and spatiotemporal evolution deriving region-valued constraints for event-window reconstruction~\cite{Jagannath2022RFFSurvey}. These verifications inherently demand calibrated scores strictly bounded by acquisition quality and environmental comparability. Concrete studies operationalize these physical-layer claims. Addressing incident manifestation, the framework in~\cite{Arcangeloni2023JammingCausal} analyzes reactive jamming utilizing external radio sensors collecting short-time received-energy traces. Separating mixed signal components via blind source separation quantifies directed influence using all-versus-one transfer-entropy statistics to capture jammer reaction patterns. This explicitly reports detection probabilities and false-alarm trade-offs under severe shadowing and collision conditions. Investigating source association, the research in~\cite{Zhao2024GANRXA} adopts \textcolor{black}{generative adversarial networks (GANs)} to adversarially suppress receiver-identifiable cues while strictly preserving transmitter separability. Applying open-set decision rules rejecting previously unseen emitters on unseen receivers yields physical-layer identity evidence explicitly conditioned on receiver calibration and operating thresholds.

\emph{\textbf{3) Limitations and Adversarial Threats:}}
Physical-layer evidence stability suffers from non-adversarial limitations and active adversarial manipulations. Intrinsic propagation dynamics reshape waveform statistics to produce severe feature drift. Evidence sparsity increases estimation variance and destabilizes learned physical-layer fingerprints. Calibration drift gradually alters observables through temperature fluctuations and equipment aging~\cite{Soltanieh2020Review}. Adversaries actively exploit these physical-layer vulnerabilities through three primary vectors. Active waveform manipulation suppresses discriminative structures while maintaining communication link viability. Learning-pipeline poisoning implants stealthy backdoors into physical-layer classifiers. Signature forgery utilizes synthesized impairment patterns for device impersonation~\cite{Zhang2025Fingerprinting}. These combined factors successfully degrade physical-layer attribution traces without causing obvious network-layer denial of service anomalies~\cite{qu2026secure}. Concrete studies illustrate how these vulnerabilities restrict physical-layer wireless investigations. Investigating calibration drift, the study in~\cite{Koyuncu2022Temperature} trains ResNet50 architectures on physical-layer transient regions at standard room temperature. Testing these models at extreme temperatures ranging from -40 to 80 degrees Celsius reveals severe accuracy degradation. Investigating learning-pipeline poisoning, the authors in~\cite{Tang2024TMCStealthyTrigger} generate stealthy triggers tailored to physical-layer wireless temporal dynamics. This specific attack achieves a 99.2\% success rate while limiting clean data classification degradation to less than 0.6\%. These severe vulnerabilities force investigators to report calibrated confidence regions and explicitly verify training provenance.

\emph{\textbf{4) Mitigations and Cross-checks:}}
Physical-layer mitigations strengthen independent verifiability and bound alternative explanations under adversarial variability. Mitigation strategies primarily include acquisition-chain hardening, uncertainty-aware reporting, and cross-layer state reconciliation~\cite{Xie2021PLASurvey}. Addressing acquisition-chain hardening, the system in~\cite{Merlo2023PicosecondSync} synchronizes distributed receivers using wireless two-way time transfer. A two-step procedure resolves matched-filter ambiguity while frequency locking actively limits hardware drift. This mechanism yields a 2.26 picoseconds timing precision. This extreme precision establishes quantitatively interpretable region-valued constraints for physical-layer multi-vantage agreement tests. Operationalizing uncertainty-aware reporting, the authors in~\cite{Ma2024AIoTAmbiguityCP} introduce a conformal-prediction module wrapping pretrained physical-layer fingerprint classifiers. The system calibrates nonconformity scores on a dedicated held-out set. This mathematically forces ambiguous observations to trigger multiple hypotheses with coverage-controlled validity rather than overconfident single-label attribution. 

\begin{figure*}[t]
    \centering
    \includegraphics[width= 0.9\linewidth]{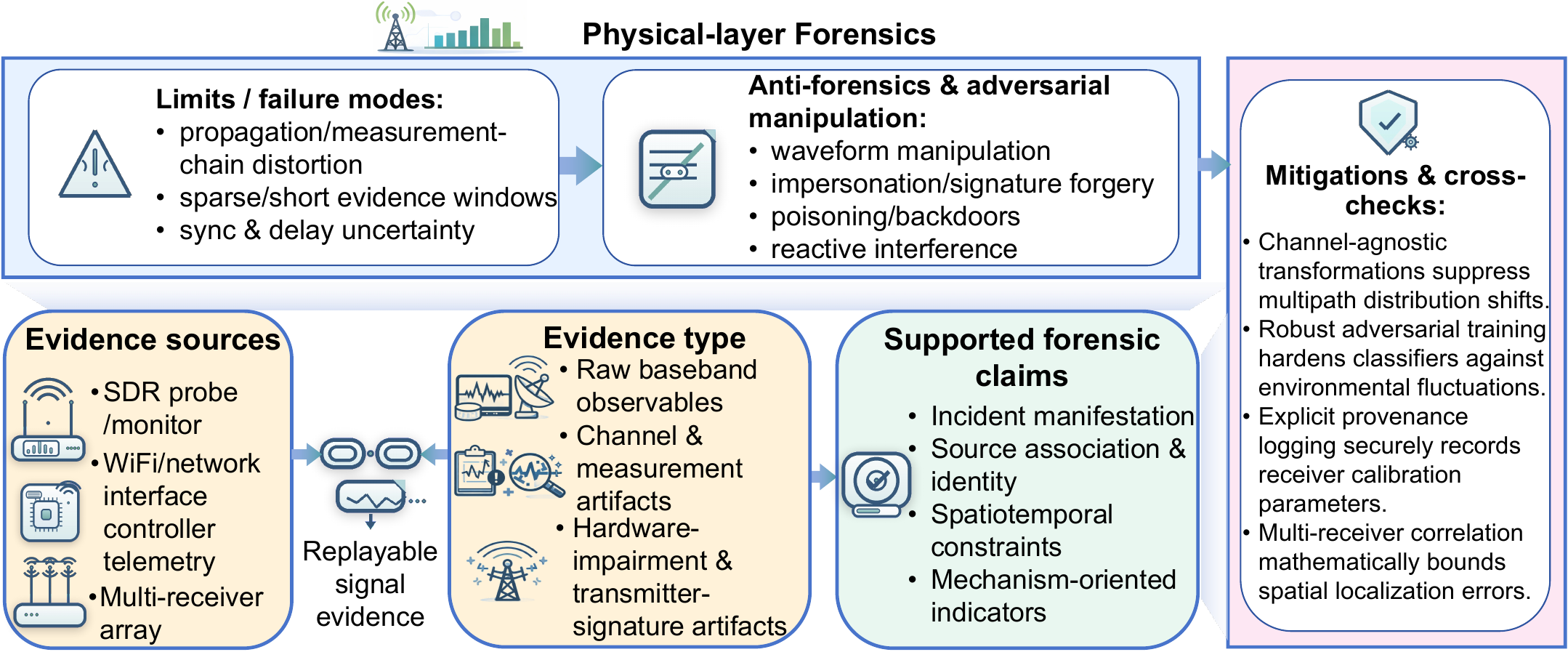}
    \caption{Physical-layer forensics framework from signal evidence to defensible claims. It illustrates provenance-bound signal capture across multiple observation points, and summarizes the supported forensic claims, representative failure modes and adversarial manipulation, and key mitigation principles~\cite{Xie2021PLASurvey}.}
    \label{fig:physical_layer_forensics}
    \vspace{-0.5cm}
\end{figure*}

\subsection{Device-layer Forensics}
\label{subsec:device_forensics}
Device-layer wireless forensics focuses on endpoint components that terminate RF links and execute wireless protocols, including baseband processors, radio interface layers, and so forth. Extracting these artifacts links over-the-air physical-layer phenomena to concrete device-state transitions~\cite{barmpatsalou2018current}.

\emph{\textbf{1) Evidence Types:}}
Device-layer wireless evidence encompasses artifacts generated by the endpoint radio protocol stack. The primary categories involve baseband firmware interfaces and radio interface layers execution states. These artifacts explain cellular \textcolor{black}{networks} and wireless anomalies beyond standard operating system logs. Concrete studies operationalize the extraction of these specialized wireless artifacts. Investigating baseband and firmware interfaces, the work in~\cite{Li2024CCSBaseMirror} demonstrates that radio interface layers binaries can be mined to recover opaque vendor-proprietary command semantics. The authors implement BaseMirror using bidirectional taint analysis seeded from baseband interaction system \textcolor{black}{application programming interface (API)}. The system resolves C++ virtual calls to recover indirect call targets and filters out non-baseband channels via associated system paths. This provides investigators with crucial device-layer explanations for physical-layer cellular behaviors. Highlighting radio interface layers execution states, the work in~\cite{Wen2023RILDefender} demonstrates that inserting monitors at the cellular modem boundary yields actionable evidence about control path abuse. The authors encode wireless attack signatures as configurable rules and parse protocol data units to extract fields such as protocol identifier and data coding scheme. The system correlates this message context to distinguish various cellular attack classes. 

\emph{\textbf{2) Supported Forensic Claims:}}
Device-layer wireless evidence primarily supports three specialized forensic claims. Radio stack behavioral attribution connects network-layer signaling events to explicit device-layer baseband execution states. Hardware-\textcolor{black}{based} identity integrity ensures cryptographic materials and subscriber identity modules remain uncompromised. Baseband protocol conformance validates cellular modem adherence to standard state machine rules against vulnerability-driven deviations. These explicit boundaries critically anchor physical-layer phenomena to verifiable device-layer execution states~\cite{CrackingTheCore2024}. Concrete studies operationalize these device-layer forensic claims using highly specific technical methodologies. Highlighting hardware-backed identity integrity, the survey in~\cite{Menetrey2022SysTEXAttestation} analyzes remote attestation across \textcolor{black}{trusted execution environments (TEEs)}. This process extracts a cryptographic hash of the application code measurement. The device-layer system cryptographically signs these claims to form verifier-checkable evidence. Investigating baseband protocol conformance, the work in~\cite{Kim2019TouchingUntouchables} introduces \textcolor{black}{long term evolution (LTE)}-based approach to dynamically test network-layer control-plane procedures. The authors utilize open-source software to generate crafted stimuli including invalid plain requests and replayed messages. A decision tree classifies device-layer baseband behaviors by tracking radio link failures from device-side logs. One attack vector skips the \textcolor{black}{radio resource control (RRC)} Security Mode Command. This omission forces the baseband to establish a data radio bearer via a plain RRC Connection Reconfiguration.

\emph{\textbf{3) Limitations and Adversarial Threats:}}
Device-layer forensics suffers from severe structural constraints encompassing privileged mediation isolating radio stack artifacts from main application processors, vendor-specific firmware opacity obscuring internal cellular state machines, and critical visibility gaps preventing physical-layer anomalies from propagating into device-layer operating system logs~\cite{Ayers2014NISTSP800101r1}. Adversaries actively exploit these vulnerabilities through log evasion injecting physical-layer binary payloads into subscriber identity modules to completely bypass logging and baseband state desynchronization utilizing crafted network-layer stimuli to force anomalous modem states without triggering application-layer alarms~\cite{Hernandez2022FirmWire}. Concrete analyses illustrate these weaponized constraints. Addressing log evasion, the system in~\cite{Hussain2023NDSSRILDefender} exploits transmitting non-interactive physical-layer binary messages via short message service channels to execute commands directly on subscriber identity modules. The intercept of calls from the Android radio interface layer successfully mitigates 19 distinct attacks and 11 malware samples with only a 1\% hourly battery overhead. Investigating baseband state desynchronization, the analysis in~\cite{Kim2019TouchingUntouchables} utilizes malformed physical-layer paging requests to force device-layer basebands into silently dropping radio resource control connections. Processing these rejections internally without upper-layer operating system notification directly enables completely untraceable denial of service operations.

\emph{\textbf{4) Mitigations and Cross-checks:}}
Device-layer mitigation maintains endpoint evidence availability despite baseband opacity through three primary strategies. Privilege-boundary interception captures communication between the application processor and the cellular modem before baseband obfuscation~\cite{Hussain2023NDSSRILDefender}. Diagnostic telemetry extraction leverages proprietary interfaces to expose physical-layer radio messages to device-layer auditors. Cross-layer state reconciliation validates isolated device-layer baseband narratives against independent network-layer signaling traces. These practices ground forensic conclusions in reproducible wireless artifacts~\cite{ISO27043_2015}. Addressing diagnostic telemetry extraction, the system in~\cite{Li2016MobileInsight} interfaces directly with the device-layer chipset diagnostic mode. The software emulates an external logger within the operating system user space via virtual hardware paths. Investigators utilize this side channel to pull raw hexadecimal logs directly from the baseband interface. Decoding these binary streams extracts granular physical-layer payloads tracking radio resource control and non-access stratum state transitions. This verifiable ground truth prevents attackers from hiding physical-layer manipulations behind opaque device-layer firmware boundaries.

\begin{figure*}[t]
    \centering
    \includegraphics[width= 0.85\linewidth]{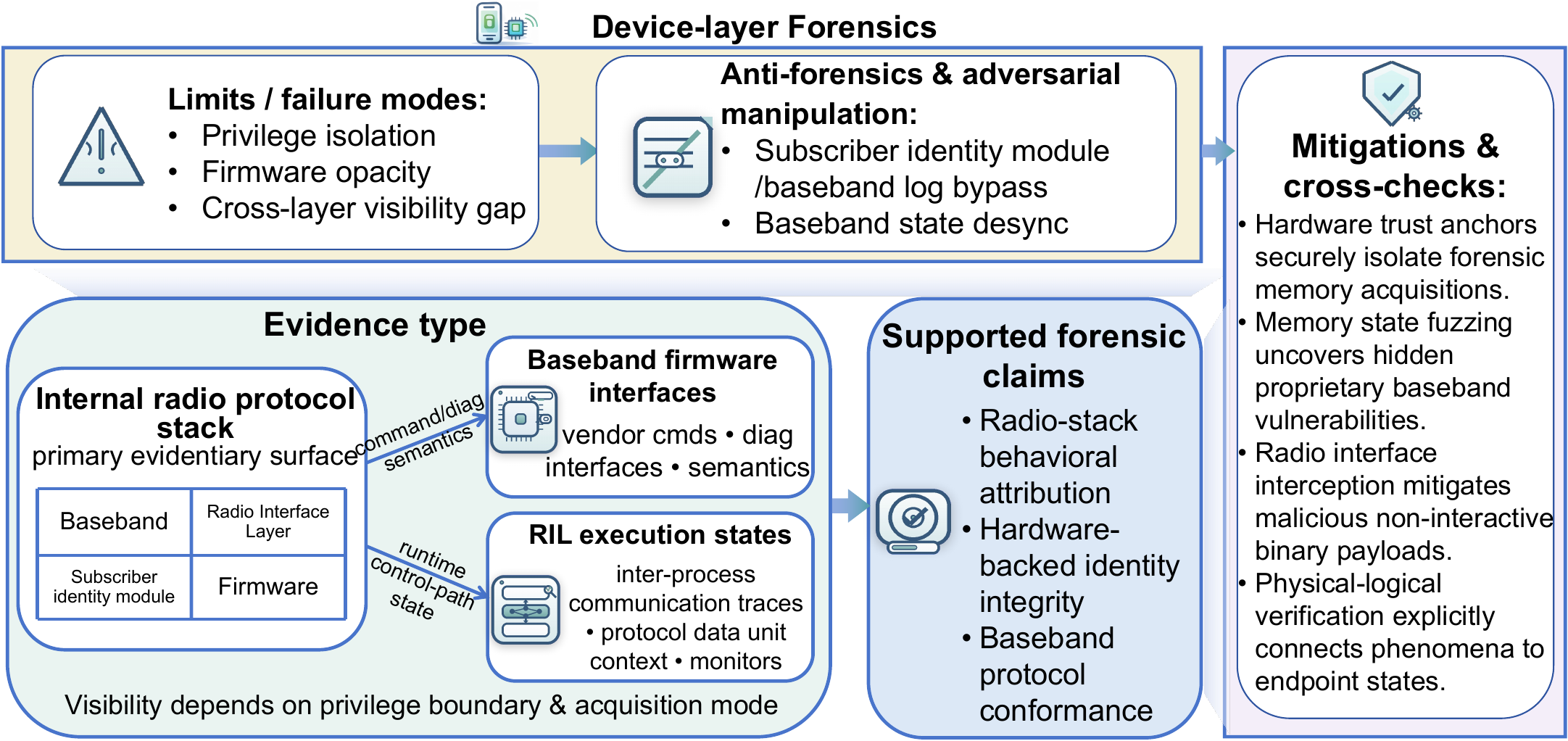}
    \caption{Device-layer forensics framework for radio-stack evidence and defensible claims. The framework depicts endpoint radio-stack evidence under privilege boundaries and acquisition provenance, and summarizes the supported forensic claims, representative limitations, adversarial manipulation, and key mitigation and cross-check measures~\cite{Hussain2023NDSSRILDefender, ISO27043_2015}.}
    \label{fig:device_layer_forensics}
    \vspace{-0.7cm}
\end{figure*}

\subsection{Network-layer Forensics}
\label{subsec:network_forensics}

Network-layer forensics utilizes network-visible traces to reconstruct wireless incidents. Packet, flow, and control-plane observations establish interaction structure even when payload visibility is limited. The evidential value of these traces relies on capture provenance. Vantage points, sampling policies, and timestamping accuracy strictly shape the claim boundary~\cite{Rizvi2022AIForNetworkForensics}. This dependency is critical in modern wireless architectures. 

\emph{\textbf{1) Evidence Types:}}
Network-layer forensics utilizes packet traces, flow records, control-plane logs, and management telemetry to reconstruct wireless incidents. Collection provenance including vantage points, sampling configurations, and transformation histories strictly bounds independent forensic verification~\cite{Lyle2022NISTIR8354}. Concrete case studies operationalize these evidence categories. Addressing packet traces, the system in~\cite{Meng2021TIFS_TIG_DApp} partitions encrypted network-layer traffic into 5-tuple flows to construct traffic interaction graphs. By linking intra-burst packets and inter-burst sequences, a graph neural network classifier trains on decentralized application measurements to support structural reconstruction claims without application-layer payload inspection. Investigating control-plane logs, the framework in~\cite{PROV5GC2024WiSec} instruments \textcolor{black}{fifth generation (5G) }core network functions to capture service-based architecture messages outside transport-layer encryption boundaries. Aggregating these plaintext records builds call flow provenance graphs that explicitly track transformations across subscription permanent identifiers, subscription concealed identifiers, 5G globally unique temporary identifiers, \textcolor{black}{internet protocol (IP) }addresses, and network function identities. This state tracking enables precise cross-entity correlation where passive network-layer monitors fail.

\emph{\textbf{2) Supported Forensic Claims:}}
Network-layer evidence primarily supports four specialized forensic claims encompassing encrypted service identification, flow-signature intrusion detection, virtualized event timeline reconstruction, and control-plane correctness analysis. These verifications require strict provenance data to definitively bind network-layer symptoms to specific device-layer or application-layer activities~\cite{Sheppard2022AIAccessNetworkForensicsSurvey}. Concrete studies operationalize these claims using highly specific methodologies. Addressing intrusion detection, the system in~\cite{Erlacher2022TDSC_IPFIXIDS} compiles Snort rule semantics into predicates over IP flow information export features. Matching exported flow records instead of raw packets increases system throughput from 2.5 Gbit/s to 9.5 Gbit/s while fully preserving signature auditability. Investigating timeline reconstruction in virtualized 5G environments, the framework in~\cite{Abouelkhair2024CNSM5GProvGen} instruments containerized core components to collect operating-system provenance events. Orchestrating these labeled scenarios generates structured provenance graphs to enable replayable cause-effect reconstruction across dispersed network-layer services.

\emph{\textbf{3) Limitations and Adversarial Threats:}}
Network-layer forensics suffers from structural incompleteness via sampled flow exports and limited packet captures, estimation uncertainty requiring strict error boundaries for telemetry metrics, and semantic ambiguity decoupling network-layer routing tuples from application-layer payloads via encryption and address translation~\cite{Hofstede2014FlowMonitoringExplained}. Adversaries weaponize these vulnerabilities through traffic obfuscation manipulating packet directions and timing patterns to defeat flow analysis. Attackers simultaneously utilize learning-pipeline poisoning to inject stealthy triggers into network-layer classifiers~\cite{Muehlberg2025BaseBridge}. Concrete studies illustrate these severe vulnerabilities. Addressing structural incompleteness, the system in~\cite{Estan2004BetterNetflow} utilizes multi-stage filters to identify heavy hitters using constrained \textcolor{black}{static random access memory (SRAM)}. Hashing collisions inherently induce false positive and negative trade-offs during flow promotion. Investigating learning-pipeline poisoning, the research in~\cite{Severi2023Poisoning} embeds explanation-guided backdoors into flow classifiers. Injecting specific packet-length sequences forces models to misclassify malicious flows as benign without impacting clean traffic, explicitly demanding rigorous training provenance verification for model-based network forensics.

\emph{\textbf{4) Mitigations and Cross-checks:}}
Network-layer mitigation maintains evidence availability and interpretability under strict privacy constraints through three primary strategies. Privacy-preserving telemetry export balances network-layer visibility with data minimization at capture points. Uncertainty-aware reporting mathematically forces machine learning inferences to output calibrated confidence intervals. Cross-layer state reconciliation validates isolated network-layer traces against independent control-plane narratives~\cite{Khan2016NetworkForensics}. These strategies collectively ensure that defensible conclusions rest on preserved artifacts with known provenance~\cite{Lyle2022NIST}. Concrete systems illustrate how these mitigation categories strengthen network-layer evidence. Addressing privacy-preserving telemetry, the work in~\cite{Xu2024InfoSensitiveINT} proposes the \textcolor{black}{in-band network telemetry (INT)}-based system. The authors formulate telemetry item selection as a 0-1 Knapsack optimization problem. They implement this logic to selectively export high-value metadata while suppressing sensitive attributes. This approach maintains network-layer monitoring accuracy above 95\% while reducing bandwidth overhead by approximately 40\%. Highlighting uncertainty-aware reporting, the study in~\cite{Talpini2024Trustworthiness} addresses the overconfidence of standard flow classifiers. The system applies Bayesian approximation techniques including Monte Carlo dropout and flipout to quantify epistemic uncertainty in intrusion detection. Establishing a threshold on predictive entropy allows the system to distinguish known attack patterns from out-of-distribution anomalies. Filtering predictions based on these uncertainty scores significantly increases network-layer evidence reliability. 

\begin{figure*}[t]
    \centering
    \includegraphics[width= 0.85\linewidth]{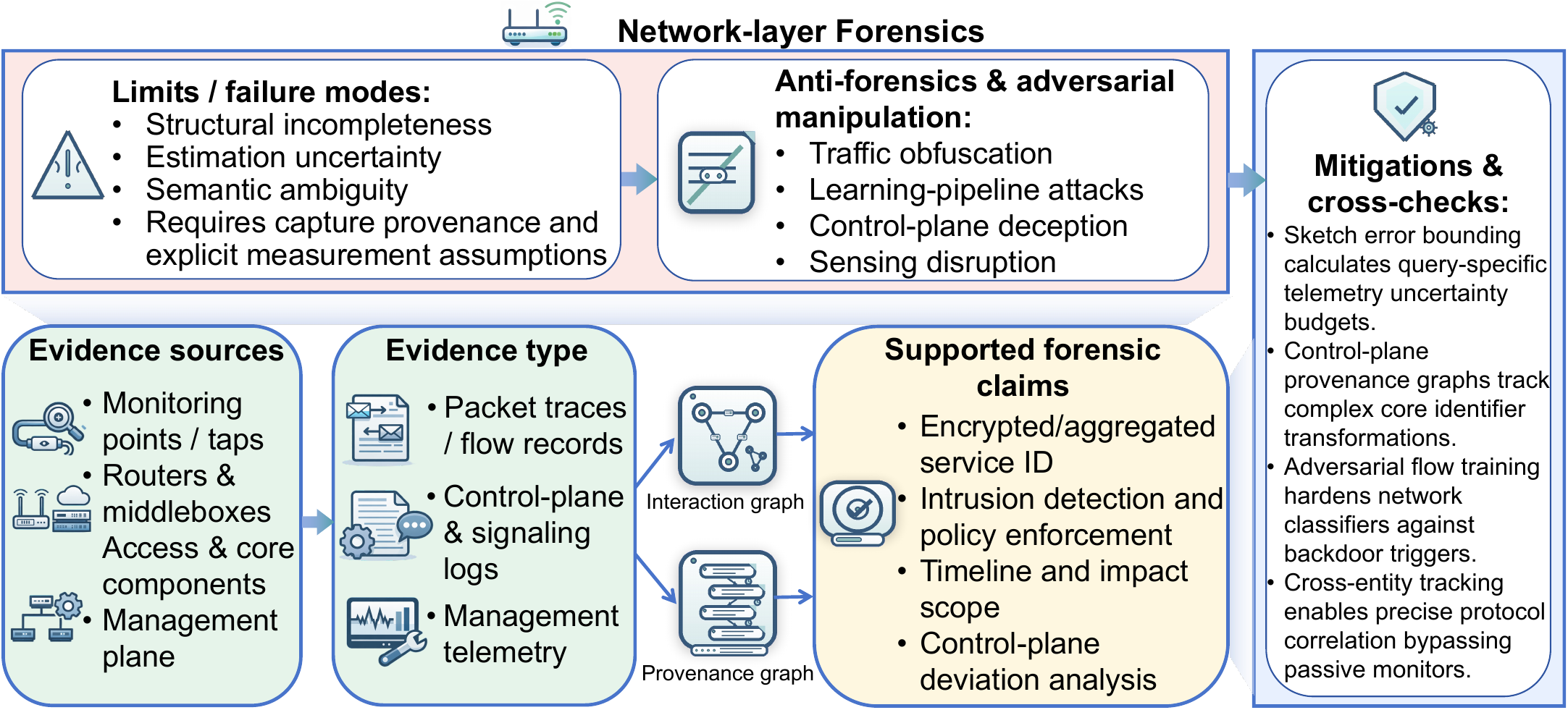}
    \caption{The network-layer forensics framework for provenance-aware traffic and signaling analysis. Depict infrastructure evidence interpreted under explicit capture provenance, and summarize the supported forensic claims, representative failure modes, adversarial manipulation, and mitigation and cross check measures based on provenance preservation, uncertainty aware reporting, and data plane and control plane reconciliation~\cite{Duan2025AdaptiveSurvey}.}
    \label{fig:network_layer_forensics}
    \vspace{-0.7cm}
\end{figure*}

\subsection{Cross-layer Fusion}
Cross-layer wireless forensics combines over-the-air measurements, device-side artifacts, and infrastructure telemetry to reconstruct incidents and support verifiable claims. This is important because evidence from a single layer is often incomplete, whereas cross-layer alignment can strengthen reconstruction by linking identity, timing, and context across heterogeneous logs and measurements~\cite{Xiong2013SecureArray}.

\emph{\textbf{1) Evidence Types and Supported Claims:}}
Cross-layer investigations extract hybrid evidentiary artifacts encompassing behavioral consistency metrics and multi-modal feature vectors~\cite{Yao2021AMI}. These integrated artifacts explicitly support specialized forensic claims encompassing identity-physical binding and operational provenance validation~\cite{Xiong2013SecureArray}. Identity-physical binding strictly verifies claimed device-layer identities against actual physical-layer RF sources. Operational provenance validation confirms network-layer sessions and application-layer payloads originate from physically legitimate transmitters rather than spoofed nodes. Concrete studies operationalize these integrated cross-layer verifications. Addressing identity-physical binding, the system in~\cite{Wang2024BlindTag} dynamically superimposes physical-layer authentication signals as covert tags onto source message waveforms by adaptively adjusting tag parameters according to real-time wireless channel states. Blindly extracting these superimposed features from noisy receptions via hypothesis testing frameworks achieves a 98\% identification rate under a -15 dB signal-to-noise ratio. This explicitly binds physical-layer transmission provenance to claimed device-layer identities without requiring application-layer cryptographic keys. Investigating operational provenance validation, the framework in~\cite{Xie2022HighCompatibility} superimposes physical-layer authentication tags onto device-layer Wi-Fi frame modulation constellations. Utilizing tag-encoding functions minimizes application-layer source message modification ratios while maintaining legitimate wireless link bit error rates within normal tolerances. This transparent integration successfully neutralizes device-layer mimicry attacks without disrupting standard network-layer processing or upper-layer communications~\cite{Zhang2020GaussianTag}.

\emph{\textbf{2) Limitations and Adversarial Threats:}}
Cross-layer forensic reliability suffers from severe decoupling factors encompassing environment-induced feature drift, encryption-induced opacity, and inherent semantic gaps~\cite{Altaibek2025CrossLayerSurvey}. Adversaries actively exploit these vulnerabilities through generative feature fabrication synthesizing physical-layer signatures to match legitimate device-layer identities, environmental reshaping cloning spatial fingerprints, and adversarial evasion forcing fusion model misclassification on malicious network-layer streams to disrupt required inter-layer correlations despite single-layer plausibility~\cite{Adesina2024AdversarialReview}. Concrete studies illustrate these weaponized cross-layer boundaries. Addressing encryption-induced opacity, the analysis in~\cite{Zhou2024TLSAnalysis} demonstrates modern \textcolor{black}{transport layer security (TLS) }1.3 protocols fully encrypting ClientHello messages and obfuscating handshake patterns via zero round trip time session resumption to completely decorrelate physical-layer traffic shapes from application-layer semantics. Investigating generative feature fabrication, the research in~\cite{Xie2024WGANEncoder} evaluates attackers utilizing Wasserstein GANs to mimic legitimate physical-layer phase difference distributions. Calculating anomaly scores from discriminator feature residual errors and image reconstruction errors successfully detects synthetic spoofs perfectly aligning with legitimate device-layer training data.

\emph{\textbf{3) Mitigations and Cross-checks:}}
Cross-layer mitigation counters decoupling threats through active challenge-response verification injecting probes to elicit unforgeable physical-layer responses, robust adversarial training hardening fusion models against gradient-based perturbations, and semantics-aware traffic recovery mining statistical flow correlations to reconstruct encrypted application-layer contexts~\cite{Duan2025AdaptiveSurvey}. Concrete studies operationalize these cross-layer defenses. Addressing active verification, the protocol in~\cite{Piana2025DroneCR} utilizes zero-sum game strategies commanding device-layer drones to traverse specific waypoints. Matching measured physical-layer channel gain statistics against unique location multipath profiles successfully establishes authenticity. This multi-round spatial challenge forces attackers to solve intractable inverse problems to completely neutralize mimicry attacks. Combating adversarial manipulation, the framework in~\cite{Lyu2025PFRTF} integrates phase flipping data augmentation to simulate diverse physical-layer signal distortions. Employing soft label training strategies to smooth decision boundaries directly improves classification accuracy by 18.12\% against optimization-based Carlini-Wagner attacks.

\subsection{Lessons Learned}
\label{subsec:chap2_lessons_learned}

Synthesizing these evidentiary dimensions reveals fundamental challenges in bridging severe cross-layer semantic gaps. Physical-layer investigations provide hardware-grounded truth bypassing upper-layer network spoofing. This fundamental advantage remains severely fragile under environmental mobility requiring picosecond-level synchronization and extensive provenance metadata~\cite{ Xin2024HighPrecision}. Device-layer mitigations expose stealthy baseband desynchronization attacks utilizing hardware-backed trust anchors. This deep visibility faces extreme vendor dependence requiring reverse-engineered diagnostic interfaces instead of standardized \textcolor{black}{forensic telemetry interfaces}~\cite{Zhang2025SCRUTINIZER}. Network-layer synthesis successfully contextualizes raw baseband samples into actionable protocol signaling sequences. Payload encryption inherently restricts this network visibility strictly to control-plane metadata~\cite{Xin2024HighPrecision}. Producing defensible and replayable wireless evidence ultimately demands overcoming opaque hardware processing delays to guarantee strict cross-layer temporal alignment.

\section{Wireless Forensics Workflow: Traditional vs. AI-based Methods}

\subsection{Forensic Readiness \& Preservation-by-design}

\emph{\textbf{1) Traditional Approaches:}} \textcolor{black}{ Traditional forensic readiness and preservation-by-design aim to ensure that a wireless system can produce evidence that is preservable, attributable, and reproducible~\cite{sadineni2021ready}. For example, the authors in~\cite{lee2021k} introduce a readiness-as-a-service view, which organizes readiness into three process groups and further refines them into 11 processes covering policy design, forensic classification, asset identification, and so forth. \textcolor{black}{They propose} centralized forensic logging, conversion of backups into logical image files, chained hash-based integrity protection, and a \textcolor{black}{capability maturity model integration (CMMI)}-style evaluation model for continuous improvement. Besides, another direction strengthens preservation infrastructure by securing how evidence is stored and accessed. In~\cite{singh2022secure}, the authors propose a digital-forensic-readiness storage model that combines \textcolor{black}{representational state transfer (REST)}-based ingestion, metadata sanitization, integrity hashing, encryption, randomized filenames, read-only access, and \textcolor{black}{two-factor authentication (2FA)}-protected download to reduce tampering and unauthorized access. 
In~\cite{jimenez2024filtering}, the authors study how to trigger forensic evidence collection in software-defined networks by integrating an evolved digital forensics and incident response architecture with an unusual-traffic detector and an unexpected-behavior detector. }

\emph{\textbf{2) AI-based Approaches:}} Traditional triggers often rely on static intrusion detection system (IDS) alerts which are prone to high false-positive rates in dynamic 5G/6G environments. Learned triggers utilize \textcolor{black}{machine learning (ML)} and signal processing to recognize subtle behavioral or physical anomalies as precursors to an attack. The study in~\cite{defenderdeng} uses \textcolor{black}{Channel State Information (CSI) }as a passive sensor and highlights the AI part through data-driven anomaly recognition and motion-pattern discrimination for radio-silent drones. Specifically, it combines Gaussian mixture model-based anomaly detection with \textcolor{black}{short-time Fourier Transform (STFT)}-based time-frequency analysis, template matching for drone-specific shifting–moving patterns, and fixed-frequency enhancement for propeller-spinning signatures, thereby turning raw CSI fluctuations into an intelligent physical-layer trigger. Experimentally, the \textcolor{black}{power spectrum (PS) and time-frequency distribution (TFD)} modules achieve 96.8\% and 93.5\% true positive rates even at 10 m, while the overall system reaches 95.65\% at 5 m and still maintains 60\% at 7 m. Beyond physical sensing, system-level intelligence is required to manage the trade-off between forensic visibility and resource consumption. The study in~\cite{LiveBoxYu} introduces an AI-based autonomic forensic-readiness design for drones, where a loop dynamically adjusts reporting interval and geolocation precision according to contextual risk. By combining blockchain-based tamper-proof storage with a prediction-driven, locality-sensitive approximation mechanism, it preserves verification correctness while reducing communication cost. In a case with 100 simulated flight paths, the method achieves up to 46\% reduction in transmitted geolocation digits with zero verification error, using a 15 s sampling baseline. 
Moreover, integrating forensic intelligence into wireless infrastructure allows for a more centralized and proactive readiness posture. The study in~\cite{PalmeseWi-Fi} details the design of a forensic-ready Wi-Fi access point (AP) to secure heterogeneous IoT ecosystems. In this architecture, the AP serves as a centralized coordinator for evidence provenance, analyzing cross-device traffic to identify coordinated botnet behaviors. By offloading the learned trigger logic to the infrastructure layer, the framework ensures the capture of distributed attack traces that individual, resource-constrained IoT nodes are unable to record independently.

Besides using AI to detect and interpret forensic-relevant events, another emerging direction is adaptive retention, where AI is used to transition from rigid periodic deletion to dynamic, value-aware lifecycle management. From terminal-side, resource-constrained nodes such as IoT devices and edge servers often struggle with the storage of comprehensive forensic logs. The study in~\cite{Rizvi2024Pushing} addresses these storage bottlenecks by pushing lightweight attack-detection models directly onto resource-constrained IoT/edge devices, so that forensic readiness is triggered locally through real-time classification rather than cloud-side analysis. Specifically, the proposed framework evaluates 9 ML/\textcolor{black}{deep learning(DL)} models for on-device multiclass attack recognition and then uses the detected attack type to drive evidence collection and preservation. The results show that AI can remain both accurate and lightweight in this setting, with multiclass accuracy reaching 99.60\% on CICIoT2023 dataset and 99.98\% on IoT-23 dataset, highlighting that the main contribution lies in resource-aware on-edge intelligence for forensic readiness. From network perspective, a more specialized advancement involves integrating intelligence within the network infrastructure to manage evidence persistence across various layers. For instance, on the access layer, the authors in~\cite{Palmese2025Resource} propose to extract traffic features over tunable aggregation windows and feeds them to pre-trained machine learning classifiers, while jointly optimizing the window length, the number of features, and the quantization bits under storage and \textcolor{black}{central processing unit (CPU)} constraints. The framework models forensic accuracy, storage cost, and computational load, and then solves a nonlinear optimization problem to allocate resources across device identification, human activity recognition from encrypted traffic, and smart-speaker interaction detection. Experiments on a Raspberry Pi 3B+ show that even under a 1 KB/s storage limit and a 10\% CPU budget, the system still maintains 84.68\% minimum accuracy with 30 active tasks. Such work shows that the main value of AI in adaptive retention lies not only in later classification, but also in deciding how much traffic evidence should be preserved, at what granularity, and for which forensic task under tight edge-side constraints.

{\arrayrulecolor{black}\color{black}

\begin{table}[tpb]
\centering
\caption{Representative studies on forensic readiness and preservation-by-design.}
\vspace{-0.25cm}
\label{tab:readiness_preservation}
\tiny
\renewcommand{\arraystretch}{1}
\setlength{\tabcolsep}{3pt}
\begin{tabular}{m{1.45cm}||c|>{\centering\arraybackslash}m{1.8cm}|>{\raggedright\arraybackslash}m{4.4cm}|>{\raggedright\arraybackslash}m{3.1cm}}
\hline
\hline
\rowcolor{gray!15}
\textbf{Category} & \textbf{Ref.} & \textbf{Focus} & \textbf{Core idea} & \textbf{Takeaway} \\
\hline

\multirow{3}{*}{\shortstack{Traditional\\ approaches}}
& \cite{lee2021k}
& \shortstack{Readiness\\ workflow}
& Readiness-as-a-service with policy design, forensic classification, asset identification, centralized logging, and chained integrity protection.
& Systematic readiness modeling, but limited adaptivity to dynamic wireless environments. \\
\cline{2-5}

~ & \cite{singh2022secure}
& \shortstack{Secure\\ preservation}
& REST-based evidence ingestion with sanitization, hashing, encryption, randomized filenames, read-only access, and 2FA.
& Strong storage-side protection with low idle overhead, but large-file latency remains noticeable. \\
\cline{2-5}

~ & \cite{jimenez2024filtering}
& \shortstack{Evidence\\ triage}
& SDN-based triggered evidence collection using unusual-traffic and unexpected-behavior detectors.
& Avoids store-everything logging and achieves about 97.3\% accuracy/F1, but depends on detector quality. \\
\hline

\multirow{5}{*}{\shortstack{AI-based\\ approaches}}
& \cite{defenderdeng}
& \shortstack{Physical-layer\\ trigger}
& CSI-based anomaly recognition with GMM/EM, STFT, and drone-pattern matching for radio-silent drone detection.
& Learns subtle physical-layer triggers; detection is strong at short range but degrades with distance. \\
\cline{2-5}

~ & \cite{LiveBoxYu}
& \shortstack{Autonomic\\ readiness}
& Risk-aware adjustment of reporting interval and geolocation precision, combined with tamper-resistant storage.
& Reduces reporting cost by up to 46\% with zero verification error, but stronger integrity lowers throughput. \\
\cline{2-5}

~ & \cite{PalmeseWi-Fi}
& \shortstack{Infrastructure-\\ side readiness}
& A forensic-ready Wi-Fi AP centrally coordinates provenance and cross-device traffic analysis in heterogeneous IoT.
& Captures distributed attack traces beyond individual IoT nodes, but is mainly AP-centric. \\
\cline{2-5}

~ & \cite{Rizvi2024Pushing}
& \shortstack{On-device\\ retention trigger}
& Lightweight ML/DL models run on IoT/edge devices to trigger local evidence collection after attack recognition.
& Enables accurate and lightweight edge-side readiness, but depends on model generalization. \\
\cline{2-5}

~ & \cite{Palmese2025Resource}
& \shortstack{Adaptive\\ retention}
& Joint optimization of aggregation window, feature number, and quantization bits under storage and CPU constraints.
& Supports value-aware evidence persistence under tight edge budgets, at the cost of higher optimization complexity. \\
\hline
\hline
\end{tabular}
\vspace{-0.45cm}
\end{table}
}

\subsection{Acquisition}
\textcolor{black}{ Evidence acquisition links physical wireless incidents to digital forensic reconstruction by collecting heterogeneous traces with integrity and provenance preserved. This section reviews sensor infrastructure, deterministic collection policies, and emerging AI-driven context-aware capture.}

\emph{\textbf{1) Tools and Placements:}} At the extraction stage, wireless forensics relies on specialized tools to access measurements that are normally hidden by commodity radio stacks. For instance, In Wi-Fi systems, the Intel 5300 CSI tool first exposed per-packet channel state through modified firmware, but only for 30 grouped subcarriers over 20/40 MHz channels, which limits delay and multipath resolution~\cite{halperin2011tool}. After that, Nexmon pushes this deeper into Broadcom devices by using mechanisms such as flashpatch-based \textcolor{black}{read-only memory (ROM)} redirection, thereby enabling monitor mode, raw frame injection, and programmable firmware behavior on commodity phones and IoT hardware~\cite{gringoli2019free}. In cellular networks, low-level measurements can also be extracted from commodity smartphones. The study in~\cite{Li2016MobileInsight} \textcolor{black}{achieves} this by using the modem’s diagnostic interface to collect raw binary logs and decode fine-grained wireless messages, while incurring modest overhead, with runtime cost of about 1–7\% CPU and 30 MB memory. Beyond capability, reliable evidence collection depends on how tools are placed. For instance, the authors in~\cite{xu2019redundant} model channel- dependent capture probability explicitly and shows that assigning multiple sniffers to the same channel can improve packet-capture reliability under fading, hardware failure, and other imperfect- monitoring conditions. \textcolor{black}{For wireless local area network (WLAN)}, the authors in~\cite{sheth2006mojo} shows that sniffer placement should be optimized where faults are likely to occur, rather than only where AP coverage exists. This is because anomalies such as hidden terminals, often need multiple correlated viewpoints for reliable diagnosis. In its design, sniffers are placed near clients and configured to report physical-layer summaries, such as noise-floor samples, to a central inference engine every 10 s for joint analysis.

\emph{\textbf{2) Policy-based Acquisition:}} While sensors provide the physical capability for evidence capture, acquisition policies define the operational logic governing their activation and depth. Traditional forensic readiness models formalize collection procedures as repeat- able service functions based on predefined trigger rules~\cite{lee2021k}. For instance, authors in~\cite{sadineni2021ready} propose a trigger-based provenance collection pipeline for low-power IoT networks, where suspicious link-layer conditions activate selective evidence logging instead of continuous full-trace retention. It is implemented on a 16-node network with 1 \textcolor{black}{user datagram protoco (UDP)} server, 15 clients, and 4 hopping channels, and is evaluated under 4 jamming variants and 3 synchronization attacks. Once triggered, the system 
stores the resulting graphs, and derives a Link-IoT dataset with 8 raw and 15 derived features for later analysis. To mitigate storage exhaustion in high-speed links, policy-based designs often employ tiered sampling strategies that balance storage footprints with reconstruction accuracy~\cite{chen2013efficient}. Building on this idea, the authors in~\cite{monteiro2023adaptive} extend tiered retention to microservice observability by adaptively switching among metrics, logs, and traces according to forensic value. They use a game-theoretic model of malicious users and forensic-ready microservices, together with a runtime \textcolor{black}{monitor-analyze-plan-execute (MAPE)} loop, to decide when higher-fidelity evidence is worth preserving for subsequent event analysis. On the 41-microservice TrainTicket benchmark~\cite{zhou2018poster}, across 9 uncertainty scenarios with 10\%–90\% malicious users, the approach improves F-measure by 26.97\%–42.50\% over sampling observability and by 3.1\%–38.96\% over full observability. This result shows that policy-based acquisition can move beyond static sampling and instead preserve high-fidelity evidence. 

\emph{\textbf{3) AI-based Acquisition: Intelligent Execution and Active Interaction}}
AI-based acquisition extends wireless forensics beyond static, rule-based thresholds toward adaptive and context-aware evidence collection. Rather than relying on fixed-rate recording, learning-based execution dynamically allocates sensing effort according to evidentiary value, observation uncertainty, and signal ambiguity, thereby improving both acquisition efficiency and forensic fidelity.

\textit{Confidence-aware and Uncertainty-quantified Sampling:} Unlike traditional multi-tier sampling policies, AI-driven acquisition adjusts sampling density according to uncertainty estimates. Prior work on uncertainty-aware wireless sensing and Bayesian data collection shows that physical variations can be mapped into confidence-aware acquisition policies~\cite{wang2025uncertainty,ruah2023bayesian}. \textcolor{black}{In forensic settings, this enables adaptive escalation from low-overhead routine monitoring to evidentiary-grade capture: systems such as uncertainty-based monitoring frameworks continuously assess the confidence of extracted physical-layer artifacts, triggering full-resolution raw \textcolor{black}{in-phase and quadrature (I/Q)} capture and denser cross-layer logging when hardware fingerprints become unreliable under fading or adversarial obfuscation, while permitting sparse sampling when confidence remains high ~\cite{tian2025cats,stahlke2024uncertainty}.}

\textit{Active Sensing for Evidence Disambiguation:} AI-based acquisition can also move beyond passive observation toward active interaction with the wireless environment. This is particularly important when ambiguous phenomena, such as impulsive attacks and environmental interference, cannot be reliably separated from passive traces alone~\cite{landa2022wip}. Emerging 5G/6G capabilities, including \textcolor{black}{integrated sensing and communication (ISAC) and reconfigurable intelligent surface (RIS)}, allow forensic agents to probe the environment through controlled pilots, adaptive beamforming, and reflection-path reconfiguration. Recent studies show that such active sensing can expose otherwise hidden threats, including passive eavesdroppers, by eliciting scenario-specific responses from suspicious nodes~\cite{wen2025exploring,yu2025learning,he2023sencom}.

\textit{AI-driven Evidence Reconstruction and Super-resolution:} When hardware limits, packet loss, or severe channel impairments fragment the evidence trail, AI models can help reconstruct incomplete observations and enhance low-resolution measurements. Building on CSI reconstruction and super-resolution methods, \textcolor{black}{recent work shows that generative and multitask learning models, e.g., Transformer,} can recover missing channel fingerprints, upscale sensing data, and restore degraded electromagnetic leakage signals~\cite{wen2018deep,jin2025channel,wang2023super,shen2023deep,nam2024data}. Such outputs can improve downstream analysis when continuous high-resolution capture is infeasible, but they should remain clearly identified as derived rather than original evidence.

\textit{The Privacy--Visibility Paradox:} At the operational level, improving evidentiary fidelity often conflicts with privacy regulation. Fine-grained protocol telemetry and raw physical-layer captures may inadvertently expose sensitive payloads, identifiers, or location information, creating a privacy--visibility paradox for forensic-ready networks. Future acquisition architectures should therefore embed privacy-by-design mechanisms, \textcolor{black}{such as edge-side signal separation, zero-knowledge telemetry proofs, and real-time payload sanitization  \cite{Xu2024InfoSensitiveINT,yang2024generative},} so that actionable forensic observables can be preserved without retaining raw user data.

\begin{figure*}[t]
    \centering
    \includegraphics[width= 0.85\linewidth]{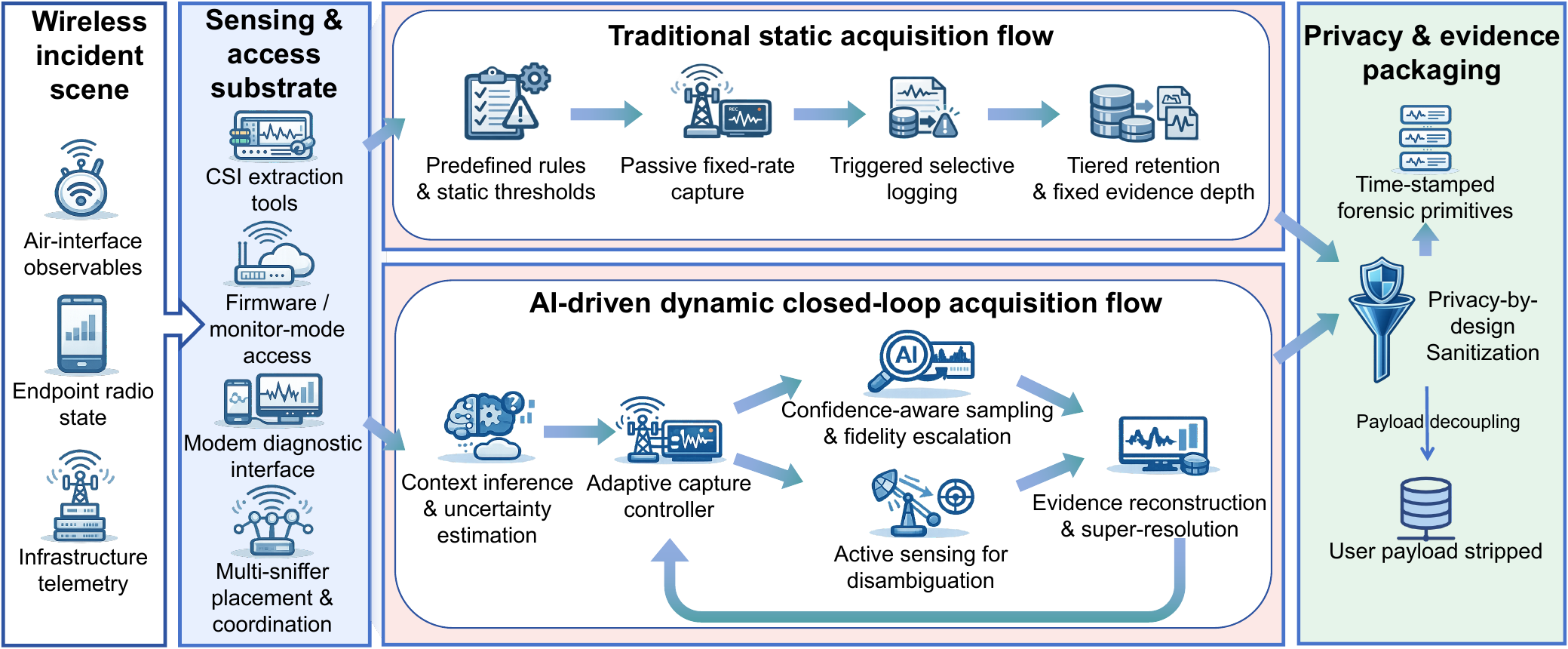}
    \caption{The acquisition framework for provenance-aware wireless evidence collection. It organizes a shared wireless observation substrate for both static rule-based acquisition and AI-driven closed-loop acquisition, and summarizes key operations including selective triggering, confidence-aware sampling, active sensing for disambiguation, and evidence reconstruction~\cite{vanini2024clock,xiong2015tonetrack}.}
    \label{fig:fig_3_2}
    \vspace{-0.5cm}
\end{figure*}

\subsection{Correlation \& Analysis}
Acquisition improves visibility, but the resulting traces remain fragmented across sensors, protocol layers, and timescales. Therefore, correlation and analysis can form the analytical core of wireless forensics, which can transform heterogeneous observations into defensible causal claims by aligning evidence across sources, disentangling device-dependent effects from channel distortion, and connecting low-level measurements to high-level incident narratives.

\textcolor{black}{\textbf{\textit{\textcolor{black}{1) Evidence Alignment and Synchronization:}}} Correlation begins by mapping heterogeneous captures into a common temporal, spectral, and logical reference frame. This is more difficult in wireless environments than in host-centric forensics because clocks drift across \textcolor{black}{software-defined radios (SDRs)}, monitors, devices, and cloud-native functions, while identifiers change across mobility, protocol transitions, encryption boundaries, and address translations. Hence, defensible cross-source analysis requires not only timestamp normalization, but also preservation of synchronization metadata, including clock sources, drift compensation, frequency-offset correction, and transformation history. For example, the study in \cite{hilburn2018sigmf} illustrates the above principle by pairing raw RF recordings with machine-readable metadata about timing, sampling, hardware context, and annotations, while the work in ~\cite{vanini2024clock} show that event reconstruction must rely on explicit time anchors rather than assumed clock correctness.}

\textcolor{black}{At the physical layer, alignment is typically performed at sample, frame, or burst granularity. Independent receivers must be reconciled through timing-offset estimation, carrier-frequency correction, and landmark matching before they can be treated as observations of the same transmission. The authors in ~\cite{Brik2008Radiometric} demonstrate this in radiometric identification through burst synchronization and offset compensation, while the study in \cite{xiong2015tonetrack} shows that receptions from multiple access points can be associated through time-of-arrival processing and cross-receiver matching. At higher layers, control-plane messages, telemetry, and protocol records must likewise be normalized and anchored to preserve causal order. For example, the work in ~\cite{Li2016MobileInsight} exposes fine-grained cellular signaling on commodity smartphones, and the study in ~\cite{pacherkar2024prov5gc} further reconstructs 5G core activity as provenance graphs while explicitly tracking identity transformations across IP addresses and network functions.}


\textcolor{black}{\textbf{\textit{\textcolor{black}{2) Rule-based Correlation and Signal Disentanglement:}}} Once alignment is established, traditional correlation reconstructs incident structure through deterministic association, statistical testing, and model-based interpretation.} 
\textcolor{black}{For example, signal disentanglement can be achieved through correlation at the physical layer, the observed waveforms often combine transmitter impairments with fading, interference, mobility, and receiver artifacts. The study in ~\cite{alshawabka2020exposing} shows that RF fingerprinting degrades sharply in realistic channels, illustrating how device-specific signatures become entangled with propagation and noise.  Therefore, classical signal processing remains central: synchronization refinement, matched filtering, blind source separation, sparse decomposition, spectral segmentation, and parameter estimation are used to isolate evidentiary components from environmental variability. For instance, the authors in ~\cite{Brik2008Radiometric} extract radiometric signatures through burst detection, alignment, and feature extraction, and the authors in ~\cite{yan2022rrf} further emphasize the need to account for channel diversity in practical RF identification. Beyond detection, these methods also recover interpretable intermediate artifacts, such as delay estimates, occupancy masks, and impairment descriptors, that can be cross-checked with higher-layer traces.}

\textcolor{black}{However, deterministic methods become fragile under encrypted services, identifier churn, mobility, missing logs, and large-scale heterogeneity. Public cellular datasets and measurement toolchains remain limited~\cite{amini2024where}, while operational cellular traffic is often encrypted and integrity protected~\cite{tan2022cellulariot}, reducing the completeness of cross-layer reconstruction. Signal models are also vulnerable to replay-style impersonation and channel mismatch~\cite{danev2010attacks,alshawabka2020exposing}. In this regard, traditional correlation can remain a defensible baseline, but not a sufficient solution for dynamic and partially observed wireless environments.}

\textcolor{black}{\textbf{\textit{\textcolor{black}{3) AI-assisted Analysis and Causal Inference:}}} AI-assisted analysis extends wireless forensics beyond explicit rule matching toward learned representation, multimodal fusion, and forensic hypothesis ranking. Rather than manually defining correspondences across sensors and protocol layers, learning-based models project heterogeneous evidence into shared latent spaces, enabling stronger association of temporally adjacent, behaviorally consistent, or causally related observations.} 
\textcolor{black}{For instance, beyond correlation, AI can also support causal reconstruction by helping explain mechanism, ordering, and competing alternatives rather than merely assigning anomaly scores. One example is that the authors in ~\cite{yen2022gnnrca} use graph neural networks for root-cause analysis over multivariate network \textcolor{black}{key performance indicators (KPI)}, allowing hidden dependencies among network elements to be inferred directly. Similarly, the authors in ~\cite{wang2022multiroot} show that learned models can rank multiple concurrent root causes in the specific environment settings, while the study in ~\cite{bajpai2024anomgraphadv} uses temporal graph networks to capture spatiotemporal dependencies in wireless traces for anomaly and intrusion analysis. In this sense, AI is most valuable not as a final judge, but as an auditable tool for ranking plausible forensic explanations.}

\textcolor{black}{Additionally, AI can mitigate partial observability by imputing missing traces, denoising corrupted captures, and fusing incomplete evidence windows. Taking an example, the authors in ~\cite{huang2023diffar} use diffusion-based modeling to repair missing CSI and improve downstream inference. However, such derived outputs must remain explicitly separated from original evidence. Learned models may otherwise rely on dataset artifacts, receiver-specific shortcuts, or site-dependent bias rather than mechanism-relevant signals. Channel shift alone may degrade RF fingerprinting performance~\cite{alshawabka2020exposing}, and recent work further demonstrates adversarial, backdoor, and practical attack vulnerabilities in learned wireless signal analysis and RF identification systems~\cite{kim2022channelaware,ma2023whitebox,zhao2025backdoor,li2024practicalwifi}. In this regard, AI should be incorporated only with provenance tracking, uncertainty reporting, and preservation of intermediate outputs for independent review.}

\begin{figure*}[t]
    \centering
    \includegraphics[width= 0.85\linewidth]{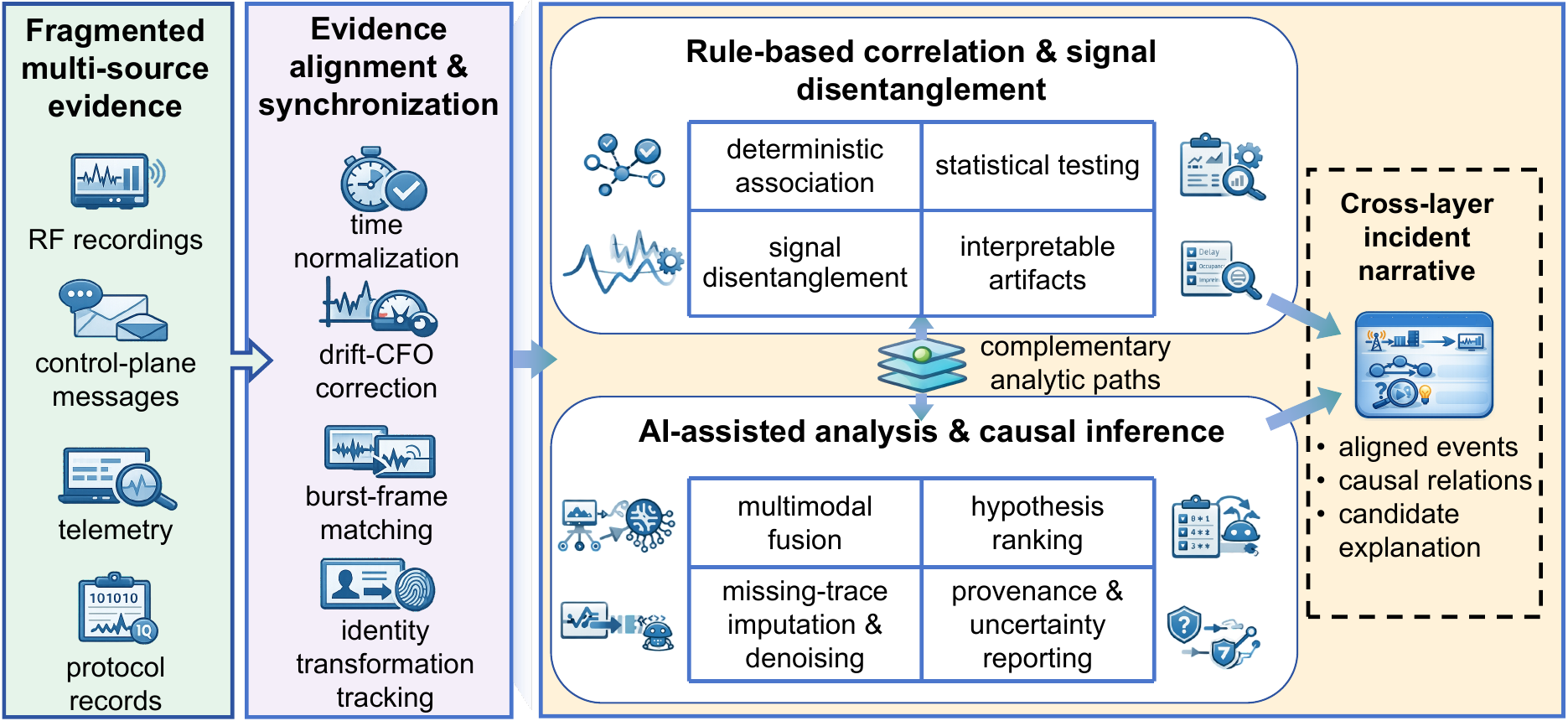}
    \caption{The correlation-and-analysis workflow for turning fragmented wireless traces into defensible cross-layer incident narratives. Depict multi-source evidence aligned in a common temporal, spectral, and logical frame, and summarize how rule-based correlation and signal disentanglement, together with AI-assisted fusion, hypothesis ranking, and trace recovery, support cross-source association, causal reconstruction, and incident interpretation with provenance and uncertainty awareness~\cite{yen2022gnnrca}.}
    \label{fig:fig_3_3}
    \vspace{-0.7cm}
\end{figure*}

\subsection{Reporting \& Reproducibility}
\textcolor{black}{The value of wireless forensics depends not only on whether evidence is captured and analyzed, but also on whether conclusions can be communicated, challenged, and independently replayed. Because wireless claims often depend on synchronization quality, calibration state, observation coverage, preprocessing choices, and model assumptions, reporting must preserve the linkage from raw evidence to intermediate transformations and final conclusions, while clearly separating direct observation from probabilistic inference and analyst interpretation. Hence, reporting and reproducibility are the mechanisms through which wireless forensic results become reviewable, transferable, and, where necessary, admissible.}

\textcolor{black}{\textbf{\textit{\textcolor{black}{1) Calibrated Reporting and Evidence Packaging:}}} Traditional forensic reporting emphasizes explicit claim-to-evidence linkage and preservation of acquisition context; in wireless settings, this discipline must be strengthened to account for environmental uncertainty and cross-layer ambiguity. The works in ~\cite{garfinkel2012dfxml,casey2017case,hilburn2018sigmf} have illustrated this principle in digital and RF settings by pairing evidence with provenance, tool, and metadata records. Therefore, a defensible wireless report should state not only what was observed, but also how it was captured, under which assumptions it was interpreted, and which alternatives remain unresolved. Capture provenance, sensor placement, synchronization status, calibration state, preprocessing steps, encryption-related visibility limits, and privilege constraints should all be treated as first-class reporting elements.}

\textcolor{black}{A practical evidence package should contain three layers: preserved raw evidence, condensed analytical views, and explicit uncertainty information. To be specific, raw evidence may include I/Q traces, CSI matrices, packet captures, baseband logs, and control-plane records, with \textcolor{black}{signal metadata format (SigMF)} serving as a portable substrate for machine-readable RF preservation~\cite{hilburn2018sigmf}. In addition, as illustrated by works in \cite{pacherkar2024prov5gc,tabiban2022vincidecoder}, analytical views such as timelines, provenance graphs, localization regions, and protocol-state summaries support triage without discarding causal structure. Finally, uncertainty information should further disclose confidence bounds, missing observations, coverage limits, and rejected hypotheses; this is consistent with both confidence calibration concerns in modern learning systems and time-anchor-based reasoning about timestamp validity~\cite{guo2017calibration,vanini2024clock}. In summary, such layering allows reviewers to examine the same case at different abstraction levels without conflating model outputs or narrative summaries with equally strong forms of evidence.}

\textcolor{black}{\textbf{\textit{\textcolor{black}{2) Reproducible Pipelines and Re-execution Artifacts:}}} Reproducibility requires preserving not only the data, but also the computational pathway that transformed it into conclusions. In wireless forensics, this includes software and firmware versions, feature-extraction code, model checkpoints, parameter settings, random seeds, calibration files, schema mappings, synchronization procedures, and transformation logs. Even with identical captures, small differences in preprocessing, decoder options, or toolchains may materially change attribution outcomes. Prior work on RF fingerprinting likewise shows that performance is inseparable from collection, preprocessing, training, and deployment choices~\cite{kuzdeba2022systems,alhazbi2024rffdl}. Therefore, the analytical pipeline should be preserved as a versioned forensic artifact rather than treated as invisible infrastructure.}

\textcolor{black}{Machine-readable manifests and replayable execution bundles are practical mechanisms for this goal. Manifests can enumerate sensor identities, clock references, observation geometry, decoding parameters, model hashes, and intermediate outputs, while containerized runtimes, executable notebooks, and scripted replay packages allow third parties to regenerate figures, rerun fusion pipelines, and test sensitivity to thresholds or alignment assumptions. For example, SigMF provides such metadata structure for RF recordings, and platforms such as POWDER, Colosseum, and Sionna illustrate replayable wireless experimentation under controlled conditions~\cite{hilburn2018sigmf,breen2021powder,bonati2021colosseum,hoydis2022sionna}. For AI-assisted workflows especially, such re-execution artifacts are essential, which is consistent with the broader emphasis of datasheets for datasets and model cards on documenting dataset lineage, evaluation conditions, and model behavior~\cite{gebru2021datasheets,mitchell2019modelcards}.}

\textcolor{black}{\textbf{\textit{\textcolor{black}{3) AI-assisted Reporting and Human Review:}}} As wireless forensic pipelines increasingly incorporate learned models, reporting must distinguish among observed measurements, extracted features, inferred associations, and final human-endorsed conclusions. An anomaly score, class label, or causal ranking should never be presented as self-sufficient evidence. Instead, reports should explain why the preferred interpretation is more credible than competing alternatives and disclose the calibration limits of the underlying model. This requirement is supported both by evidence that neural confidence scores are often miscalibrated and by documentation frameworks arguing that model outputs must be contextualized rather than reported in isolation~\cite{guo2017calibration,mitchell2019modelcards}. Supporting materials such as saliency maps, attention traces, feature attribution, and counterfactual comparisons may assist interpretation, but only when their provenance and uncertainty are explicitly bounded~\cite{xu2024explainability,zhao2024interpretable}.}

\textcolor{black}{According to the above analysis, human review remains indispensable. Analysts must determine whether a model relied on mechanism-relevant evidence or on incidental correlates tied to vendor, site, or collection procedure. This is particularly important in RF fingerprinting, where prior studies show that failures often originate from data collection, preprocessing, and deployment mismatch rather than model architecture alone~\cite{kuzdeba2022systems,alhazbi2024rffdl}. Reports should thus preserve analyst annotations, hypothesis revisions, and adjudication decisions alongside automated outputs, maintaining an auditable boundary between machine suggestion and forensic conclusion. In high-stakes settings, AI-assisted reporting should not replace expert judgment, but make the relationship among preserved evidence, learned inference, and human reasoning open to independent scrutiny~\cite{tabiban2022vincidecoder,mitchell2019modelcards}.}

\begin{figure*}[t]
    \centering
    \includegraphics[width= 0.85\linewidth]{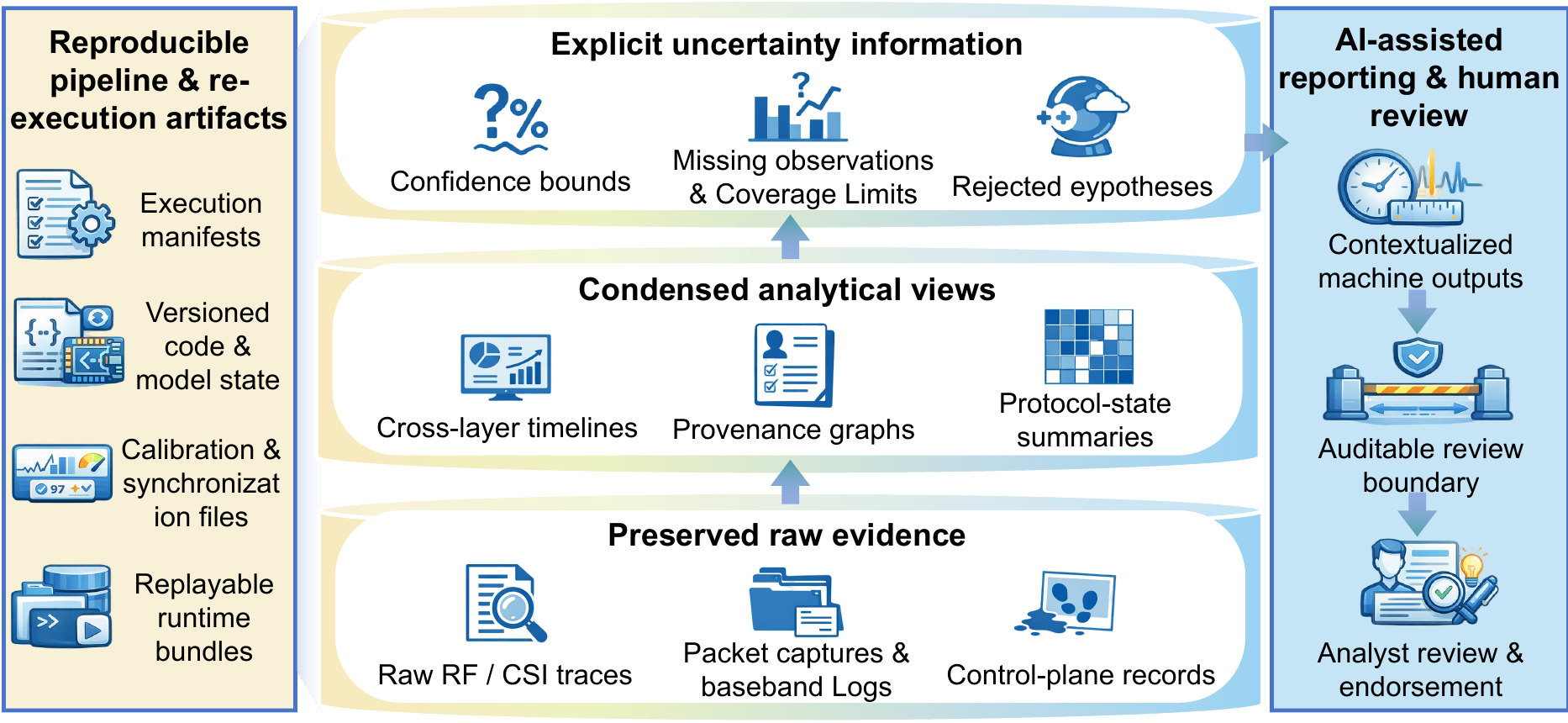}
    \caption{The framework for turning analyzed wireless evidence into reviewable and replayable forensic conclusions. It summarizes a versioned computational pathway and how calibrated reporting, replayable execution bundles. It also presents the AI-assisted reporting with an auditable machine-to-human review boundary support independent scrutiny, transferable interpretation, and human-endorsed forensic conclusions~\cite{alhazbi2024rffdl}.}
    \label{fig:fig_3_4}
    \vspace{-0.5cm}
\end{figure*}

\subsection{Lessons Learned}
Based on the above presented content, the comparison between traditional and AI-based methods shows that wireless forensics should evolve toward a hybrid workflow rather than treating the two paradigms as substitutes. Traditional methods remain indispensable because they provide explicit evidence-preservation logic, interpretable correlation paths, and reproducible reporting structures; \textcolor{black}{for example, forensic-readiness models in \cite{lee2021k} explicitly define evidence preservation and management procedures, and provenance-graph-based analysis in \cite{pacherkar2024prov5gc} supports auditable cross-layer correlation and structured attribution.} However, their effectiveness is often limited by partial observability, encrypted traffic, identifier churn, mobility, and large-scale heterogeneity. AI-based methods improve this situation by enabling adaptive triggering, uncertainty-aware acquisition, multimodal fusion, causal hypothesis ranking, and evidence reconstruction under resource and visibility constraints; \textcolor{black}{for example, the work in \cite{defenderdeng} uses CSI-driven learning to proactively detect radio-silent drones, and the study in \cite{LiveBoxYu} dynamically adapts evidence collection for forensic-ready drone services under resource constraints. However, AI also introduces new risks, including dataset bias, poor calibration, adversarial manipulation, and reduced interpretability \cite{Rizvi2022AIForNetworkForensics,guo2017calibration}.} Therefore, future wireless forensics should combine the procedural rigor of traditional methods with the adaptivity of AI-based methods; the former provides the foundation for trust, provenance, and reproducibility, while the latter enhances efficiency, resilience, and analytical reach under complex wireless conditions.

\section{Forensic Applications}
\textcolor{black}{This section analyzes the application of forensics in different wireless domains. At the same time, we provide some use cases to further demonstrate the specific use of forensics.}

\subsection{Detection \& Anomaly Discovery }
Detection and anomaly discovery is typically the first always-on forensic function in a wireless system. Beyond raising alarms, it must operate under limited storage/visibility by deciding \emph{what} evidence to preserve and \emph{where} to investigate next. Concretely, it should localize anomalies to suspicious time windows, frequency regions, cells, or links, and retain sufficient artifacts so later stages can reproduce the finding and test alternative explanations.

\emph{\textbf{1) Radio and spectrum based anomaly discovery:}} Many anomalies are visible only at the radio interface (e.g., jamming, unexpected emitters, abnormal spectral occupancy). Since exhaustive labeling is impractical, unsupervised or weakly supervised learning is widely used to learn normality and flag deviations~\cite{rajendran2019unsupervised}. A shared feature space can make heterogeneous sensors comparable, for example, in~\cite{rajendran2019crowdsourced}, the authors build a common feature space across sensors and identify anomalies as outliers, with a closed loop that incorporates expert feedback. Technically, a global encoder with sensor-specific decoders forms an adversarial-autoencoder–based detector, producing anomaly scores from reconstruction/discriminator losses. To make results actionable and shareable, they perform interactive clustering on learned features and then apply semi-supervised outlier detection with label propagation, separating sparse-but-normal events from frequent-but-illegal events using limited feedback. Evaluations on 4 sensors over 56 days show gains over baselines such as one class \textcolor{black}{support vector machine (SVM)}, and also indicate thresholding remains difficult without a semi-supervised phase.
Similarly, reconstruction-based detectors can be implemented directly on spectrogram evidence. In~\cite{zhou2021radio}, received signals are converted to short-time Fourier Transform spectrogram images, then a modified GAN augmented with an encoder enables spectrogram reconstruction. Anomalies are scored using a weighted combination of pixel-level reconstruction error and discriminator loss, with the threshold derived from normal training scores. Simulations show consistent gains over~\cite{rajendran2019unsupervised}, improving detection sensitivity by up to 10 dB. 

\emph{\textbf{2) KPI based anomaly discovery for service accountability:}} 
In modern wireless systems, many incidents first appear as \emph{service} anomalies (e.g., throughput drops, handover failure bursts, control plane instability) rather than raw spectral changes~\cite{alves2023machine}. Thus, forensic anomaly discovery for service accountability often relies on multi KPI time series collected by base stations (e.g., reference signal received power, access, scheduling, link robustness)~\cite{ahasan2022supervised}. Since these KPIs are high-dimensional and strongly influenced by external factors, modeling periodicity and context is critical. For example,~\cite{huang2022cellular} proposes an unsupervised detector combining time-series decomposition with an \textcolor{black}{long short-term memory (LSTM)}-based GAN. It removes seasonal components to train an LSTM-GAN on normal-only data so the generator captures normal KPI dynamics, and at inference performs an inverse search to find generator inputs that best reproduce an observed KPI window; the reconstruction residual is used as the anomaly score. On the CellPAD KPI dataset, it outperforms ARIMA and TCN-AE, achieving F1/AUPRC of 0.94/0.99 on sudden-drop point anomalies and 0.91/0.96 on segment anomalies.


\emph{\textbf{3) Rogue infrastructure and identity anomalies:}}
Rogue infrastructure denotes adversarial radios impersonating legitimate entities (e.g., fake base stations, deceptive access points) to extract identifiers or enable follow-on manipulation. Forensic anomaly discovery must convert these short-lived interactions into interpretable evidences (e.g., trace segments) that can be corroborated across layers and stakeholders. For example, protocol- and trace-level artifacts enable end-user detection. In~\cite{mubasshir2025gotta}, an on-device learning framework detects fake base stations and multi-step attacks from cellular protocol traces: first, a packet-level stateful LSTM with attention flags suspicious packets by modeling long-range sequence dependencies; Then, the traces are represented as directed graphs of packet transitions, and graph-based matching/learning identifies reshaped or previously unseen attacks. It reports 96\% detection accuracy with 2.96\% false positives, 86\% accuracy on unseen variants, and a lightweight footprint of about 835 KB memory with power below 2 mW. 
Moreover, cross-user aggregation and external consistency checks strengthen rogue-infrastructure evidence. For instance, the crowdsourced system in~\cite{li2017fbs} has \textcolor{black}{user equipments (UEs) }upload lightweight message/context logs, then the backend correlates evidence using complementary high-precision detectors: unusually strong serving-cell \textcolor{black}{received signal strength indicator (RSSI)}, invalid \textcolor{black}{base station (BS)} identifier syntax, an \textcolor{black}{short message service (SMS)}-content bag-of-words SVM, and BS–WiFi inconsistency checks (inferring location from nearby WiFi and comparing with carrier coverage). It unions these detectors for conservative \textcolor{black}{fake base station (FBS)} attribution and performs localization, reporting 4.7\% suspicious messages attributable to the FBS and median localization accuracy of 11 m. 


\subsection{Attribution \& Identification}
Infrastructure-side attribution aims to determine which service, slice, or application class was responsible for the observed wireless behavior using only infrastructure traces, such as radio access network (RAN) controllers~\cite{riyaz2018deep}. The goal is to produce labels that support accountability, triage, and later reconstruction when payload inspection is infeasible due to privacy constraints.

\emph{\textbf{1) Infrastructure-side attribution in edge networks:}} 
A first line of work performs attribution close to the radio edge by mapping RAN telemetry to service labels that can be logged and replayed. In~\cite{groen2024tractor}, the framework closes the loop from traffic/slice classification to physical resource block optimization, avoiding packet-level inspection. It releases an Open-RAN-compliant KPI dataset (about 2.83 GB from 447 minutes of real 5G traffic) and trains slice classifiers on sliding windows that stack consecutive KPI snapshots; each snapshot contains 17 KPI features and a new decision is produced every 250 ms. The classifier reaches up to 99\% offline classification accuracy and up to 92\% online classification accuracy for specific slices, enabling near-real-time UE reassignment across slices. 
A second line attributes responsibility using core/edge visibility into flows, service functions, and protocol metadata. 
Taking an example, for encrypted user-plane traffic,~\cite{huoh2022flow} models each bi-directional flow as a graph: packets are nodes with raw byte sequence features, edges encode packet order and inter-arrival time, and flow-level statistics are added as global attributes. A Graph Nets–style message-passing pipeline avoids fixed-length padding and achieves 97.56\% accuracy on the ISCXVPN2016 VPN subset, outperforming 96.00\% raw-bytes-only and 95\% fixed-length \textcolor{black}{convolutional neural network (CNN)}/attention-LSTM baselines. 

\emph{\textbf{2) Signal-native device identification:}}
Signal-native identification links a transmission to a specific user device using signal-level artifacts when higher-layer identifiers are missing, spoofed, or encrypted. A representative approach treats raw I/Q (or closely related low-level bytes) as the primary evidence object. In~\cite{akbari2021look}, an encrypted web traffic classification pipeline keeps only ClientHello/ServerHello handshake header bytes while masking fields (e.g., cipher info) to avoid server-name lookup effects. Each flow is encoded as a three-channel time-series so hundreds of packets are modeled without exploding input size, and standard flow statistics from a flow meter are appended. These features are processed by a tripartite network and dense layers with strong dropout, achieving about 95.6\% accuracy/F1, outperforming 91\% of CNN/CNN-LSTM baselines trained on raw TLS bytes.
Beyond general radios, identification at scale appears in back-scatter systems. 
In practical scenarios, because large labeled datasets are rarely available per device and signatures drift across days or power cycles, robustness hinges on transfer and label efficiency. The study in~\cite{li2022radionet} shows that preprocessing choices directly affect performance and applies adversarial domain adaptation to improve cross-day device classification accuracy from 8.41\% to 43.17\% on one dataset and from 25.98\% to 65.24\% on another, reducing recollection cost. 


\subsection{Provenance \& Localization}
While infrastructure-side attribution labels the responsible service or traffic class from edge-visible traces, provenance and localization emphasize the \emph{spatial} evidence chain: each location/trajectory claim must be tied to underlying wireless measurements so results remain reproducible and defensible. In wireless environments, location evidence is often inferred from noisy, transient RSS and CSI observations collected across distributed receivers~\cite{kotaru2015spotfi}. Thus, this section focuses on extracting auditable spatial evidence to localize the transmitter/source area, reconstruct trajectories, and report confidence regions suitable for dispute resolution.

\emph{\textbf{1) Provenance-aware localization and source finding:}}
Provenance-aware localization and source finding seeks where a suspicious transmission originated while keeping the source claim verifiable from preserved evidence. For example,~\cite{mitchell2022deep} rasterizes \textcolor{black}{received signal strength (RSS)}, samples into a grid and applies a \textcolor{black}{U-shaped network (U-Net) }to output a continuous likelihood heatmap over candidate transmitter positions. It then thresholds the heatmap and uses connected components to produce discrete estimates with explicit missed-detection and false-alarm accounting. Reported localization error ranges from 0.7 m to 12.4 m, with runtime about 1--14 ms, fitting preserve-then-reconstruct workflows.
Complementary to purely data-driven inference, \cite{nardin2023jamming} augments a standard path-loss model with a lightweight learned correction to keep localization interpretable while adapting to mismatch. Using about 10,000 received-power observations over a 1 km$^2$ area and only the 15 strongest observations, the augmented model can approach the Cramér--Rao bound in the path-loss scenario, offering a reference for evidentiary soundness when only partial high-value measurements are preserved.

\emph{\textbf{2) Trajectory reconstruction with confidence regions:}}
Beyond localization, trajectory reconstruction must describe how the target moves and how certain each step is, so investigators can report a path with defensible confidence regions. A practical approach is to treat confidence as a first-class output and propagate it over time. In~\cite{tedeschini2024real}, a cooperative tracking pipeline uses a teacher-student Bayesian neural network to predict location and uncertainty, and injects learned epistemic uncertainty into the multi base station tracking update via the likelihood function. This down-weights out-of-distribution channel conditions and sparse training regions, achieving 46 cm median error and $<1$ m error in 87\% of cases, outperforming temporal CNN baselines. Reported epistemic uncertainty is about 10 cm in dense regions, rises to about 50 cm in sparse regions, and exceeds 2 m in extreme low-data areas, enabling admissible uncertainty bounds for trajectory reporting.
Moreover, for lightweight uncertainty quantification,~\cite{sadr2021uncertainty} applies Monte Carlo dropout to deep learning based mmWave \textcolor{black}{multiple-input multiple-output (MIMO)} localization as a low-complexity approximation to Bayesian neural network inference, explicitly producing confidence interval bounds. It runs many stochastic forward passes to obtain predictive mean/variance and constructs a confidence ellipse scaled by the chi square distribution. Simulations use a 28-GHz urban ray-tracing model, a beamforming fingerprint codebook with $M=32$, and a dense $401\times401$ grid; uncertainty uses 1,000 Monte Carlo dropout samples per position with dropout 0.2. Empirically, ellipse shape depends on maximum received power, becoming eccentric in low-power regions and tightening at higher power, explaining why some trajectory segments are inherently less defensible despite plausible point estimates. Overall, trajectory reconstruction is more forensic-friendly when each update includes a calibrated confidence region and the system flags when that region becomes unreliable.

\emph{\textbf{3) Case study:}} In indoor non-line-of-sight (NLoS) investigations, 
the forensic objective is to convert time-stamped CSI into defensible spatiotemporal evidence of where a non-cooperative moving target was likely located within a time window~\cite{wang2023through}. 
To support provenance-aware localization, the method first stabilizes evidence formation and then derives physical parameters. A dedicated reference channel captures the direct signal and suppresses interference and phase errors in surveillance channels, thereby revealing target-induced reflections. Next, a packet-aggregation two-dimensional matrix pencil method jointly estimates path-length-change-rate and absolute time of flight (ToF) under low \textcolor{black}{signal-to-noise ratio (SNR)}. Presence is detected from changes in the ToF distribution, and localization is obtained through geometric reasoning: multi-antenna ToF measurements form ellipse constraints whose intersection determines the target position.
The prototype is built on IEEE 802.11ac. A directional-antenna transmitter operates on channel 161 (5.805 GHz) with 80 MHz bandwidth at 600 Hz, while a Broadcom 4366C0 receiver extracts CSI using Nexmon~\cite{schulz2017nexmon}. The setup includes one directional reference channel and two omni surveillance channels, with offline MATLAB processing on a workstation. Reported timing shows that detection and localization can be updated at about one-second granularity in the tested configuration.

\begin{figure*}[htbp]
\centering
\subfloat[]{%
  \includegraphics[width=0.28\textwidth]{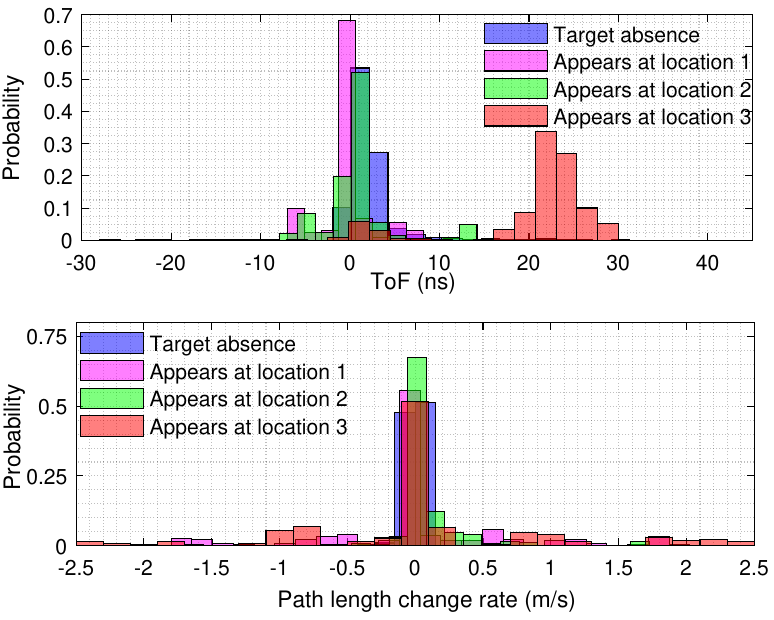}%
}\hfill
\subfloat[]{%
  \includegraphics[width=0.28\textwidth]{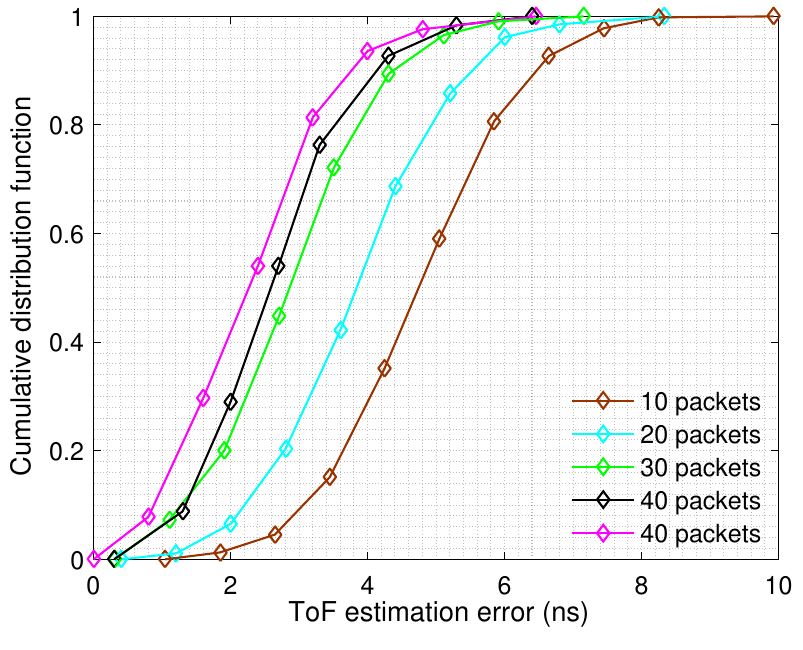}%
}\hfill
\subfloat[]{%
  \includegraphics[width=0.28\textwidth]{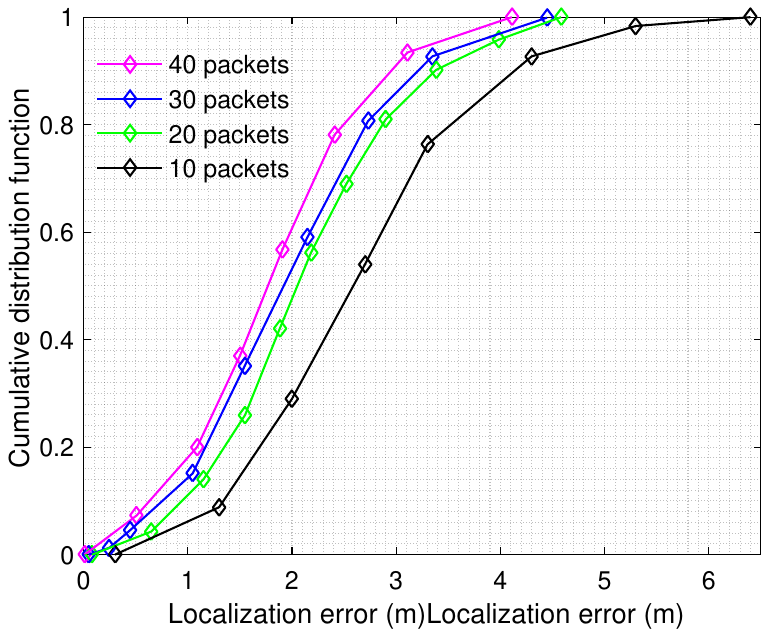}%
}
\vspace{-0.3cm}
\caption{The results of the case for provenance-aware detection and localization in wireless forensics \cite{wang2023through}. (a) Parameter distribution comparison. (b) The ToF estimation accuracy. (c) The localization accuracy.}
\vspace{-0.5cm}
\label{C111}
\end{figure*}

The brick wall scenario in Fig.~\ref{C111} provides several forensic indicators. Fig.~\ref{C111}a shows measurable shifts in estimated ToF and path-length-change-rate between the empty room and different target locations, indicating that CSI logs contain usable NLoS evidence after proper preprocessing. Fig.~\ref{C111}b shows that ToF estimation remains effective and improves as more packets are aggregated, confirming higher evidence fidelity under low SNR. Fig.~\ref{C111}c further shows localization improvement with 10, 20, 30, and 40 packets, yielding median errors of 2.53 m, 2.05 m, 1.90 m, and 1.78 m, respectively. Additional results in~\cite{wang2023through} show that narrower bandwidth degrades localization, whereas more packets improve it, making the tradeoff among measurement cost, update rate, and uncertainty explicit.
Overall, this case study highlights a practical forensic pattern for provenance-aware localization: reliable location claims require controlled evidence formation, physically meaningful intermediate parameters such as ToF, and a geometric reporting process that can be replayed.

\subsection{Authenticity and Anti-forgery}
Authenticity and anti-forgery are prerequisites for using wireless traces as defensible forensic evidence. Authenticity asks whether preserved evidence can be credibly bound to a claimed transmitter, time window, and operating context, while anti-forgery asks whether an adversary could manufacture or alter those evidences to create a misleading narrative without being detected~\cite{jiang2013rejecting}. This becomes critical when encryption blocks payload inspection and evidence is increasingly signal-native and distributed, motivating physical-layer tags/watermarks for validation.

\emph{\textbf{1) Authentication at the signal level:}}
A common method is to embed an authentication tag into the physical waveform so verification remains possible even under payload encryption~\cite{zhang2023tag}. In tag-based physical-layer authentication, the transmitter superimposes a structured tag and the receiver verifies it using shared secrets and statistical tests, producing a lightweight authenticity hook that can be preserved for later forensic checking~\cite{hoang2024physical}. For example, \cite{xie2022physical} proposes encoded tag-based physical-layer authentication for single block (ET-SB) and multiple blocks (ET-MB): ET-SB uses the polarity relationship of two sequential tag symbols to build a binary mapping vector that decides whether each message symbol is tagged or unchanged, while ET-MB extends the mapping across multiple tag blocks to improve robustness at low SNR while increasing compatibility. Receiver verification is formalized as Neyman--Pearson hypothesis testing~\cite{grunwald2024beyond}, with an optimal test statistic tailored to partial superimposition (unlike classic statistics that assume tagging everywhere). They report compatibility improvements of 15.63\% and 23.45\%, and security gains of 6 dB and 12 dB at SNR 10 dB versus the baseline that superimposes the tag on the entire modulated message block.
However, a practical barrier is that some tag-based designs
require broadcasting tag parameters or tuning them empirically,
which can create overhead and leakage opportunities. 
Compared with the baseline using channel characteristics instead of distributing keys to generate authentication tags~\cite{an2021tag}, the overhead gap grows from 0.25 KBytes to 4.75 KBytes and the latency gap from 0.46 ms to 7.64 ms as message transmissions increase. Third, \cite{zhang2023tag} leverages the RIS as an authenticity actuator: RIS configurations deliberately reshape the channel into a verifiable pattern. Results are reported as \textcolor{black}{receiver operating characteristic (ROC) }curves for false-alarm probabilities from $10^{-1}$ to $10^{-4}$; when false-alarm is pushed below $10^{-3}$, detection probability can drop below 0.92, which is useful for forensics because admissible authenticity claims are tied to an explicit operating point.

\emph{\textbf{2) Replay-resistant validation for wireless evidence:}} Replay can forge convincing wireless evidence without synthesizing signals, so replay resistance is essential for forensic validation. For instance,~\cite{tomasin2022challenge} proposes a challenge response physical layer authentication scheme over partially controllable channels. The receiver uses a controllable propagation component, such as RIS, to issue a physical challenge by randomly selecting a new RIS configuration before transmission. A packet is accepted only if the CSI estimated from it matches the expected channel state information under that configuration, so a previously recorded waveform becomes invalid after reconfiguration. With an RIS of 80/100 elements, they apply a likelihood-ratio test and evaluate misdetection at target false-alarm probabilities from $10^{-4}$ to $10^{-2}$, showing that stronger RIS randomization reduces spoofing success but also lowers spectral efficiency. They report that reducing spectral efficiency by about 3 bits/s/Hz can push misdetection to roughly $10^{-5}$--$10^{-3}$, while the no-randomization baseline stays close to one under strong attacker assumptions.

\begin{figure*}[htbp]
\centering
\subfloat[]{%
  \includegraphics[width=3.5cm]{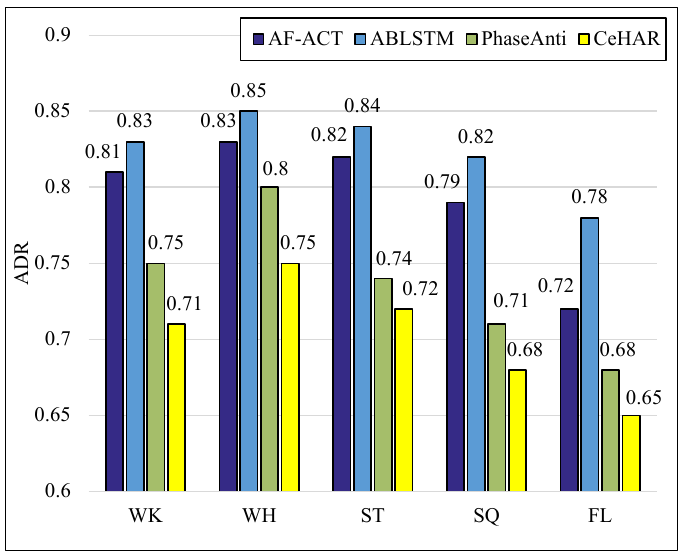}%
} \hspace{1.5cm}
\subfloat[]{%
  \includegraphics[width=3.5cm]{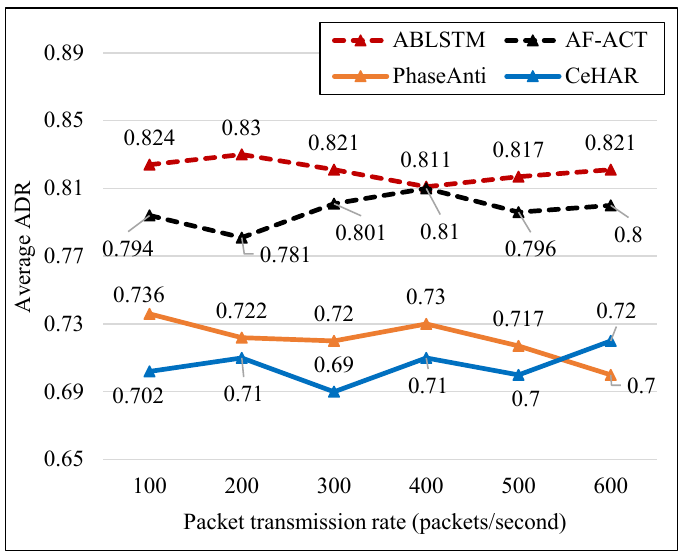}%
}
\vspace{-0.3cm}
\caption{Results of the case for authenticity and anti-forgery in wireless forensics \cite{wang2025generative}. (a) Accuracy degradation rate of 5 activities for different systems. WK indicates walking, WH is waving hand, ST is sitting, SQ means squatting, and FL represents falling. (b) The impact of communication speed on ADR performance.}
\vspace{-0.2cm}
\label{444}
\end{figure*}

\emph{\textbf{3) Case Study:}}
In ISAC systems, CSI from routine communications enables sensing but also exposes physical environments to illegitimate nodes, creating an authenticity and anti-forgery tension: legitimate sensing must remain reliable/auditable while unauthorized receivers may conduct covert surveillance or manipulate outcomes. A practical authenticity carrier is the pilot. Instead of a standardized pilot, the transmitter forges it by embedding extra side information (a safeguarding signal)~\cite{wang2025generative}.
Accordingly, the method forges a new pilot by generating a protection signal and modulating it onto the pilot. It includes three modules: (i) a discrete conditional diffusion model generates an activation graph for ISAC link/node selection conditioned on device deployment and user location to reduce sensing network cost; (ii) a continuous conditional diffusion model generates a safeguarding signal conditioned on the activated links/nodes and the legitimate node; (iii) the safeguarding signal is embedded into the pilot amplitude to mask user-induced CSI fluctuations. 
Further speaking, the evaluation uses a software-defined-radio testbed (USRP N321 with external clock and GNU Radio Tx/Rx chains) in two indoor scenarios (office room for training/evaluation and meeting room for evaluation). Using accuracy degradation rate on AF-ACT~\cite{zhang2022csi}, ABLSTM~\cite{chen2018wifi}, PhaseAnti~\cite{huang2020towards}, and CeHAR~\cite{lu2022cehar} over five actions, Fig.~\ref{444}(a) reports average \textcolor{black}{accuracy degradation rate (ADR)} of 0.82, 0.79, 0.74, and 0.70, showing forged pilots make CSI an authorization-bound artifact.

We also investigate the impact of packet transmission rate on our method. As Fig.~\ref{444}(b) shows, the average ADR of each system remains nearly stable across different transmission rates. For instance, when the transmission rate increases from 100 to 600, the ADRs of AF-ACT are 0.79, 0.78, 0.80, 0.81, 0.80, and 0.80. This capability is vital for anti-forgery, as a practical attacker may try to manipulate the system by operating at a different traffic rate to recover stable sensing features. In this regard, a key authenticity takeaway is to enforce authenticity at the signal interface: authentic sensing evidence should be a CSI stream recoverable only via an authorized replay procedure, not something any listener can derive from the same over-the-air pilot~\cite{kong2024csi}. Therefore, forensic readiness should preserve and report the authorization-bound decoding context as first-class evidence, storing sensing outputs together with provenance fields (capture descriptors, tool versions, calibration/synchronization status) so a third party can replay authorized decoding and reject forged or replayed outputs lacking a valid integrity and provenance chain.



\subsection{Timeline Reconstruction \& Event Correlation}
Timeline reconstruction in wireless forensics turns heterogeneous traces into a coherent, time-ordered narrative of what happened when, while event correlation links cross-layer symptoms to root actions. Because investigations combine discrete events (logs/telemetry) and continuous streams (RF/CSI), defensible reconstruction requires explicit timelines, reproducible parsing, and correlation rules that can be replayed.

\emph{\textbf{1) Cross-domain telemetry correlation for reconstructing timelines:}}
Recent systems reconstruct incident timelines by correlating heterogeneous telemetry across layers, emphasizing reliable correlation keys and causal/provenance structure beyond raw timestamps. For instance, cross-layer telemetry (CLT)~\cite{iurman2021towards} reconstructs end-to-end timelines by joining application distributed tracing with in-band network telemetry: it binds application trace context to IPv6 hop-by-hop evidence and correlates them at collectors so each application span is explained by per-hop network measurements. Their evaluation shows the enhanced span exposes router queue growth, attributing the issue to network congestion rather than application logic, and reports \textcolor{black}{round-trip time (RTT)} distributions remain unchanged relative to the congested baseline because overhead is dominated by a lightweight netlink call. 
Building on correlated events, other works~\cite{ashok2024traceweaver} assemble reconstructed incident paths using distributed tracing and provenance graphs~\cite{toslali2021automating}. For example,~\cite{shen2023network} reconstructs causal request paths from network-observable remote procedure call boundaries without code instrumentation: it adds lightweight in-network/host-side capture at application boundary interfaces and assembles traces via a storage/indexing pipeline for high-rate events. Reported per-event overhead is 277--889 ns, a single-trace reconstruction query takes $\sim$1 s, and a 15-minute window search returns in $\sim$0.06 s, supporting rapid forensic pivoting. 

\emph{\textbf{2) Wireless sensing driven timeline reconstruction:}}
While cross-domain telemetry correlation restores missing linkages among logs for cyber incident timelines, wireless sensing links RF traces to physical-world actions by converting continuous measurements into timestamped event primitives (e.g., counts, identity cues) that can be fused with telemetry. For example, \cite{korany2021counting} treats WiFi sensing as event-stream construction: it extracts fidgeting and silent intervals from CSI and maps them into an occupancy estimate, producing a time-ordered event sequence alignable with access logs, camera timestamps, and alarm logs. Technically, it computes CSI phase differences across multiple Rx antennas, selects reliable subcarriers via SNR-based calibration, and denoises using principal component analysis (PCA). Crowd fidgeting is detected via spectral energy outside the breathing band, and seated-people counts are inferred by maximum a posteriori (MAP) estimation under a fidget-to-occupancy model. Reported performance includes 96.3\% average counting accuracy, mean absolute error 0.44, and normalized mean square error 0.015.


\emph{\textbf{3) Case Study:}}
For crowd-related incidents (e.g., stations), investigators need a defensible, time-ordered account of how flows formed and split/merged. Wireless sensing can leverage existing infrastructure without relying on cameras~\cite{wang2024generative}: routine downlink signals are timestamped, logged, and parsed into structured flow primitives, producing a replayable timeline artifact for correlation with operational logs and incident reports. This case uses synchronized capture of timestamped downlink CSI and generative-AI-assisted inference to extract flow events. Concretely, 
downlink CSI is captured at an infrastructure-side receiver and segmented into short sliding windows, where each window is one time slice on the reconstructed timeline. For each window, it derives a velocity-acceleration spectrum to estimate target count, then applies a weighted conditional diffusion model to denoise/sharpen the spectrum for stable estimates. In parallel, antenna-array CSI estimates \textcolor{black}{direction of arrival (DoA)} and ToF, and diffusion enhances the DoA spectrum when spacing exceeds half a wavelength. Velocity, DoA, and ToF are fused and clustered to infer subflow count and subflow size, then stitched over time to form the flow timeline.

\begin{figure*}[htbp]
  \centering
  \begin{subfigure}[t]{0.26\textwidth}
    \centering
    \includegraphics[width=\linewidth]{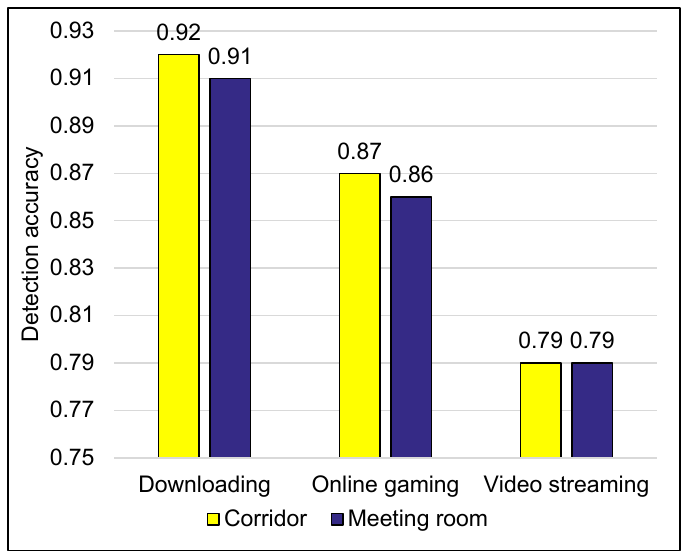}
    \caption{}
  \end{subfigure}
  \hfill
  \begin{subfigure}[t]{0.26\textwidth}
    \centering
    \includegraphics[width=\linewidth]{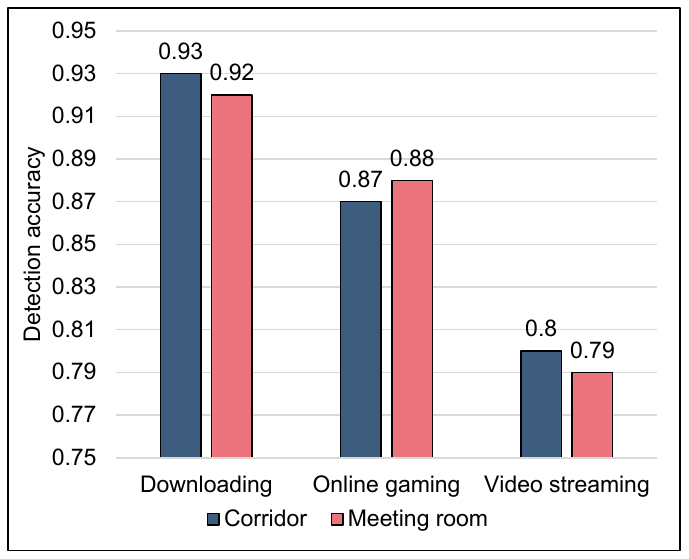}
    \caption{}
  \end{subfigure}
  \hfill
  \begin{subfigure}[t]{0.26\textwidth}
    \centering
    \includegraphics[width=\linewidth]{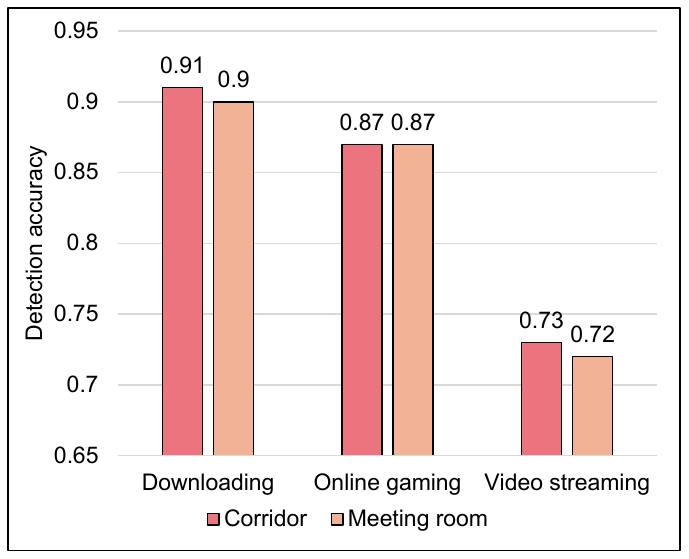}
    \caption{}
  \end{subfigure}
  \vspace{-0.3cm}
  \caption{The results of the case for flow detection in wireless forensics \cite{wang2024generative}. (a) The performance of detecting the number of targets. (b) The performance of detecting the number of subflows. (c) The performance of detecting the subflow size.}
  \label{C1}
  \vspace{-0.7cm}
\end{figure*}

The experiments in a corridor and meeting room use two APs and one UE: one AP transmits, and the other captures CSI via Nexmon~\cite{schafer2021human}. The receiver has four RF channels (one directional reference plus three-antenna ULA with one-wavelength spacing) at 5.805 GHz and 80 MHz. With UE activities (downloading, online gaming, video streaming), target-count accuracies in the corridor are 92\%, 87\%, and 79\% (Fig.~\ref{C1}), while subflow-size accuracies are 91\%, 87\%, 73\% (corridor) and 90\%, 87\%, 72\% (meeting room). Overall, this case shows routine downlink CSI can be converted into replayable primitives (target count, subflow count, subflow size) for timeline reconstruction, but reliability depends on capture conditions—packet rate and environment complexity affect estimation and event ordering—so reporting should preserve raw CSI, timestamps, and capture context controlling sampling density and spectrum resolution.


\subsection{\textcolor{black}{Lessons Learned}}
Across Sections 4.1--4.5, the key lesson is that defensible cyber-(physical) detection depends not on isolated alerts, but on replayable, attributable, and uncertainty-aware evidence that remains credible across heterogeneous devices, networks, and sensing modalities. In \textit{Detection \& Anomaly Discovery}, learning-based methods over radio and spectrum measurements support continuous anomaly discovery and help identify where evidence should be preserved under label scarcity~\cite{huang2022cellular,islam2022deep}. However, they remain limited by threshold sensitivity, drift, uncertainty, and periodicity shifts, and often stop at anomaly scores rather than preservation decisions. In \textit{Attribution \& Identification}, infrastructure-visible traces and flow-level metadata can map incidents to accountable services, slices, or applications~\cite{riyaz2018deep}, but remain sensitive to traffic mixing, site dependence, and weak confidence calibration. In \textit{Provenance \& Localization}, auditable intermediates such as heatmaps and ellipse constraints, help turn RSS and CSI measurements into defensible spatial claims~\cite{kotaru2015spotfi,wang2023through}, although calibration dependence and incomplete provenance reporting still limit cross-site reproducibility. For \textit{Authenticity and Anti-forgery}, signal-interface authentication, including physical-layer tags and RIS challenge mechanisms, can strengthen evidentiary credibility by binding observations to claimed transmitters or contexts~\cite{zhang2023tag,kong2024csi}. However, operational overhead and incomplete preservation of decoding context hinder third-party replay. In \textit{Timeline Reconstruction \& Event Correlation}, telemetry fusion, provenance graphs, and CSI/RF sensing improve causal reconstruction and align RF activity with system logs~\cite{he2023sencom}, but remain fragile under clock mismatch, caching, concurrency, and capture-dependent conditions. Overall, evaluations report false alarms, overhead, scalability, and sometimes uncertainty or domain shift~\cite{tedeschini2024real,stahlke2024uncertainty}, but unified evidence-package standards and key reporting details, such as calibration and synchronization, remain lacking.

\section{Open Challenges \& Future Directions}
\subsection{Generalization and Domain Shift}
Generalization remains a core challenge because cross-site transfer, mobility, and calibration drift can alter the mapping from wireless measurements (RSS/CSI/KPI/trace) to forensic conclusions, causing out-of-distribution error growth and weakening evidentiary strength unless calibration state and robustness diagnostics are preserved as provenance metadata~\cite{mitchell2023learning}. Future work should therefore pursue audit-friendly adaptation that separates portable inference from site-specific calibration.

\subsection{Resource-aware Forensics}
Resource-aware forensics must balance evidentiary fidelity against edge constraints, because continuous RF/CSI/KPI capture consumes storage, bandwidth, compute, and latency, and may affect operational performance; the key challenge is to design cost-aware pipelines that decide what to capture, at what resolution, and for how long, while preserving replayable reconstruction and responsive correlation~\cite{wang2023through}. Promising directions include anomaly-driven adaptive capture, multi-tier processing that retains lightweight timestamped primitives broadly.

\subsection{Benefits and Risks of Generative Evidence}
Generative models can improve forensic usability by denoising and sharpening intermediate evidence such as motion spectra and DoA spectra, thereby turning continuous RF/CSI into stable, timestamped event primitives for timeline reconstruction and correlation~\cite{yang2024generative}. However, they also introduce admissibility risks because enhanced artifacts may be challenged as synthetic, may suppress rare but real features, or may be exploited to fabricate narratives. A defensible direction is therefore to treat raw traces as primary evidence and generative outputs as derived artifacts.

\section{Conclusion}
This survey reviewed intelligent wireless forensics in 5G/6G and beyond. Unlike prior surveys that mainly emphasize detection, attribution, or defense, this work systematized wireless evidence sources across physical, device, network, and cross-layer views, and unified the forensic pipeline from readiness and acquisition to correlation, analysis, and reporting. We also showed how AI can support evidence capture, cross-layer interpretation, and reproducible reconstruction, while highlighting that defensible conclusions still depend on provenance, calibration, synchronization, and replayability. Looking ahead, key challenges include domain shift, cost-aware evidence preservation, missing evidence-package standards, and the admissibility risks of generative evidence enhancement. By bridging wireless evidence, signal processing, networking, and AI, this survey outlines a practical path toward reproducible, auditable, and trustworthy forensic reconstruction in future wireless systems.

\vspace{-0.2cm}
\bibliographystyle{ACM-Reference-Format}
\bibliography{sample-base}


\begin{thebibliography}{182}


\ifx \showCODEN    \undefined \def \showCODEN     #1{\unskip}     \fi
\ifx \showDOI      \undefined \def \showDOI       #1{#1}\fi
\ifx \showISBNx    \undefined \def \showISBNx     #1{\unskip}     \fi
\ifx \showISBNxiii \undefined \def \showISBNxiii  #1{\unskip}     \fi
\ifx \showISSN     \undefined \def \showISSN      #1{\unskip}     \fi
\ifx \showLCCN     \undefined \def \showLCCN      #1{\unskip}     \fi
\ifx \shownote     \undefined \def \shownote      #1{#1}          \fi
\ifx \showarticletitle \undefined \def \showarticletitle #1{#1}   \fi
\ifx \showURL      \undefined \def \showURL       {\relax}        \fi
\providecommand\bibfield[2]{#2}
\providecommand\bibinfo[2]{#2}
\providecommand\natexlab[1]{#1}
\providecommand\showeprint[2][]{arXiv:#2}

\bibitem[Akinbi(2023)]%
        {akinbi2023digital}
\bibfield{author}{\bibinfo{person}{Alex~Olushola Akinbi}.} \bibinfo{year}{2023}\natexlab{}.
\newblock \showarticletitle{Digital forensics challenges and readiness for 6G Internet of Things (IoT) networks}.
\newblock \bibinfo{journal}{\emph{Wiley Interdisciplinary Reviews: Forensic Science}} \bibinfo{volume}{5}, \bibinfo{number}{6} (\bibinfo{year}{2023}), \bibinfo{pages}{e1496}.
\newblock


\bibitem[Nelson et~al\mbox{.}(2025)]%
        {nelson2025incident}
\bibfield{author}{\bibinfo{person}{Alex Nelson}, \bibinfo{person}{Sanjay Rekhi}, \bibinfo{person}{Murugiah Souppaya}, {and} \bibinfo{person}{Karen Scarfone}.} \bibinfo{year}{2025}\natexlab{}.
\newblock \showarticletitle{Incident Response Recommendations and Considerations for Cybersecurity Risk Management}.
\newblock  (\bibinfo{year}{2025}).
\newblock


\bibitem[Wang et~al\mbox{.}(2022)]%
        {wang2022cnn}
\bibfield{author}{\bibinfo{person}{Ziyue Wang} {et~al\mbox{.}}} \bibinfo{year}{2022}\natexlab{}.
\newblock \showarticletitle{CNN-and GAN-based classification of malicious code families: A code visualization approach}.
\newblock \bibinfo{journal}{\emph{Int. J. Intell. Syst.}} \bibinfo{volume}{37}, \bibinfo{number}{12} (\bibinfo{year}{2022}), \bibinfo{pages}{12472--12489}.
\newblock


\bibitem[Khan et~al\mbox{.}(2019)]%
        {khan2019survey}
\bibfield{author}{\bibinfo{person}{Rabia Khan} {et~al\mbox{.}}} \bibinfo{year}{2019}\natexlab{}.
\newblock \showarticletitle{A survey on security and privacy of 5G technologies: Potential solutions, recent advancements, and future directions}.
\newblock \bibinfo{journal}{\emph{IEEE Commun. Surv. Tutor.}} \bibinfo{volume}{22}, \bibinfo{number}{1} (\bibinfo{year}{2019}), \bibinfo{pages}{196--248}.
\newblock


\bibitem[Rizvi et~al\mbox{.}(2022)]%
        {rizvi2022application}
\bibfield{author}{\bibinfo{person}{Syed Rizvi}, \bibinfo{person}{Mark Scanlon}, \bibinfo{person}{Jimmy McGibney}, {and} \bibinfo{person}{John Sheppard}.} \bibinfo{year}{2022}\natexlab{}.
\newblock \showarticletitle{Application of artificial intelligence to network forensics: Survey, challenges and future directions}.
\newblock \bibinfo{journal}{\emph{Ieee Access}}  \bibinfo{volume}{10} (\bibinfo{year}{2022}), \bibinfo{pages}{110362--110384}.
\newblock


\bibitem[Stoyanova et~al\mbox{.}(2020)]%
        {stoyanova2020survey}
\bibfield{author}{\bibinfo{person}{Maria Stoyanova} {et~al\mbox{.}}} \bibinfo{year}{2020}\natexlab{}.
\newblock \showarticletitle{A survey on the internet of things (IoT) forensics: challenges, approaches, and open issues}.
\newblock \bibinfo{journal}{\emph{IEEE Commun. Surv. Tutor.}} \bibinfo{volume}{22}, \bibinfo{number}{2} (\bibinfo{year}{2020}), \bibinfo{pages}{1191--1221}.
\newblock


\bibitem[Yang et~al\mbox{.}(2026)]%
        {yang2026data}
\bibfield{author}{\bibinfo{person}{Yaoqi Yang} {et~al\mbox{.}}} \bibinfo{year}{2026}\natexlab{}.
\newblock \showarticletitle{Data Freshness Performance Analysis and Optimization in Timely and Secure Low Altitude Economics}.
\newblock \bibinfo{journal}{\emph{IEEE Trans. Cognit. Commun. Networking}}  \bibinfo{volume}{12} (\bibinfo{year}{2026}), \bibinfo{pages}{6016--6030}.
\newblock


\bibitem[Nelson et~al\mbox{.}(2024)]%
        {nelson2024incident}
\bibfield{author}{\bibinfo{person}{Alexander Nelson} {et~al\mbox{.}}} \bibinfo{year}{2024}\natexlab{}.
\newblock \bibinfo{booktitle}{\emph{Incident response recommendations and considerations for cybersecurity risk management: a CSF 2.0 community profile}}.
\newblock \bibinfo{type}{{T}echnical {R}eport}. \bibinfo{institution}{National Institute of Standards and Technology}.
\newblock


\bibitem[Alrajeh et~al\mbox{.}(2017)]%
        {alrajeh2017evidence}
\bibfield{author}{\bibinfo{person}{Dalal Alrajeh}, \bibinfo{person}{Liliana Pasquale}, {and} \bibinfo{person}{Bashar Nuseibeh}.} \bibinfo{year}{2017}\natexlab{}.
\newblock \showarticletitle{On evidence preservation requirements for forensic-ready systems}. In \bibinfo{booktitle}{\emph{Proceedings of the 2017 11th joint meeting on foundations of software engineering}}. \bibinfo{pages}{559--569}.
\newblock


\bibitem[Casino et~al\mbox{.}(2022)]%
        {casino2022research}
\bibfield{author}{\bibinfo{person}{Fran Casino} {et~al\mbox{.}}} \bibinfo{year}{2022}\natexlab{}.
\newblock \showarticletitle{Research trends, challenges, and emerging topics in digital forensics: A review of reviews}.
\newblock \bibinfo{journal}{\emph{Ieee Access}}  \bibinfo{volume}{10} (\bibinfo{year}{2022}), \bibinfo{pages}{25464--25493}.
\newblock


\bibitem[Pilli et~al\mbox{.}(2010)]%
        {pilli2010network}
\bibfield{author}{\bibinfo{person}{Emmanuel~S Pilli}, \bibinfo{person}{Ramesh~C Joshi}, {and} \bibinfo{person}{Rajdeep Niyogi}.} \bibinfo{year}{2010}\natexlab{}.
\newblock \showarticletitle{Network forensic frameworks: Survey and research challenges}.
\newblock \bibinfo{journal}{\emph{digital investigation}} \bibinfo{volume}{7}, \bibinfo{number}{1-2} (\bibinfo{year}{2010}), \bibinfo{pages}{14--27}.
\newblock


\bibitem[Khan et~al\mbox{.}(2016)]%
        {khan2016network}
\bibfield{author}{\bibinfo{person}{Suleman Khan} {et~al\mbox{.}}} \bibinfo{year}{2016}\natexlab{}.
\newblock \showarticletitle{Network forensics: Review, taxonomy, and open challenges}.
\newblock \bibinfo{journal}{\emph{Journal of Network and Computer Applications}}  \bibinfo{volume}{66} (\bibinfo{year}{2016}), \bibinfo{pages}{214--235}.
\newblock


\bibitem[Quick and Choo(2014)]%
        {quick2014impacts}
\bibfield{author}{\bibinfo{person}{Darren Quick} {and} \bibinfo{person}{Kim-Kwang~Raymond Choo}.} \bibinfo{year}{2014}\natexlab{}.
\newblock \showarticletitle{Impacts of increasing volume of digital forensic data: A survey and future research challenges}.
\newblock \bibinfo{journal}{\emph{Digital Investigation}} \bibinfo{volume}{11}, \bibinfo{number}{4} (\bibinfo{year}{2014}), \bibinfo{pages}{273--294}.
\newblock


\bibitem[van Baar et~al\mbox{.}(2014)]%
        {van2014digital}
\bibfield{author}{\bibinfo{person}{Ruud~B van Baar}, \bibinfo{person}{Harm~MA van Beek}, {and} \bibinfo{person}{EJ Van~Eijk}.} \bibinfo{year}{2014}\natexlab{}.
\newblock \showarticletitle{Digital forensics as a service: A game changer}.
\newblock \bibinfo{journal}{\emph{Digital Investigation}}  \bibinfo{volume}{11} (\bibinfo{year}{2014}), \bibinfo{pages}{S54--S62}.
\newblock


\bibitem[Pichan et~al\mbox{.}(2015)]%
        {pichan2015cloud}
\bibfield{author}{\bibinfo{person}{Ameer Pichan}, \bibinfo{person}{Mihai Lazarescu}, {and} \bibinfo{person}{Sie~Teng Soh}.} \bibinfo{year}{2015}\natexlab{}.
\newblock \showarticletitle{Cloud forensics: technical challenges, solutions and comparative analysis}.
\newblock \bibinfo{journal}{\emph{Digital investigation}}  \bibinfo{volume}{13} (\bibinfo{year}{2015}), \bibinfo{pages}{38--57}.
\newblock


\bibitem[Manral et~al\mbox{.}(2019)]%
        {manral2019systematic}
\bibfield{author}{\bibinfo{person}{Bharat Manral} {et~al\mbox{.}}} \bibinfo{year}{2019}\natexlab{}.
\newblock \showarticletitle{A systematic survey on cloud forensics challenges, solutions, and future directions}.
\newblock \bibinfo{journal}{\emph{ACM Comput. Surv.}} \bibinfo{volume}{52}, \bibinfo{number}{6} (\bibinfo{year}{2019}), \bibinfo{pages}{1--38}.
\newblock


\bibitem[Yaqoob et~al\mbox{.}(2019)]%
        {yaqoob2019internet}
\bibfield{author}{\bibinfo{person}{Ibrar Yaqoob} {et~al\mbox{.}}} \bibinfo{year}{2019}\natexlab{}.
\newblock \showarticletitle{Internet of things forensics: Recent advances, taxonomy, requirements, and open challenges}.
\newblock \bibinfo{journal}{\emph{Future Generation Computer Systems}}  \bibinfo{volume}{92} (\bibinfo{year}{2019}), \bibinfo{pages}{265--275}.
\newblock


\bibitem[Atlam et~al\mbox{.}(2020)]%
        {atlam2020internet}
\bibfield{author}{\bibinfo{person}{Hany~F Atlam} {et~al\mbox{.}}} \bibinfo{year}{2020}\natexlab{}.
\newblock \showarticletitle{Internet of things forensics: A review}.
\newblock \bibinfo{journal}{\emph{Internet of Things}}  \bibinfo{volume}{11} (\bibinfo{year}{2020}), \bibinfo{pages}{100220}.
\newblock


\bibitem[Servida and Casey(2019)]%
        {servida2019iot}
\bibfield{author}{\bibinfo{person}{Francesco Servida} {and} \bibinfo{person}{Eoghan Casey}.} \bibinfo{year}{2019}\natexlab{}.
\newblock \showarticletitle{IoT forensic challenges and opportunities for digital traces}.
\newblock \bibinfo{journal}{\emph{Digital Investigation}}  \bibinfo{volume}{28} (\bibinfo{year}{2019}), \bibinfo{pages}{S22--S29}.
\newblock


\bibitem[Danev et~al\mbox{.}(2012)]%
        {danev2012physical}
\bibfield{author}{\bibinfo{person}{Boris Danev}, \bibinfo{person}{Davide Zanetti}, {and} \bibinfo{person}{Srdjan Capkun}.} \bibinfo{year}{2012}\natexlab{}.
\newblock \showarticletitle{On physical-layer identification of wireless devices}.
\newblock \bibinfo{journal}{\emph{ACM Comput. Surv.}} \bibinfo{volume}{45}, \bibinfo{number}{1} (\bibinfo{year}{2012}), \bibinfo{pages}{1--29}.
\newblock


\bibitem[Xu et~al\mbox{.}(2015)]%
        {xu2015device}
\bibfield{author}{\bibinfo{person}{Qiang Xu} {et~al\mbox{.}}} \bibinfo{year}{2015}\natexlab{}.
\newblock \showarticletitle{Device fingerprinting in wireless networks: Challenges and opportunities}.
\newblock \bibinfo{journal}{\emph{IEEE Commun. Surv. Tutorials}} \bibinfo{volume}{18}, \bibinfo{number}{1} (\bibinfo{year}{2015}), \bibinfo{pages}{94--104}.
\newblock


\bibitem[Jagannath et~al\mbox{.}(2022)]%
        {jagannath2022comprehensive}
\bibfield{author}{\bibinfo{person}{Anu Jagannath} {et~al\mbox{.}}} \bibinfo{year}{2022}\natexlab{}.
\newblock \showarticletitle{A comprehensive survey on radio frequency (RF) fingerprinting: Traditional approaches, deep learning, and open challenges}.
\newblock \bibinfo{journal}{\emph{Computer Networks}}  \bibinfo{volume}{219} (\bibinfo{year}{2022}), \bibinfo{pages}{109455}.
\newblock


\bibitem[Mitchell and Chen(2014)]%
        {mitchell2014survey}
\bibfield{author}{\bibinfo{person}{Robert Mitchell} {and} \bibinfo{person}{Ray Chen}.} \bibinfo{year}{2014}\natexlab{}.
\newblock \showarticletitle{A survey of intrusion detection in wireless network applications}.
\newblock \bibinfo{journal}{\emph{Computer Communications}}  \bibinfo{volume}{42} (\bibinfo{year}{2014}), \bibinfo{pages}{1--23}.
\newblock


\bibitem[Heijden et~al\mbox{.}(2018)]%
        {van2018survey}
\bibfield{author}{\bibinfo{person}{Van~Der Heijden} {et~al\mbox{.}}} \bibinfo{year}{2018}\natexlab{}.
\newblock \showarticletitle{Survey on misbehavior detection in cooperative intelligent transportation systems}.
\newblock \bibinfo{journal}{\emph{IEEE Commun. Surv. Tutorials}} \bibinfo{volume}{21}, \bibinfo{number}{1} (\bibinfo{year}{2018}), \bibinfo{pages}{779--811}.
\newblock


\bibitem[Stoyanova et~al\mbox{.}(2020)]%
        {Stoyanova2020IoTForensics}
\bibfield{author}{\bibinfo{person}{Maria Stoyanova} {et~al\mbox{.}}} \bibinfo{year}{2020}\natexlab{}.
\newblock \showarticletitle{A Survey on the Internet of Things (IoT) Forensics: Challenges, Approaches, and Open Issues}.
\newblock \bibinfo{journal}{\emph{IEEE Commun. Surv. Tutorials}} \bibinfo{volume}{22}, \bibinfo{number}{2} (\bibinfo{year}{2020}), \bibinfo{pages}{1191--1221}.
\newblock


\bibitem[Lyle et~al\mbox{.}(2022)]%
        {Lyle2022NISTIR8354}
\bibfield{author}{\bibinfo{person}{James~R. Lyle} {et~al\mbox{.}}} \bibinfo{year}{2022}\natexlab{}.
\newblock \bibinfo{booktitle}{\emph{Digital Investigation Techniques: A NIST Scientific Foundation Review}}.
\newblock \bibinfo{type}{{T}echnical {R}eport} NIST IR 8354. \bibinfo{institution}{National Institute of Standards and Technology}.
\newblock


\bibitem[Han et~al\mbox{.}(2023)]%
        {han2023smart}
\bibfield{author}{\bibinfo{person}{Zhaoyang Han} {et~al\mbox{.}}} \bibinfo{year}{2023}\natexlab{}.
\newblock \showarticletitle{Smart optimization solution for channel access attack defense under UAV-aided heterogeneous network}.
\newblock \bibinfo{journal}{\emph{IEEE Internet Things J.}} \bibinfo{volume}{10}, \bibinfo{number}{21} (\bibinfo{year}{2023}), \bibinfo{pages}{18890--18897}.
\newblock


\bibitem[Jagannath et~al\mbox{.}(2022)]%
        {Jagannath2022RFFSurvey}
\bibfield{author}{\bibinfo{person}{Anu Jagannath} {et~al\mbox{.}}} \bibinfo{year}{2022}\natexlab{}.
\newblock \showarticletitle{A Comprehensive Survey on Radio Frequency (RF) Fingerprinting: Traditional Approaches, Deep Learning, and Open Challenges}.
\newblock \bibinfo{journal}{\emph{Computer Networks}}  \bibinfo{volume}{219} (\bibinfo{year}{2022}), \bibinfo{pages}{109455}.
\newblock


\bibitem[Hilburn et~al\mbox{.}(2017)]%
        {Hilburn2017SigMF}
\bibfield{author}{\bibinfo{person}{Ben Hilburn} {et~al\mbox{.}}} \bibinfo{year}{2017}\natexlab{}.
\newblock \showarticletitle{SigMF: The Signal Metadata Format}.
\newblock \bibinfo{journal}{\emph{Proceedings of the GNU Radio Conference (GRCon)}} (\bibinfo{year}{2017}).
\newblock
\newblock
\shownote{Available via GNU Radio Conference Proceedings}.


\bibitem[Halperin et~al\mbox{.}(2011)]%
        {Halperin2011CSITool}
\bibfield{author}{\bibinfo{person}{Daniel Halperin}, \bibinfo{person}{Wenjun Hu}, \bibinfo{person}{Anmol Sheth}, {and} \bibinfo{person}{David Wetherall}.} \bibinfo{year}{2011}\natexlab{}.
\newblock \showarticletitle{Tool Release: Gathering 802.11n Traces with Channel State Information}.
\newblock \bibinfo{journal}{\emph{ACM SIGCOMM Computer Communication Review}} \bibinfo{volume}{41}, \bibinfo{number}{1} (\bibinfo{year}{2011}), \bibinfo{pages}{53}.
\newblock


\bibitem[Brik et~al\mbox{.}(2008)]%
        {Brik2008Radiometric}
\bibfield{author}{\bibinfo{person}{Vladimir Brik} {et~al\mbox{.}}} \bibinfo{year}{2008}\natexlab{}.
\newblock \showarticletitle{Wireless Device Identification with Radiometric Signatures}. In \bibinfo{booktitle}{\emph{Proceedings of the 14th ACM International Conference on Mobile Computing and Networking}}. \bibinfo{publisher}{Association for Computing Machinery}, \bibinfo{pages}{116--127}.
\newblock


\bibitem[Arcangeloni et~al\mbox{.}(2023)]%
        {Arcangeloni2023JammingCausal}
\bibfield{author}{\bibinfo{person}{Luca Arcangeloni} {et~al\mbox{.}}} \bibinfo{year}{2023}\natexlab{}.
\newblock \showarticletitle{Detection of Jamming Attacks via Source Separation and Causal Inference}.
\newblock \bibinfo{journal}{\emph{IEEE Trans. Commun.}} \bibinfo{volume}{71}, \bibinfo{number}{8} (\bibinfo{year}{2023}), \bibinfo{pages}{4793--4806}.
\newblock
\urldef\tempurl%
\url{https://doi.org/10.1109/TCOMM.2023.3281467}
\showDOI{\tempurl}


\bibitem[Zhao et~al\mbox{.}(2024)]%
        {Zhao2024GANRXA}
\bibfield{author}{\bibinfo{person}{Tianyi Zhao}, \bibinfo{person}{Shamik Sarkar}, \bibinfo{person}{Enes Krijestorac}, {and} \bibinfo{person}{Danijela Cabric}.} \bibinfo{year}{2024}\natexlab{}.
\newblock \showarticletitle{GAN-RXA: A Practical Scalable Solution to Receiver-Agnostic Transmitter Fingerprinting}.
\newblock \bibinfo{journal}{\emph{arXiv preprint}} (\bibinfo{year}{2024}).
\newblock
\showeprint[arxiv]{2303.14312}~[eess.SP]
\newblock
\shownote{arXiv:2303.14312}.


\bibitem[Soltanieh et~al\mbox{.}(2020)]%
        {Soltanieh2020Review}
\bibfield{author}{\bibinfo{person}{Naeimeh Soltanieh} {et~al\mbox{.}}} \bibinfo{year}{2020}\natexlab{}.
\newblock \showarticletitle{A Review of Radio Frequency Fingerprinting Techniques}.
\newblock \bibinfo{journal}{\emph{IEEE Journal of Radio Frequency Identification}} \bibinfo{volume}{4}, \bibinfo{number}{3} (\bibinfo{year}{2020}), \bibinfo{pages}{222--233}.
\newblock


\bibitem[Zhang et~al\mbox{.}(2025)]%
        {Zhang2025Fingerprinting}
\bibfield{author}{\bibinfo{person}{Junqing Zhang} {et~al\mbox{.}}} \bibinfo{year}{2025}\natexlab{}.
\newblock \showarticletitle{Physical Layer-Based Device Fingerprinting for Wireless Security: From Theory to Practice}.
\newblock \bibinfo{journal}{\emph{IEEE T INF FOREN SEC.}}  \bibinfo{volume}{20} (\bibinfo{year}{2025}), \bibinfo{pages}{5296--5325}.
\newblock
\urldef\tempurl%
\url{https://doi.org/10.1109/TIFS.2025.3570118}
\showDOI{\tempurl}


\bibitem[Qu et~al\mbox{.}(2026)]%
        {qu2026secure}
\bibfield{author}{\bibinfo{person}{Yanfeng Qu} {et~al\mbox{.}}} \bibinfo{year}{2026}\natexlab{}.
\newblock \showarticletitle{Secure and privacy-preserving issues in integrated sensing and communication-enabled wireless networks: a survey}.
\newblock \bibinfo{journal}{\emph{EURASIP Journal on Advances in Signal Processing}} \bibinfo{volume}{2026}, \bibinfo{number}{1} (\bibinfo{year}{2026}), \bibinfo{pages}{4}.
\newblock


\bibitem[Yılmaz and Yazıcı(2022)]%
        {Koyuncu2022Temperature}
\bibfield{author}{\bibinfo{person}{Özkan Yılmaz} {and} \bibinfo{person}{Mehmet~Akif Yazıcı}.} \bibinfo{year}{2022}\natexlab{}.
\newblock \showarticletitle{The Effect of Ambient Temperature On Device Classification Based On Radio Frequency Fingerprint Recognition}.
\newblock \bibinfo{journal}{\emph{Sakarya University Journal of Computer and Information Sciences}}  \bibinfo{volume}{5} (\bibinfo{date}{08} \bibinfo{year}{2022}), \bibinfo{pages}{233--245}.
\newblock


\bibitem[Tang et~al\mbox{.}(2024)]%
        {Tang2024TMCStealthyTrigger}
\bibfield{author}{\bibinfo{person}{Zijie Tang} {et~al\mbox{.}}} \bibinfo{year}{2024}\natexlab{}.
\newblock \showarticletitle{RF Domain Backdoor Attack on Signal Classification Via Stealthy Trigger}.
\newblock \bibinfo{journal}{\emph{IEEE Trans. Mob. Comput.}} (\bibinfo{year}{2024}).
\newblock
\urldef\tempurl%
\url{https://doi.org/10.1109/TMC.2024.3404341}
\showDOI{\tempurl}
\newblock
\shownote{Early Access}.


\bibitem[Xie et~al\mbox{.}(2021)]%
        {Xie2021PLASurvey}
\bibfield{author}{\bibinfo{person}{Ning Xie}, \bibinfo{person}{Shilian Li}, {and} \bibinfo{person}{Zhongwen Tan}.} \bibinfo{year}{2021}\natexlab{}.
\newblock \showarticletitle{A Survey of Physical-Layer Authentication in Wireless Communications}.
\newblock \bibinfo{journal}{\emph{IEEE Commun. Surv. Tutorials}} \bibinfo{volume}{23}, \bibinfo{number}{1} (\bibinfo{year}{2021}), \bibinfo{pages}{282--310}.
\newblock
\urldef\tempurl%
\url{https://doi.org/10.1109/COMST.2020.3042188}
\showDOI{\tempurl}


\bibitem[Merlo et~al\mbox{.}(2023)]%
        {Merlo2023PicosecondSync}
\bibfield{author}{\bibinfo{person}{Jason~M. Merlo} {et~al\mbox{.}}} \bibinfo{year}{2023}\natexlab{}.
\newblock \showarticletitle{Wireless Picosecond Time Synchronization for Distributed Antenna Arrays}.
\newblock \bibinfo{journal}{\emph{IEEE T Microw Theory}} \bibinfo{volume}{71}, \bibinfo{number}{4} (\bibinfo{year}{2023}), \bibinfo{pages}{1720--1731}.
\newblock
\urldef\tempurl%
\url{https://doi.org/10.1109/TMTT.2022.3227878}
\showDOI{\tempurl}


\bibitem[Ma et~al\mbox{.}(2024)]%
        {Ma2024AIoTAmbiguityCP}
\bibfield{author}{\bibinfo{person}{Junwei Ma} {et~al\mbox{.}}} \bibinfo{year}{2024}\natexlab{}.
\newblock \showarticletitle{Navigating Uncertainty: Ambiguity Quantification in Fingerprinting-based Indoor Localization}. In \bibinfo{booktitle}{\emph{Proceedings of the 2024 IEEE Annual Congress on Artificial Intelligence of Things}}. \bibinfo{publisher}{IEEE}.
\newblock


\bibitem[Barmpatsalou et~al\mbox{.}(2018)]%
        {barmpatsalou2018current}
\bibfield{author}{\bibinfo{person}{Konstantia Barmpatsalou}, \bibinfo{person}{Tiago Cruz}, \bibinfo{person}{Edmundo Monteiro}, {and} \bibinfo{person}{Paulo Simoes}.} \bibinfo{year}{2018}\natexlab{}.
\newblock \showarticletitle{Current and future trends in mobile device forensics: A survey}.
\newblock \bibinfo{journal}{\emph{ACM Comput. Surv.}} \bibinfo{volume}{51}, \bibinfo{number}{3} (\bibinfo{year}{2018}), \bibinfo{pages}{1--31}.
\newblock


\bibitem[Li et~al\mbox{.}(2024)]%
        {Li2024CCSBaseMirror}
\bibfield{author}{\bibinfo{person}{Wenqiang Li}, \bibinfo{person}{Haohuang Wen}, {and} \bibinfo{person}{Zhiqiang Lin}.} \bibinfo{year}{2024}\natexlab{}.
\newblock \showarticletitle{BaseMirror: Automatic Reverse Engineering of Baseband Commands from Android's Radio Interface Layer}. In \bibinfo{booktitle}{\emph{Proceedings of the 2024 on ACM SIGSAC Conference on Computer and Communications Security (CCS 2024)}}. \bibinfo{publisher}{ACM}, \bibinfo{pages}{2311--2325}.
\newblock
\urldef\tempurl%
\url{https://doi.org/10.1145/3658644.3690254}
\showDOI{\tempurl}


\bibitem[Wen et~al\mbox{.}(2023)]%
        {Wen2023RILDefender}
\bibfield{author}{\bibinfo{person}{Haohuang Wen} {et~al\mbox{.}}} \bibinfo{year}{2023}\natexlab{}.
\newblock \showarticletitle{Thwarting Smartphone SMS Attacks at the Radio Interface Layer}. In \bibinfo{booktitle}{\emph{Network and Distributed System Security Symposium (NDSS)}}. \bibinfo{publisher}{The Internet Society}.
\newblock
\urldef\tempurl%
\url{https://doi.org/10.14722/ndss.2023.24432}
\showDOI{\tempurl}


\bibitem[Muñoz(2024)]%
        {CrackingTheCore2024}
\bibfield{author}{\bibinfo{person}{Antonio Muñoz}.} \bibinfo{year}{2024}\natexlab{}.
\newblock \showarticletitle{Cracking the Core: Hardware Vulnerabilities in Android Devices Unveiled}.
\newblock \bibinfo{journal}{\emph{Electronics}} \bibinfo{volume}{13}, \bibinfo{number}{21} (\bibinfo{year}{2024}), \bibinfo{pages}{4269}.
\newblock


\bibitem[M{\'e}n{\'e}trey et~al\mbox{.}(2022)]%
        {Menetrey2022SysTEXAttestation}
\bibfield{author}{\bibinfo{person}{J{\"a}mes M{\'e}n{\'e}trey}, \bibinfo{person}{Christian G{\"o}ttel}, \bibinfo{person}{Marcelo Pasin}, \bibinfo{person}{Pascal Felber}, {and} \bibinfo{person}{Valerio Schiavoni}.} \bibinfo{year}{2022}\natexlab{}.
\newblock \showarticletitle{An exploratory study of attestation mechanisms for trusted execution environments}.
\newblock \bibinfo{journal}{\emph{arXiv preprint arXiv:2204.06790}} (\bibinfo{year}{2022}).
\newblock


\bibitem[Kim et~al\mbox{.}(2019)]%
        {Kim2019TouchingUntouchables}
\bibfield{author}{\bibinfo{person}{Hongil Kim} {et~al\mbox{.}}} \bibinfo{year}{2019}\natexlab{}.
\newblock \showarticletitle{Touching the Untouchables: Dynamic Security Analysis of the LTE Control Plane}. In \bibinfo{booktitle}{\emph{2019 IEEE Symposium on Security and Privacy (SP)}}. \bibinfo{publisher}{IEEE}, \bibinfo{pages}{1153--1168}.
\newblock
\urldef\tempurl%
\url{https://doi.org/10.1109/SP.2019.00038}
\showDOI{\tempurl}


\bibitem[Ayers et~al\mbox{.}(2014)]%
        {Ayers2014NISTSP800101r1}
\bibfield{author}{\bibinfo{person}{Rick Ayers}, \bibinfo{person}{Sam Brothers}, {and} \bibinfo{person}{Wayne Jansen}.} \bibinfo{year}{2014}\natexlab{}.
\newblock \bibinfo{booktitle}{\emph{Guidelines on Mobile Device Forensics}}.
\newblock \bibinfo{type}{NIST Special Publication} 800-101 Revision 1. \bibinfo{institution}{National Institute of Standards and Technology}.
\newblock
\urldef\tempurl%
\url{https://doi.org/10.6028/NIST.SP.800-101r1}
\showDOI{\tempurl}


\bibitem[Hernandez et~al\mbox{.}(2022)]%
        {Hernandez2022FirmWire}
\bibfield{author}{\bibinfo{person}{Grant Hernandez} {et~al\mbox{.}}} \bibinfo{year}{2022}\natexlab{}.
\newblock \showarticletitle{FirmWire: Transparent Dynamic Analysis for Cellular Baseband Firmware}. In \bibinfo{booktitle}{\emph{Proceedings of the Network and Distributed System Security Symposium (NDSS)}}.
\newblock


\bibitem[Wen et~al\mbox{.}(2023)]%
        {Hussain2023NDSSRILDefender}
\bibfield{author}{\bibinfo{person}{Haohuang Wen} {et~al\mbox{.}}} \bibinfo{year}{2023}\natexlab{}.
\newblock \showarticletitle{Thwarting Smartphone SMS Attacks at the Radio Interface Layer}. In \bibinfo{booktitle}{\emph{Network and Distributed System Security Symposium (NDSS)}}.
\newblock


\bibitem[ISO(2015)]%
        {ISO27043_2015}
 \bibinfo{year}{2015}\natexlab{}.
\newblock \bibinfo{title}{Information technology --- Security techniques --- Incident investigation principles and processes}.
\newblock
\newblock
\urldef\tempurl%
\url{https://www.iso.org/standard/44407.html}
\showURL{%
\tempurl}


\bibitem[Li et~al\mbox{.}(2016)]%
        {Li2016MobileInsight}
\bibfield{author}{\bibinfo{person}{Yuanjie Li} {et~al\mbox{.}}} \bibinfo{year}{2016}\natexlab{}.
\newblock \showarticletitle{MobileInsight: Extracting and Analyzing Cellular Network Information on Smartphones}. In \bibinfo{booktitle}{\emph{Proceedings of the 22nd Annual International Conference on Mobile Computing and Networking (MobiCom)}}.
\newblock


\bibitem[Rizvi et~al\mbox{.}(2022)]%
        {Rizvi2022AIForNetworkForensics}
\bibfield{author}{\bibinfo{person}{Syed Rizvi} {et~al\mbox{.}}} \bibinfo{year}{2022}\natexlab{}.
\newblock \showarticletitle{Application of Artificial Intelligence to Network Forensics: Survey, Challenges and Future Directions}.
\newblock \bibinfo{journal}{\emph{IEEE Access}}  \bibinfo{volume}{10} (\bibinfo{year}{2022}), \bibinfo{pages}{110362--110384}.
\newblock
\urldef\tempurl%
\url{https://doi.org/10.1109/ACCESS.2022.3214086}
\showDOI{\tempurl}


\bibitem[Shen et~al\mbox{.}(2021)]%
        {Meng2021TIFS_TIG_DApp}
\bibfield{author}{\bibinfo{person}{Meng Shen} {et~al\mbox{.}}} \bibinfo{year}{2021}\natexlab{}.
\newblock \showarticletitle{Accurate Decentralized Application Identification via Encrypted Traffic Analysis Using Graph Neural Networks}.
\newblock \bibinfo{journal}{\emph{IEEE T INF FOREN SEC.}}  \bibinfo{volume}{16} (\bibinfo{year}{2021}), \bibinfo{pages}{2367--2380}.
\newblock
\urldef\tempurl%
\url{https://doi.org/10.1109/TIFS.2021.3059654}
\showDOI{\tempurl}


\bibitem[Pacherkar and Yan(2024)]%
        {PROV5GC2024WiSec}
\bibfield{author}{\bibinfo{person}{Harsh~Sanjay Pacherkar} {and} \bibinfo{person}{Guanhua Yan}.} \bibinfo{year}{2024}\natexlab{}.
\newblock \showarticletitle{PROV5GC: Hardening 5G Core Network Security with Attack Detection and Attribution Based on Provenance Graphs}. In \bibinfo{booktitle}{\emph{Proceedings of the 17th ACM Conference on Security and Privacy in Wireless and Mobile Networks (WiSec)}}. \bibinfo{publisher}{ACM}.
\newblock
\urldef\tempurl%
\url{https://doi.org/10.1145/3643833.3656129}
\showDOI{\tempurl}


\bibitem[Sheppard et~al\mbox{.}(2022)]%
        {Sheppard2022AIAccessNetworkForensicsSurvey}
\bibfield{author}{\bibinfo{person}{John Sheppard} {et~al\mbox{.}}} \bibinfo{year}{2022}\natexlab{}.
\newblock \showarticletitle{Artificial Intelligence for Network Forensics: A Survey}.
\newblock \bibinfo{journal}{\emph{IEEE Access}}  \bibinfo{volume}{10} (\bibinfo{year}{2022}), \bibinfo{pages}{113687--113717}.
\newblock
\urldef\tempurl%
\url{https://doi.org/10.1109/ACCESS.2022.3214506}
\showDOI{\tempurl}


\bibitem[Erlacher and Dressler(2022)]%
        {Erlacher2022TDSC_IPFIXIDS}
\bibfield{author}{\bibinfo{person}{Felix Erlacher} {and} \bibinfo{person}{Falko Dressler}.} \bibinfo{year}{2022}\natexlab{}.
\newblock \showarticletitle{On High-Speed Flow-based Intrusion Detection using Snort-compatible Signatures}.
\newblock \bibinfo{journal}{\emph{IEEE Trans. Dependable Secure Comput.}} \bibinfo{volume}{19}, \bibinfo{number}{2} (\bibinfo{year}{2022}), \bibinfo{pages}{1190--1205}.
\newblock
\urldef\tempurl%
\url{https://doi.org/10.1109/TDSC.2020.3026747}
\showDOI{\tempurl}


\bibitem[Abouelkhair et~al\mbox{.}(2024)]%
        {Abouelkhair2024CNSM5GProvGen}
\bibfield{author}{\bibinfo{person}{Amr Abouelkhair} {et~al\mbox{.}}} \bibinfo{year}{2024}\natexlab{}.
\newblock \showarticletitle{5GProvGen: 5G Provenance Dataset Generation Framework}. In \bibinfo{booktitle}{\emph{2024 20th International Conference on Network and Service Management (CNSM)}}. \bibinfo{publisher}{IFIP}.
\newblock


\bibitem[Hofstede et~al\mbox{.}(2014)]%
        {Hofstede2014FlowMonitoringExplained}
\bibfield{author}{\bibinfo{person}{Rick Hofstede} {et~al\mbox{.}}} \bibinfo{year}{2014}\natexlab{}.
\newblock \showarticletitle{Flow Monitoring Explained: From Packet Capture to Data Analysis with NetFlow and {IPFIX}}.
\newblock \bibinfo{journal}{\emph{IEEE Commun. Surv. Tutorials}} \bibinfo{volume}{16}, \bibinfo{number}{4} (\bibinfo{year}{2014}), \bibinfo{pages}{2037--2064}.
\newblock
\urldef\tempurl%
\url{https://doi.org/10.1109/COMST.2014.2321898}
\showDOI{\tempurl}


\bibitem[Klischies et~al\mbox{.}(2025)]%
        {Muehlberg2025BaseBridge}
\bibfield{author}{\bibinfo{person}{Daniel Klischies} {et~al\mbox{.}}} \bibinfo{year}{2025}\natexlab{}.
\newblock \showarticletitle{BaseBridge: Bridging the Gap Between Over-the-Air and Emulation Testing for Cellular Baseband Firmware}. In \bibinfo{booktitle}{\emph{IEEE Symposium on Security and Privacy (SP) 2025}}. \bibinfo{publisher}{IEEE}, \bibinfo{pages}{1101--1119}.
\newblock
\urldef\tempurl%
\url{https://doi.org/10.1109/SP61157.2025.00142}
\showDOI{\tempurl}


\bibitem[Estan et~al\mbox{.}(2004)]%
        {Estan2004BetterNetflow}
\bibfield{author}{\bibinfo{person}{Cristian Estan} {et~al\mbox{.}}} \bibinfo{year}{2004}\natexlab{}.
\newblock \showarticletitle{Building a Better NetFlow}. In \bibinfo{booktitle}{\emph{Proceedings of the ACM SIGCOMM Conference}}. \bibinfo{pages}{245--256}.
\newblock
\urldef\tempurl%
\url{https://doi.org/10.1145/1015467.1015495}
\showDOI{\tempurl}


\bibitem[Severi et~al\mbox{.}(2023)]%
        {Severi2023Poisoning}
\bibfield{author}{\bibinfo{person}{Giorgio Severi} {et~al\mbox{.}}} \bibinfo{year}{2023}\natexlab{}.
\newblock \showarticletitle{Poisoning Network Flow Classifiers}. In \bibinfo{booktitle}{\emph{Proceedings of the 39th Annual Computer Security Applications Conference (ACSAC '23)}}. \bibinfo{publisher}{ACM}, \bibinfo{address}{Austin, TX, USA}, \bibinfo{pages}{337--351}.
\newblock
\urldef\tempurl%
\url{https://doi.org/10.1145/3627106.3627123}
\showDOI{\tempurl}


\bibitem[Khan et~al\mbox{.}(2016)]%
        {Khan2016NetworkForensics}
\bibfield{author}{\bibinfo{person}{Suleman Khan} {et~al\mbox{.}}} \bibinfo{year}{2016}\natexlab{}.
\newblock \showarticletitle{Network forensics: Review, taxonomy, and open challenges}.
\newblock \bibinfo{journal}{\emph{Journal of Network and Computer Applications}}  \bibinfo{volume}{66} (\bibinfo{year}{2016}), \bibinfo{pages}{214--235}.
\newblock
\urldef\tempurl%
\url{https://doi.org/10.1016/j.jnca.2016.03.005}
\showDOI{\tempurl}


\bibitem[Lyle et~al\mbox{.}(2022)]%
        {Lyle2022NIST}
\bibfield{author}{\bibinfo{person}{Jim~R. Lyle} {et~al\mbox{.}}} \bibinfo{year}{2022}\natexlab{}.
\newblock \bibinfo{booktitle}{\emph{Digital Investigation Techniques: A NIST Scientific Foundation Review}}.
\newblock \bibinfo{type}{Tech. Rep.} NIST IR 8354. \bibinfo{institution}{National Institute of Standards and Technology}.
\newblock
\urldef\tempurl%
\url{https://doi.org/10.6028/NIST.IR.8354}
\showDOI{\tempurl}


\bibitem[Xu et~al\mbox{.}(2024)]%
        {Xu2024InfoSensitiveINT}
\bibfield{author}{\bibinfo{person}{Zhaojin Xu}, \bibinfo{person}{Zhaoming Lu}, {and} \bibinfo{person}{Zuqing Zhu}.} \bibinfo{year}{2024}\natexlab{}.
\newblock \showarticletitle{Information-Sensitive In-Band Network Telemetry in P4-Based Programmable Data Plane}.
\newblock \bibinfo{journal}{\emph{IEEE/ACM Transactions on Networking}} \bibinfo{volume}{32}, \bibinfo{number}{1} (\bibinfo{year}{2024}), \bibinfo{pages}{568--581}.
\newblock


\bibitem[Talpini et~al\mbox{.}(2024)]%
        {Talpini2024Trustworthiness}
\bibfield{author}{\bibinfo{person}{Jacopo Talpini}, \bibinfo{person}{Fabio Sartori}, {and} \bibinfo{person}{Marco Savi}.} \bibinfo{year}{2024}\natexlab{}.
\newblock \showarticletitle{Enhancing trustworthiness in ML-based network intrusion detection with uncertainty quantification}.
\newblock \bibinfo{journal}{\emph{Journal of Reliable Intelligent Environments}} \bibinfo{volume}{10}, \bibinfo{number}{4} (\bibinfo{year}{2024}), \bibinfo{pages}{501--520}.
\newblock
\urldef\tempurl%
\url{https://doi.org/10.1007/s40860-024-00238-8}
\showDOI{\tempurl}


\bibitem[Duan et~al\mbox{.}(2025)]%
        {Duan2025AdaptiveSurvey}
\bibfield{author}{\bibinfo{person}{Zhuoying Duan} {et~al\mbox{.}}} \bibinfo{year}{2025}\natexlab{}.
\newblock \showarticletitle{Adaptive Strategies in Enhancing Physical Layer Security: A Comprehensive Survey}.
\newblock \bibinfo{journal}{\emph{ACM Comput. Surv.}} \bibinfo{volume}{57}, \bibinfo{number}{7} (\bibinfo{year}{2025}), \bibinfo{pages}{1--36}.
\newblock


\bibitem[Xiong and Jamieson(2013)]%
        {Xiong2013SecureArray}
\bibfield{author}{\bibinfo{person}{Jie Xiong} {and} \bibinfo{person}{Kyle Jamieson}.} \bibinfo{year}{2013}\natexlab{}.
\newblock \showarticletitle{Securearray: Improving wifi security with fine-grained physical-layer information}. In \bibinfo{booktitle}{\emph{Proceedings of the 19th annual international conference on Mobile computing \& networking}}. \bibinfo{pages}{441--452}.
\newblock


\bibitem[Yao et~al\mbox{.}(2021)]%
        {Yao2021AMI}
\bibfield{author}{\bibinfo{person}{Ruizhe Yao} {et~al\mbox{.}}} \bibinfo{year}{2021}\natexlab{}.
\newblock \showarticletitle{Intrusion Detection System in the Advanced Metering Infrastructure: A Cross-Layer Feature-Fusion CNN-LSTM-Based Approach}.
\newblock \bibinfo{journal}{\emph{Sensors}} \bibinfo{volume}{21}, \bibinfo{number}{2} (\bibinfo{year}{2021}), \bibinfo{pages}{626}.
\newblock


\bibitem[Wang et~al\mbox{.}(2024)]%
        {Wang2024BlindTag}
\bibfield{author}{\bibinfo{person}{Chen Wang}, \bibinfo{person}{Mingrui Sha}, \bibinfo{person}{Wei Xiong}, \bibinfo{person}{Ning Xie}, \bibinfo{person}{Rui Mao}, \bibinfo{person}{Peichang Zhang}, {and} \bibinfo{person}{Lei Huang}.} \bibinfo{year}{2024}\natexlab{}.
\newblock \showarticletitle{Blind Tag-Based Physical-Layer Authentication}.
\newblock \bibinfo{journal}{\emph{IEEE/ACM Transactions on Networking}} \bibinfo{volume}{32}, \bibinfo{number}{1} (\bibinfo{year}{2024}), \bibinfo{pages}{4735–4748}.
\newblock


\bibitem[Xie et~al\mbox{.}(2022)]%
        {Xie2022HighCompatibility}
\bibfield{author}{\bibinfo{person}{Ning Xie} {et~al\mbox{.}}} \bibinfo{year}{2022}\natexlab{}.
\newblock \showarticletitle{Physical Layer Authentication With High Compatibility Using an Encoding Approach}.
\newblock \bibinfo{journal}{\emph{IEEE Trans. Commun.}} \bibinfo{volume}{70}, \bibinfo{number}{12} (\bibinfo{year}{2022}), \bibinfo{pages}{8270--8285}.
\newblock


\bibitem[Zhang et~al\mbox{.}(2020)]%
        {Zhang2020GaussianTag}
\bibfield{author}{\bibinfo{person}{Ning Zhang} {et~al\mbox{.}}} \bibinfo{year}{2020}\natexlab{}.
\newblock \showarticletitle{Physical-Layer Authentication for Internet of Things via WFRFT-Based Gaussian Tag Embedding}.
\newblock \bibinfo{journal}{\emph{IEEE Internet Things J.}} \bibinfo{volume}{7}, \bibinfo{number}{9} (\bibinfo{year}{2020}), \bibinfo{pages}{9001--9010}.
\newblock


\bibitem[Altaibek et~al\mbox{.}(2025)]%
        {Altaibek2025CrossLayerSurvey}
\bibfield{author}{\bibinfo{person}{Mamyr Altaibek} {et~al\mbox{.}}} \bibinfo{year}{2025}\natexlab{}.
\newblock \showarticletitle{A Survey of Cross-Layer Security for Resource-Constrained IoT Devices}.
\newblock \bibinfo{journal}{\emph{Applied Sciences}} \bibinfo{volume}{15}, \bibinfo{number}{17} (\bibinfo{year}{2025}), \bibinfo{pages}{9691}.
\newblock


\bibitem[Adesina et~al\mbox{.}(2023)]%
        {Adesina2024AdversarialReview}
\bibfield{author}{\bibinfo{person}{Damilola Adesina} {et~al\mbox{.}}} \bibinfo{year}{2023}\natexlab{}.
\newblock \showarticletitle{Adversarial Machine Learning in Wireless Communications Using RF Data: A Review}.
\newblock \bibinfo{journal}{\emph{IEEE Access}}  \bibinfo{volume}{25} (\bibinfo{year}{2023}), \bibinfo{pages}{77–100}.
\newblock


\bibitem[Zhou et~al\mbox{.}(2024)]%
        {Zhou2024TLSAnalysis}
\bibfield{author}{\bibinfo{person}{Jiuxing Zhou}, \bibinfo{person}{Wei Fu}, \bibinfo{person}{Wei Hu}, \bibinfo{person}{Zhihong Sun}, \bibinfo{person}{Tao He}, {and} \bibinfo{person}{Zhihong Zhang}.} \bibinfo{year}{2024}\natexlab{}.
\newblock \showarticletitle{Challenges and Advances in Analyzing TLS 1.3-Encrypted Traffic: A Comprehensive Survey}.
\newblock \bibinfo{journal}{\emph{Electronics}} \bibinfo{volume}{13}, \bibinfo{number}{20} (\bibinfo{year}{2024}), \bibinfo{pages}{4000}.
\newblock


\bibitem[Xie et~al\mbox{.}(2024)]%
        {Xie2024WGANEncoder}
\bibfield{author}{\bibinfo{person}{Wei Xie} {et~al\mbox{.}}} \bibinfo{year}{2024}\natexlab{}.
\newblock \showarticletitle{A Novel PHY-Layer Spoofing Attack Detection Scheme Based on WGAN-Encoder Model}.
\newblock \bibinfo{journal}{\emph{IEEE T INF FOREN SEC.}}  \bibinfo{volume}{19} (\bibinfo{year}{2024}), \bibinfo{pages}{8616--8629}.
\newblock


\bibitem[Piana et~al\mbox{.}(2025)]%
        {Piana2025DroneCR}
\bibfield{author}{\bibinfo{person}{Mattia Piana}, \bibinfo{person}{Francesco Ardizzon}, {and} \bibinfo{person}{Stefano Tomasin}.} \bibinfo{year}{2025}\natexlab{}.
\newblock \showarticletitle{Challenge-Response to Authenticate Drone Communications: A Game Theoretic Approach}.
\newblock \bibinfo{journal}{\emph{IEEE T INF FOREN SEC.}}  \bibinfo{volume}{20} (\bibinfo{year}{2025}), \bibinfo{pages}{4890--4903}.
\newblock


\bibitem[Bao et~al\mbox{.}(2025)]%
        {Lyu2025PFRTF}
\bibfield{author}{\bibinfo{person}{Zhida Bao} {et~al\mbox{.}}} \bibinfo{year}{2025}\natexlab{}.
\newblock \showarticletitle{PFRTF: A Robust Training Framework to Counter Adversarial Attacks in Signal Classification for Next-G Consumer Electronics}.
\newblock \bibinfo{journal}{\emph{IEEE Trans. Consum. Electron.}} \bibinfo{volume}{71}, \bibinfo{number}{1} (\bibinfo{date}{Feb} \bibinfo{year}{2025}), \bibinfo{pages}{1235--1248}.
\newblock


\bibitem[Xin et~al\mbox{.}(2024)]%
        {Xin2024HighPrecision}
\bibfield{author}{\bibinfo{person}{Jihao Xin} {et~al\mbox{.}}} \bibinfo{year}{2024}\natexlab{}.
\newblock \showarticletitle{High-Precision Time Difference of Arrival Estimation Method Based on Phase Measurement}.
\newblock \bibinfo{journal}{\emph{Remote Sensing}} \bibinfo{volume}{16}, \bibinfo{number}{7} (\bibinfo{year}{2024}), \bibinfo{pages}{1197}.
\newblock


\bibitem[Zhang et~al\mbox{.}(2025)]%
        {Zhang2025SCRUTINIZER}
\bibfield{author}{\bibinfo{person}{Yiming Zhang} {et~al\mbox{.}}} \bibinfo{year}{2025}\natexlab{}.
\newblock \showarticletitle{SCRUTINIZER Towards Secure Forensics on Compromised TrustZone}. In \bibinfo{booktitle}{\emph{Proceedings of the Network and Distributed System Security Symposium}}.
\newblock


\bibitem[Sadineni et~al\mbox{.}(2021)]%
        {sadineni2021ready}
\bibfield{author}{\bibinfo{person}{Lakshminarayana Sadineni}, \bibinfo{person}{Emmanuel~S Pilli}, {and} \bibinfo{person}{Ramesh~Babu Battula}.} \bibinfo{year}{2021}\natexlab{}.
\newblock \showarticletitle{Ready-iot: A novel forensic readiness model for internet of things}. In \bibinfo{booktitle}{\emph{2021 IEEE 7th World Forum on Internet of Things (WF-IoT)}}. IEEE, \bibinfo{pages}{89--94}.
\newblock


\bibitem[Lee and Kim(2021)]%
        {lee2021k}
\bibfield{author}{\bibinfo{person}{Sung~Jin Lee} {and} \bibinfo{person}{Gi~Bum Kim}.} \bibinfo{year}{2021}\natexlab{}.
\newblock \showarticletitle{K-FFRaaS: A Generic Model for Financial Forensic Readiness as a Service in Korea}.
\newblock \bibinfo{journal}{\emph{IEEE Access}}  \bibinfo{volume}{9} (\bibinfo{year}{2021}), \bibinfo{pages}{130094--130110}.
\newblock


\bibitem[Singh et~al\mbox{.}(2022)]%
        {singh2022secure}
\bibfield{author}{\bibinfo{person}{Avinash Singh}, \bibinfo{person}{Richard~Adeyemi Ikuesan}, {and} \bibinfo{person}{Hein Venter}.} \bibinfo{year}{2022}\natexlab{}.
\newblock \showarticletitle{Secure storage model for digital forensic readiness}.
\newblock \bibinfo{journal}{\emph{IEEE Access}}  \bibinfo{volume}{10} (\bibinfo{year}{2022}), \bibinfo{pages}{19469--19480}.
\newblock


\bibitem[Jim{\'e}nez et~al\mbox{.}(2024)]%
        {jimenez2024filtering}
\bibfield{author}{\bibinfo{person}{Mar{\'\i}a~B Jim{\'e}nez} {et~al\mbox{.}}} \bibinfo{year}{2024}\natexlab{}.
\newblock \showarticletitle{A filtering model for evidence gathering in an sdn-oriented digital forensic and incident response context}.
\newblock \bibinfo{journal}{\emph{IEEE Access}}  \bibinfo{volume}{12} (\bibinfo{year}{2024}), \bibinfo{pages}{75792--75808}.
\newblock


\bibitem[Deng et~al\mbox{.}(2024)]%
        {defenderdeng}
\bibfield{author}{\bibinfo{person}{Jiangyi Deng} {et~al\mbox{.}}} \bibinfo{year}{2024}\natexlab{}.
\newblock \showarticletitle{Dr. Defender: Proactive Detection of Autopilot Drones Based on {CSI}}.
\newblock \bibinfo{journal}{\emph{IEEE T INF FOREN SEC.}}  \bibinfo{volume}{19} (\bibinfo{year}{2024}), \bibinfo{pages}{194--206}.
\newblock


\bibitem[Yu et~al\mbox{.}(2019)]%
        {LiveBoxYu}
\bibfield{author}{\bibinfo{person}{Yijun Yu} {et~al\mbox{.}}} \bibinfo{year}{2019}\natexlab{}.
\newblock \showarticletitle{LiveBox: A Self-Adaptive Forensic-Ready Service for Drones}.
\newblock \bibinfo{journal}{\emph{IEEE Access}}  \bibinfo{volume}{7} (\bibinfo{year}{2019}), \bibinfo{pages}{148401--148412}.
\newblock


\bibitem[Palmese et~al\mbox{.}(2023)]%
        {PalmeseWi-Fi}
\bibfield{author}{\bibinfo{person}{Fabio Palmese} {et~al\mbox{.}}} \bibinfo{year}{2023}\natexlab{}.
\newblock \showarticletitle{Designing a Forensic-Ready Wi-Fi Access Point for the Internet of Things}.
\newblock \bibinfo{journal}{\emph{IEEE Internet Things J.}} \bibinfo{volume}{10}, \bibinfo{number}{23} (\bibinfo{year}{2023}), \bibinfo{pages}{20686--20702}.
\newblock


\bibitem[Rizvi et~al\mbox{.}(2024)]%
        {Rizvi2024Pushing}
\bibfield{author}{\bibinfo{person}{Syed Rizvi} {et~al\mbox{.}}} \bibinfo{year}{2024}\natexlab{}.
\newblock \showarticletitle{Pushing Network Forensic Readiness to the Edge: A Resource Constrained Artificial Intelligence Based Methodology}. In \bibinfo{booktitle}{\emph{2024 Cyber Research Conference - Ireland (Cyber-RCI)}}. \bibinfo{pages}{1--8}.
\newblock


\bibitem[Palmese et~al\mbox{.}(2025)]%
        {Palmese2025Resource}
\bibfield{author}{\bibinfo{person}{Fabio Palmese} {et~al\mbox{.}}} \bibinfo{year}{2025}\natexlab{}.
\newblock \showarticletitle{Resource Optimization for Evidence Collection and Preservation in IoT Forensics-Ready Access Points}.
\newblock \bibinfo{journal}{\emph{IEEE Trans. Netw. Serv. Manag.}} \bibinfo{volume}{22}, \bibinfo{number}{5} (\bibinfo{year}{2025}), \bibinfo{pages}{4495--4508}.
\newblock


\bibitem[Halperin et~al\mbox{.}(2011)]%
        {halperin2011tool}
\bibfield{author}{\bibinfo{person}{Daniel Halperin}, \bibinfo{person}{Wenjun Hu}, \bibinfo{person}{Anmol Sheth}, {and} \bibinfo{person}{David Wetherall}.} \bibinfo{year}{2011}\natexlab{}.
\newblock \showarticletitle{Tool release: Gathering 802.11 n traces with channel state information}.
\newblock \bibinfo{journal}{\emph{ACM SIGCOMM computer communication review}} \bibinfo{volume}{41}, \bibinfo{number}{1} (\bibinfo{year}{2011}), \bibinfo{pages}{53--53}.
\newblock


\bibitem[Gringoli et~al\mbox{.}(2019)]%
        {gringoli2019free}
\bibfield{author}{\bibinfo{person}{Francesco Gringoli} {et~al\mbox{.}}} \bibinfo{year}{2019}\natexlab{}.
\newblock \showarticletitle{Free your CSI: A channel state information extraction platform for modern Wi-Fi chipsets}. In \bibinfo{booktitle}{\emph{Proceedings of the 13th International Workshop on Wireless Network Testbeds, Experimental Evaluation \& Characterization}}. \bibinfo{pages}{21--28}.
\newblock


\bibitem[Xu et~al\mbox{.}(2019)]%
        {xu2019redundant}
\bibfield{author}{\bibinfo{person}{Jing Xu} {et~al\mbox{.}}} \bibinfo{year}{2019}\natexlab{}.
\newblock \showarticletitle{Redundant sniffer deployment for multi-channel wireless network forensics with unreliable conditions}.
\newblock \bibinfo{journal}{\emph{IEEE Trans. Cognit. Commun. Networking}} \bibinfo{volume}{6}, \bibinfo{number}{1} (\bibinfo{year}{2019}), \bibinfo{pages}{394--407}.
\newblock


\bibitem[Sheth et~al\mbox{.}(2006)]%
        {sheth2006mojo}
\bibfield{author}{\bibinfo{person}{Anmol Sheth} {et~al\mbox{.}}} \bibinfo{year}{2006}\natexlab{}.
\newblock \showarticletitle{MOJO: A distributed physical layer anomaly detection system for 802.11 WLANs}. In \bibinfo{booktitle}{\emph{Proceedings of the 4th international conference on Mobile systems, applications and services}}. \bibinfo{pages}{191--204}.
\newblock


\bibitem[Chen et~al\mbox{.}(2013)]%
        {chen2013efficient}
\bibfield{author}{\bibinfo{person}{Shaxun Chen}, \bibinfo{person}{Kai Zeng}, {and} \bibinfo{person}{Prasant Mohapatra}.} \bibinfo{year}{2013}\natexlab{}.
\newblock \showarticletitle{Efficient data capturing for network forensics in cognitive radio networks}.
\newblock \bibinfo{journal}{\emph{IEEE/ACM Transactions on Networking}} \bibinfo{volume}{22}, \bibinfo{number}{6} (\bibinfo{year}{2013}), \bibinfo{pages}{1988--2000}.
\newblock


\bibitem[Monteiro et~al\mbox{.}(2023)]%
        {monteiro2023adaptive}
\bibfield{author}{\bibinfo{person}{Davi Monteiro}, \bibinfo{person}{Yijun Yu}, \bibinfo{person}{Andrea Zisman}, {and} \bibinfo{person}{Bashar Nuseibeh}.} \bibinfo{year}{2023}\natexlab{}.
\newblock \showarticletitle{Adaptive observability for forensic-ready microservice systems}.
\newblock \bibinfo{journal}{\emph{IEEE Transactions on Services Computing}} \bibinfo{volume}{16}, \bibinfo{number}{5} (\bibinfo{year}{2023}), \bibinfo{pages}{3196--3209}.
\newblock


\bibitem[Zhou et~al\mbox{.}(2018)]%
        {zhou2018poster}
\bibfield{author}{\bibinfo{person}{Xiang Zhou}, \bibinfo{person}{Xin Peng}, \bibinfo{person}{Tao Xie}, \bibinfo{person}{Jun Sun}, \bibinfo{person}{Chenjie Xu}, \bibinfo{person}{Chao Ji}, {and} \bibinfo{person}{Wenyun Zhao}.} \bibinfo{year}{2018}\natexlab{}.
\newblock \showarticletitle{Poster: Benchmarking microservice systems for software engineering research. In 2018 IEEE/ACM 40th International Conference on Software Engineering: Companion (ICSE-Companion)}.
\newblock \bibinfo{journal}{\emph{IEEE, 323{\'s}324}} (\bibinfo{year}{2018}).
\newblock


\bibitem[Wang et~al\mbox{.}(2025)]%
        {wang2025uncertainty}
\bibfield{author}{\bibinfo{person}{Shixiong Wang} {et~al\mbox{.}}} \bibinfo{year}{2025}\natexlab{}.
\newblock \showarticletitle{Uncertainty awareness in wireless communications and sensing}.
\newblock \bibinfo{journal}{\emph{IEEE Communications Magazine}} (\bibinfo{year}{2025}).
\newblock


\bibitem[Ruah et~al\mbox{.}(2023)]%
        {ruah2023bayesian}
\bibfield{author}{\bibinfo{person}{Clement Ruah}, \bibinfo{person}{Osvaldo Simeone}, {and} \bibinfo{person}{Bashir~M Al-Hashimi}.} \bibinfo{year}{2023}\natexlab{}.
\newblock \showarticletitle{A Bayesian framework for digital twin-based control, monitoring, and data collection in wireless systems}.
\newblock \bibinfo{journal}{\emph{IEEE J. Sel. Areas Commun.}} \bibinfo{volume}{41}, \bibinfo{number}{10} (\bibinfo{year}{2023}), \bibinfo{pages}{3146--3160}.
\newblock


\bibitem[Tian et~al\mbox{.}(2025)]%
        {tian2025cats}
\bibfield{author}{\bibinfo{person}{Yichen Tian} {et~al\mbox{.}}} \bibinfo{year}{2025}\natexlab{}.
\newblock \showarticletitle{CATS: Towards Accurate Device-Free Tracking by Quantifying the Sensing Confidence}.
\newblock \bibinfo{journal}{\emph{IEEE Trans. Mobile Comput.}} (\bibinfo{year}{2025}).
\newblock


\bibitem[Stahlke et~al\mbox{.}(2024)]%
        {stahlke2024uncertainty}
\bibfield{author}{\bibinfo{person}{Maximilian Stahlke} {et~al\mbox{.}}} \bibinfo{year}{2024}\natexlab{}.
\newblock \showarticletitle{Uncertainty-based fingerprinting model monitoring for radio localization}.
\newblock \bibinfo{journal}{\emph{IEEE Journal of Indoor and Seamless Positioning and Navigation}}  \bibinfo{volume}{2} (\bibinfo{year}{2024}), \bibinfo{pages}{166--176}.
\newblock


\bibitem[Landa et~al\mbox{.}(2022)]%
        {landa2022wip}
\bibfield{author}{\bibinfo{person}{Iratxe Landa} {et~al\mbox{.}}} \bibinfo{year}{2022}\natexlab{}.
\newblock \showarticletitle{WIP: Impulsive noise source recognition with OFDM-WiFi signals based on channel state information using machine learning}. In \bibinfo{booktitle}{\emph{2022 IEEE 23rd International Symposium on a World of Wireless, Mobile and Multimedia Networks (WoWMoM)}}. IEEE, \bibinfo{pages}{157--160}.
\newblock


\bibitem[Wen et~al\mbox{.}(2025)]%
        {wen2025exploring}
\bibfield{author}{\bibinfo{person}{Yun Wen} {et~al\mbox{.}}} \bibinfo{year}{2025}\natexlab{}.
\newblock \showarticletitle{Exploring passive eves with self-refine sensing: A novel ISAC-aided secure communication system with STAR-RIS}.
\newblock \bibinfo{journal}{\emph{IEEE Trans. Wireless Commun.}} (\bibinfo{year}{2025}).
\newblock


\bibitem[Yu et~al\mbox{.}(2025)]%
        {yu2025learning}
\bibfield{author}{\bibinfo{person}{Xianglin Yu} {et~al\mbox{.}}} \bibinfo{year}{2025}\natexlab{}.
\newblock \showarticletitle{Learning-Based Predictive Beamforming for Secure ISAC via IRS}.
\newblock \bibinfo{journal}{\emph{IEEE Trans. Commun.}} (\bibinfo{year}{2025}).
\newblock


\bibitem[He et~al\mbox{.}(2023)]%
        {he2023sencom}
\bibfield{author}{\bibinfo{person}{Yinghui He} {et~al\mbox{.}}} \bibinfo{year}{2023}\natexlab{}.
\newblock \showarticletitle{Sencom: Integrated sensing and communication with practical wifi}. In \bibinfo{booktitle}{\emph{Proceedings of the 29th Annual International Conference on Mobile Computing and Networking}}. \bibinfo{pages}{1--16}.
\newblock


\bibitem[Wen et~al\mbox{.}(2018)]%
        {wen2018deep}
\bibfield{author}{\bibinfo{person}{Chao-Kai Wen}, \bibinfo{person}{Wan-Ting Shih}, {and} \bibinfo{person}{Shi Jin}.} \bibinfo{year}{2018}\natexlab{}.
\newblock \showarticletitle{Deep learning for massive MIMO CSI feedback}.
\newblock \bibinfo{journal}{\emph{IEEE Wireless Communications Letters}} \bibinfo{volume}{7}, \bibinfo{number}{5} (\bibinfo{year}{2018}), \bibinfo{pages}{748--751}.
\newblock


\bibitem[Jin et~al\mbox{.}(2025)]%
        {jin2025channel}
\bibfield{author}{\bibinfo{person}{Zhenzhou Jin} {et~al\mbox{.}}} \bibinfo{year}{2025}\natexlab{}.
\newblock \showarticletitle{Channel Fingerprint Construction for Massive MIMO: A Deep Conditional Generative Approach}.
\newblock \bibinfo{journal}{\emph{IEEE Trans. Wireless Commun.}} (\bibinfo{year}{2025}).
\newblock


\bibitem[Wang et~al\mbox{.}(2023)]%
        {wang2023super}
\bibfield{author}{\bibinfo{person}{Xiping Wang} {et~al\mbox{.}}} \bibinfo{year}{2023}\natexlab{}.
\newblock \showarticletitle{Super-resolution of wireless channel characteristics: A multitask learning model}.
\newblock \bibinfo{journal}{\emph{IEEE Transactions on Antennas and Propagation}} \bibinfo{volume}{71}, \bibinfo{number}{10} (\bibinfo{year}{2023}), \bibinfo{pages}{8197--8209}.
\newblock


\bibitem[Shen et~al\mbox{.}(2023)]%
        {shen2023deep}
\bibfield{author}{\bibinfo{person}{Wenhan Shen}, \bibinfo{person}{Zhijin Qin}, {and} \bibinfo{person}{Arumugam Nallanathan}.} \bibinfo{year}{2023}\natexlab{}.
\newblock \showarticletitle{Deep learning for super-resolution channel estimation in reconfigurable intelligent surface aided systems}.
\newblock \bibinfo{journal}{\emph{IEEE Trans. Commun.}} \bibinfo{volume}{71}, \bibinfo{number}{3} (\bibinfo{year}{2023}), \bibinfo{pages}{1491--1503}.
\newblock


\bibitem[Nam et~al\mbox{.}(2024)]%
        {nam2024data}
\bibfield{author}{\bibinfo{person}{Taesik Nam} {et~al\mbox{.}}} \bibinfo{year}{2024}\natexlab{}.
\newblock \showarticletitle{Data generation and augmentation method for deep learning-based VDU leakage signal restoration algorithm}.
\newblock \bibinfo{journal}{\emph{IEEE T INF FOREN SEC.}}  \bibinfo{volume}{19} (\bibinfo{year}{2024}), \bibinfo{pages}{5220--5234}.
\newblock


\bibitem[Yang et~al\mbox{.}(2024)]%
        {yang2024generative}
\bibfield{author}{\bibinfo{person}{Yaoqi Yang}, \bibinfo{person}{Bangning Zhang}, \bibinfo{person}{Daoxing Guo}, \bibinfo{person}{Hongyang Du}, \bibinfo{person}{Zehui Xiong}, \bibinfo{person}{Dusit Niyato}, {and} \bibinfo{person}{Zhu Han}.} \bibinfo{year}{2024}\natexlab{}.
\newblock \showarticletitle{Generative AI for secure and privacy-preserving mobile crowdsensing}.
\newblock \bibinfo{journal}{\emph{IEEE Wireless Communications}} \bibinfo{volume}{31}, \bibinfo{number}{6} (\bibinfo{year}{2024}), \bibinfo{pages}{29--38}.
\newblock


\bibitem[Vanini et~al\mbox{.}(2024)]%
        {vanini2024clock}
\bibfield{author}{\bibinfo{person}{C{\'e}line Vanini} {et~al\mbox{.}}} \bibinfo{year}{2024}\natexlab{}.
\newblock \showarticletitle{Was the Clock Correct? Exploring Timestamp Interpretation through Time Anchors for Digital Forensic Event Reconstruction}.
\newblock \bibinfo{journal}{\emph{Forensic Science International: Digital Investigation}}  \bibinfo{volume}{49} (\bibinfo{year}{2024}), \bibinfo{pages}{301759}.
\newblock


\bibitem[Xiong et~al\mbox{.}(2015)]%
        {xiong2015tonetrack}
\bibfield{author}{\bibinfo{person}{Jie Xiong}, \bibinfo{person}{Karthikeyan Sundaresan}, {and} \bibinfo{person}{Kyle Jamieson}.} \bibinfo{year}{2015}\natexlab{}.
\newblock \showarticletitle{ToneTrack: Leveraging Frequency-Agile Radios for Time-Based Indoor Wireless Localization}. In \bibinfo{booktitle}{\emph{Proceedings of the 21st Annual International Conference on Mobile Computing and Networking}} \emph{(\bibinfo{series}{MobiCom '15})}. \bibinfo{publisher}{ACM}, \bibinfo{pages}{537--549}.
\newblock


\bibitem[Hilburn et~al\mbox{.}(2018)]%
        {hilburn2018sigmf}
\bibfield{author}{\bibinfo{person}{Ben Hilburn}, \bibinfo{person}{Nathan West}, \bibinfo{person}{Tim O'Shea}, {and} \bibinfo{person}{Tamoghna Roy}.} \bibinfo{year}{2018}\natexlab{}.
\newblock \showarticletitle{{SigMF}: The Signal Metadata Format}.
\newblock \bibinfo{journal}{\emph{Proceedings of the GNU Radio Conference}} \bibinfo{volume}{3}, \bibinfo{number}{1} (\bibinfo{date}{Sept.} \bibinfo{year}{2018}).
\newblock


\bibitem[Pacherkar and Yan(2024)]%
        {pacherkar2024prov5gc}
\bibfield{author}{\bibinfo{person}{Harsh~Sanjay Pacherkar} {and} \bibinfo{person}{Guanhua Yan}.} \bibinfo{year}{2024}\natexlab{}.
\newblock \showarticletitle{PROV5GC: Hardening 5G core network security with attack detection and attribution based on provenance graphs}. In \bibinfo{booktitle}{\emph{Proceedings of the 17th ACM Conference on Security and Privacy in Wireless and Mobile Networks}}. \bibinfo{pages}{254--264}.
\newblock


\bibitem[Al-Shawabka et~al\mbox{.}(2020)]%
        {alshawabka2020exposing}
\bibfield{author}{\bibinfo{person}{Amani Al-Shawabka} {et~al\mbox{.}}} \bibinfo{year}{2020}\natexlab{}.
\newblock \showarticletitle{Exposing the Fingerprint: Dissecting the Impact of the Wireless Channel on Radio Fingerprinting}. In \bibinfo{booktitle}{\emph{IEEE INFOCOM 2020 -- IEEE Conference on Computer Communications}}. \bibinfo{publisher}{IEEE}, \bibinfo{pages}{646--655}.
\newblock


\bibitem[Yan et~al\mbox{.}(2022)]%
        {yan2022rrf}
\bibfield{author}{\bibinfo{person}{Wenqing Yan} {et~al\mbox{.}}} \bibinfo{year}{2022}\natexlab{}.
\newblock \showarticletitle{RRF: A Robust Radiometric Fingerprint System that Embraces Wireless Channel Diversity}. In \bibinfo{booktitle}{\emph{Proceedings of the 15th ACM Conference on Security and Privacy in Wireless and Mobile Networks}}. \bibinfo{publisher}{ACM}, \bibinfo{pages}{85--97}.
\newblock


\bibitem[Amini et~al\mbox{.}(2024)]%
        {amini2024where}
\bibfield{author}{\bibinfo{person}{Maryam Amini}, \bibinfo{person}{Razvan Stanica}, {and} \bibinfo{person}{Catherine Rosenberg}.} \bibinfo{year}{2024}\natexlab{}.
\newblock \showarticletitle{Where Are the (Cellular) Data?}
\newblock \bibinfo{journal}{\emph{Comput. Surveys}} \bibinfo{volume}{56}, \bibinfo{number}{2} (\bibinfo{year}{2024}), \bibinfo{pages}{48:1--48:25}.
\newblock


\bibitem[Tan et~al\mbox{.}(2022)]%
        {tan2022cellulariot}
\bibfield{author}{\bibinfo{person}{Zhaowei Tan}, \bibinfo{person}{Boyan Ding}, \bibinfo{person}{Jinghao Zhao}, \bibinfo{person}{Yunqi Guo}, {and} \bibinfo{person}{Songwu Lu}.} \bibinfo{year}{2022}\natexlab{}.
\newblock \showarticletitle{Breaking Cellular IoT with Forged Data-plane Signaling: Attacks and Countermeasure}.
\newblock \bibinfo{journal}{\emph{ACM Transactions on Sensor Networks}} \bibinfo{volume}{18}, \bibinfo{number}{4} (\bibinfo{year}{2022}), \bibinfo{pages}{59:1--59:26}.
\newblock


\bibitem[Danev et~al\mbox{.}(2010)]%
        {danev2010attacks}
\bibfield{author}{\bibinfo{person}{Boris Danev}, \bibinfo{person}{Heinrich Luecken}, \bibinfo{person}{Srdjan Capkun}, {and} \bibinfo{person}{Karim M.~El Defrawy}.} \bibinfo{year}{2010}\natexlab{}.
\newblock \showarticletitle{Attacks on Physical-Layer Identification}. In \bibinfo{booktitle}{\emph{Proceedings of the Third ACM Conference on Wireless Network Security}} \emph{(\bibinfo{series}{WiSec '10})}. \bibinfo{publisher}{ACM}, \bibinfo{pages}{89--98}.
\newblock


\bibitem[Yen et~al\mbox{.}(2022)]%
        {yen2022gnnrca}
\bibfield{author}{\bibinfo{person}{Chia-Cheng Yen} {et~al\mbox{.}}} \bibinfo{year}{2022}\natexlab{}.
\newblock \showarticletitle{Graph Neural Network based Root Cause Analysis Using Multivariate Time-series KPIs for Wireless Networks}. In \bibinfo{booktitle}{\emph{NOMS 2022-2022 IEEE/IFIP Network Operations and Management Symposium}}. \bibinfo{publisher}{IEEE}, \bibinfo{pages}{1--7}.
\newblock


\bibitem[Wang et~al\mbox{.}(2022)]%
        {wang2022multiroot}
\bibfield{author}{\bibinfo{person}{Weili Wang} {et~al\mbox{.}}} \bibinfo{year}{2022}\natexlab{}.
\newblock \showarticletitle{Real-Time Analysis of Multiple Root Causes for Anomalies Assisted by Digital Twin in NFV Environment}.
\newblock \bibinfo{journal}{\emph{IEEE Transactions on Network and Service Management}} \bibinfo{volume}{19}, \bibinfo{number}{2} (\bibinfo{year}{2022}), \bibinfo{pages}{905--921}.
\newblock


\bibitem[Bajpai et~al\mbox{.}(2024)]%
        {bajpai2024anomgraphadv}
\bibfield{author}{\bibinfo{person}{Supriya Bajpai} {et~al\mbox{.}}} \bibinfo{year}{2024}\natexlab{}.
\newblock \showarticletitle{AnomGraphAdv: Enhancing Anomaly and Network Intrusion Detection in Wireless Networks Using Adversarial Training and Temporal Graph Networks}. In \bibinfo{booktitle}{\emph{Proceedings of the 17th ACM Conference on Security and Privacy in Wireless and Mobile Networks}} \emph{(\bibinfo{series}{WiSec '24})}. \bibinfo{publisher}{ACM}, \bibinfo{pages}{113--122}.
\newblock


\bibitem[Huang et~al\mbox{.}(2023)]%
        {huang2023diffar}
\bibfield{author}{\bibinfo{person}{Shuokang Huang} {et~al\mbox{.}}} \bibinfo{year}{2023}\natexlab{}.
\newblock \showarticletitle{DiffAR: Adaptive Conditional Diffusion Model for Temporal-augmented Human Activity Recognition}. In \bibinfo{booktitle}{\emph{Proceedings of the Thirty-Second International Joint Conference on Artificial Intelligence}}. \bibinfo{publisher}{IJCAI}, \bibinfo{pages}{3812--3820}.
\newblock


\bibitem[Kim et~al\mbox{.}(2022)]%
        {kim2022channelaware}
\bibfield{author}{\bibinfo{person}{Brian Kim} {et~al\mbox{.}}} \bibinfo{year}{2022}\natexlab{}.
\newblock \showarticletitle{Channel-Aware Adversarial Attacks Against Deep Learning-Based Wireless Signal Classifiers}.
\newblock \bibinfo{journal}{\emph{IEEE Transactions on Wireless Communications}} \bibinfo{volume}{21}, \bibinfo{number}{6} (\bibinfo{year}{2022}), \bibinfo{pages}{3868--3880}.
\newblock


\bibitem[Ma and otehrs(2023)]%
        {ma2023whitebox}
\bibfield{author}{\bibinfo{person}{Jie Ma} {and} \bibinfo{person}{otehrs}.} \bibinfo{year}{2023}\natexlab{}.
\newblock \showarticletitle{White-Box Adversarial Attacks on Deep Learning-Based Radio Frequency Fingerprint Identification}. In \bibinfo{booktitle}{\emph{2023 IEEE International Conference on Communications (ICC)}}. \bibinfo{publisher}{IEEE}, \bibinfo{pages}{3714--3719}.
\newblock


\bibitem[Zhao et~al\mbox{.}(2025)]%
        {zhao2025backdoor}
\bibfield{author}{\bibinfo{person}{Tianya Zhao} {et~al\mbox{.}}} \bibinfo{year}{2025}\natexlab{}.
\newblock \showarticletitle{Explanation-Guided Backdoor Attacks Against Model-Agnostic RF Fingerprinting Systems}.
\newblock \bibinfo{journal}{\emph{IEEE Transactions on Mobile Computing}} \bibinfo{volume}{24}, \bibinfo{number}{3} (\bibinfo{year}{2025}), \bibinfo{pages}{2029--2042}.
\newblock


\bibitem[Li et~al\mbox{.}(2024)]%
        {li2024practicalwifi}
\bibfield{author}{\bibinfo{person}{Changming Li} {et~al\mbox{.}}} \bibinfo{year}{2024}\natexlab{}.
\newblock \showarticletitle{Practical Adversarial Attack on WiFi Sensing Through Unnoticeable Communication Packet Perturbation}. In \bibinfo{booktitle}{\emph{Proceedings of the 30th Annual International Conference on Mobile Computing and Networking}} \emph{(\bibinfo{series}{MobiCom '24})}. \bibinfo{publisher}{ACM}, \bibinfo{pages}{373--387}.
\newblock


\bibitem[Garfinkel(2012)]%
        {garfinkel2012dfxml}
\bibfield{author}{\bibinfo{person}{Simson~L. Garfinkel}.} \bibinfo{year}{2012}\natexlab{}.
\newblock \showarticletitle{Digital forensics XML and the DFXML toolset}.
\newblock \bibinfo{journal}{\emph{Digital Investigation}} \bibinfo{volume}{8}, \bibinfo{number}{3-4} (\bibinfo{year}{2012}), \bibinfo{pages}{161--174}.
\newblock


\bibitem[Casey et~al\mbox{.}(2017)]%
        {casey2017case}
\bibfield{author}{\bibinfo{person}{Eoghan Casey} {et~al\mbox{.}}} \bibinfo{year}{2017}\natexlab{}.
\newblock \showarticletitle{Advancing coordinated cyber-investigations and tool interoperability using a community developed specification language}.
\newblock \bibinfo{journal}{\emph{Digital Investigation}}  \bibinfo{volume}{22} (\bibinfo{year}{2017}), \bibinfo{pages}{14--45}.
\newblock


\bibitem[Tabiban et~al\mbox{.}(2022)]%
        {tabiban2022vincidecoder}
\bibfield{author}{\bibinfo{person}{Azadeh Tabiban} {et~al\mbox{.}}} \bibinfo{year}{2022}\natexlab{}.
\newblock \showarticletitle{VinciDecoder: Automatically Interpreting Provenance Graphs into Textual Forensic Reports with Application to OpenStack}. In \bibinfo{booktitle}{\emph{Secure IT Systems -- 27th Nordic Conference, NordSec 2022}}. \bibinfo{publisher}{Springer}, \bibinfo{pages}{346--367}.
\newblock


\bibitem[Guo et~al\mbox{.}(2017)]%
        {guo2017calibration}
\bibfield{author}{\bibinfo{person}{Chuan Guo} {et~al\mbox{.}}} \bibinfo{year}{2017}\natexlab{}.
\newblock \showarticletitle{On Calibration of Modern Neural Networks}. In \bibinfo{booktitle}{\emph{Proceedings of the 34th International Conference on Machine Learning}} \emph{(\bibinfo{series}{Proceedings of Machine Learning Research}, Vol.~\bibinfo{volume}{70})}. \bibinfo{pages}{1321--1330}.
\newblock


\bibitem[Kuzdeba et~al\mbox{.}(2022)]%
        {kuzdeba2022systems}
\bibfield{author}{\bibinfo{person}{Scott Kuzdeba} {et~al\mbox{.}}} \bibinfo{year}{2022}\natexlab{}.
\newblock \showarticletitle{Systems View to Designing RF Fingerprinting for Real-World Operations}. In \bibinfo{booktitle}{\emph{Proceedings of the 2022 ACM Workshop on Wireless Security and Machine Learning}}. \bibinfo{publisher}{ACM}, \bibinfo{pages}{33--38}.
\newblock


\bibitem[Al-Hazbi et~al\mbox{.}(2024)]%
        {alhazbi2024rffdl}
\bibfield{author}{\bibinfo{person}{Saeif Al-Hazbi} {et~al\mbox{.}}} \bibinfo{year}{2024}\natexlab{}.
\newblock \showarticletitle{Radio Frequency Fingerprinting via Deep Learning: Challenges and Opportunities}. In \bibinfo{booktitle}{\emph{20th International Wireless Communications and Mobile Computing Conference (IWCMC 2024)}}. \bibinfo{publisher}{IEEE}, \bibinfo{pages}{824--829}.
\newblock


\bibitem[Breen et~al\mbox{.}(2021)]%
        {breen2021powder}
\bibfield{author}{\bibinfo{person}{Joe Breen} {et~al\mbox{.}}} \bibinfo{year}{2021}\natexlab{}.
\newblock \showarticletitle{Powder: Platform for Open Wireless Data-driven Experimental Research}.
\newblock \bibinfo{journal}{\emph{Computer Networks}}  \bibinfo{volume}{197} (\bibinfo{year}{2021}), \bibinfo{pages}{108281}.
\newblock


\bibitem[Bonati et~al\mbox{.}(2021)]%
        {bonati2021colosseum}
\bibfield{author}{\bibinfo{person}{Leonardo Bonati} {et~al\mbox{.}}} \bibinfo{year}{2021}\natexlab{}.
\newblock \showarticletitle{Colosseum: Large-Scale Wireless Experimentation Through Hardware-in-the-Loop Network Emulation}. In \bibinfo{booktitle}{\emph{2021 IEEE International Symposium on Dynamic Spectrum Access Networks (DySPAN)}}. \bibinfo{publisher}{IEEE}, \bibinfo{pages}{105--113}.
\newblock


\bibitem[Hoydis et~al\mbox{.}(2022)]%
        {hoydis2022sionna}
\bibfield{author}{\bibinfo{person}{Jakob Hoydis} {et~al\mbox{.}}} \bibinfo{year}{2022}\natexlab{}.
\newblock \showarticletitle{Sionna: An Open-Source Library for Next-Generation Physical Layer Research}.
\newblock \bibinfo{journal}{\emph{arXiv preprint arXiv:2203.11854}} (\bibinfo{year}{2022}).
\newblock


\bibitem[Gebru et~al\mbox{.}(2021)]%
        {gebru2021datasheets}
\bibfield{author}{\bibinfo{person}{Timnit Gebru} {et~al\mbox{.}}} \bibinfo{year}{2021}\natexlab{}.
\newblock \showarticletitle{Datasheets for Datasets}.
\newblock \bibinfo{journal}{\emph{Commun. ACM}} \bibinfo{volume}{64}, \bibinfo{number}{12} (\bibinfo{year}{2021}), \bibinfo{pages}{86--92}.
\newblock


\bibitem[Mitchell et~al\mbox{.}(2019)]%
        {mitchell2019modelcards}
\bibfield{author}{\bibinfo{person}{Margaret Mitchell} {et~al\mbox{.}}} \bibinfo{year}{2019}\natexlab{}.
\newblock \showarticletitle{Model Cards for Model Reporting}. In \bibinfo{booktitle}{\emph{Proceedings of the Conference on Fairness, Accountability, and Transparency}}. \bibinfo{publisher}{ACM}, \bibinfo{pages}{220--229}.
\newblock


\bibitem[Xu et~al\mbox{.}(2024)]%
        {xu2024explainability}
\bibfield{author}{\bibinfo{person}{Bo Xu} {et~al\mbox{.}}} \bibinfo{year}{2024}\natexlab{}.
\newblock \showarticletitle{Towards explainability for AI-based edge wireless signal automatic modulation classification}.
\newblock \bibinfo{journal}{\emph{Journal of Cloud Computing}} \bibinfo{volume}{13}, \bibinfo{number}{1} (\bibinfo{year}{2024}), \bibinfo{pages}{10}.
\newblock


\bibitem[Zhao et~al\mbox{.}(2024)]%
        {zhao2024interpretable}
\bibfield{author}{\bibinfo{person}{Tianya Zhao}, \bibinfo{person}{Xuyu Wang}, {and} \bibinfo{person}{Shiwen Mao}.} \bibinfo{year}{2024}\natexlab{}.
\newblock \showarticletitle{Cross-domain, Scalable, and Interpretable RF Device Fingerprinting}. In \bibinfo{booktitle}{\emph{IEEE INFOCOM 2024 -- IEEE Conference on Computer Communications}}. \bibinfo{publisher}{IEEE}, \bibinfo{pages}{2099--2108}.
\newblock


\bibitem[Rajendran et~al\mbox{.}(2019a)]%
        {rajendran2019unsupervised}
\bibfield{author}{\bibinfo{person}{Sreeraj Rajendran}, \bibinfo{person}{Wannes Meert}, \bibinfo{person}{Vincent Lenders}, {and} \bibinfo{person}{Sofie Pollin}.} \bibinfo{year}{2019}\natexlab{a}.
\newblock \showarticletitle{Unsupervised wireless spectrum anomaly detection with interpretable features}.
\newblock \bibinfo{journal}{\emph{IEEE Trans. Cognit. Commun. Networking}} \bibinfo{volume}{5}, \bibinfo{number}{3} (\bibinfo{year}{2019}), \bibinfo{pages}{637--647}.
\newblock


\bibitem[Rajendran et~al\mbox{.}(2019b)]%
        {rajendran2019crowdsourced}
\bibfield{author}{\bibinfo{person}{Sreeraj Rajendran}, \bibinfo{person}{Vincent Lenders}, \bibinfo{person}{Wannes Meert}, {and} \bibinfo{person}{Sofie Pollin}.} \bibinfo{year}{2019}\natexlab{b}.
\newblock \showarticletitle{Crowdsourced wireless spectrum anomaly detection}.
\newblock \bibinfo{journal}{\emph{IEEE Trans. Cognit. Commun. Networking}} \bibinfo{volume}{6}, \bibinfo{number}{2} (\bibinfo{year}{2019}), \bibinfo{pages}{694--703}.
\newblock


\bibitem[Zhou et~al\mbox{.}(2021)]%
        {zhou2021radio}
\bibfield{author}{\bibinfo{person}{Xuanhan Zhou} {et~al\mbox{.}}} \bibinfo{year}{2021}\natexlab{}.
\newblock \showarticletitle{A radio anomaly detection algorithm based on modified generative adversarial network}.
\newblock \bibinfo{journal}{\emph{IEEE Wireless Communications Letters}} \bibinfo{volume}{10}, \bibinfo{number}{7} (\bibinfo{year}{2021}), \bibinfo{pages}{1552--1556}.
\newblock


\bibitem[Alves et~al\mbox{.}(2023)]%
        {alves2023machine}
\bibfield{author}{\bibinfo{person}{Pedro~VA Alves} {et~al\mbox{.}}} \bibinfo{year}{2023}\natexlab{}.
\newblock \showarticletitle{Machine learning applied to anomaly detection on 5g o-ran architecture}.
\newblock \bibinfo{journal}{\emph{Procedia Computer Science}}  \bibinfo{volume}{222} (\bibinfo{year}{2023}), \bibinfo{pages}{81--93}.
\newblock


\bibitem[Ahasan et~al\mbox{.}(2022)]%
        {ahasan2022supervised}
\bibfield{author}{\bibinfo{person}{Md~Rakibul Ahasan} {et~al\mbox{.}}} \bibinfo{year}{2022}\natexlab{}.
\newblock \showarticletitle{Supervised learning based mobile network anomaly detection from key performance indicator (KPI) data}. In \bibinfo{booktitle}{\emph{2022 IEEE Region 10 Symposium (TENSYMP)}}. IEEE, \bibinfo{pages}{1--6}.
\newblock


\bibitem[Huang et~al\mbox{.}(2022)]%
        {huang2022cellular}
\bibfield{author}{\bibinfo{person}{Jiajia Huang}, \bibinfo{person}{Ernest Kurniawan}, {and} \bibinfo{person}{Sumei Sun}.} \bibinfo{year}{2022}\natexlab{}.
\newblock \showarticletitle{Cellular kpi anomaly detection with gan and time series decomposition}. In \bibinfo{booktitle}{\emph{ICC 2022-IEEE International Conference on Communications}}. IEEE, \bibinfo{pages}{4074--4079}.
\newblock


\bibitem[Mubasshir and other(2025)]%
        {mubasshir2025gotta}
\bibfield{author}{\bibinfo{person}{Kazi~Samin Mubasshir} {and} \bibinfo{person}{other}.} \bibinfo{year}{2025}\natexlab{}.
\newblock \showarticletitle{Gotta Detect 'Em All: Fake Base Station and Multi-Step Attack Detection in Cellular Networks}. In \bibinfo{booktitle}{\emph{Proceedings of the 34th USENIX Security Symposium}}. \bibinfo{publisher}{USENIX Association}, \bibinfo{address}{Seattle, WA, USA}.
\newblock


\bibitem[Li and otherso(2017)]%
        {li2017fbs}
\bibfield{author}{\bibinfo{person}{Zhenhua Li} {and} \bibinfo{person}{otherso}.} \bibinfo{year}{2017}\natexlab{}.
\newblock \showarticletitle{FBS-Radar: Uncovering Fake Base Stations at Scale in the Wild.}. In \bibinfo{booktitle}{\emph{NDSS}}.
\newblock


\bibitem[Riyaz et~al\mbox{.}(2018)]%
        {riyaz2018deep}
\bibfield{author}{\bibinfo{person}{Shamnaz Riyaz}, \bibinfo{person}{Kunal Sankhe}, \bibinfo{person}{Stratis Ioannidis}, {and} \bibinfo{person}{Kaushik Chowdhury}.} \bibinfo{year}{2018}\natexlab{}.
\newblock \showarticletitle{Deep learning convolutional neural networks for radio identification}.
\newblock \bibinfo{journal}{\emph{IEEE Communications Magazine}} \bibinfo{volume}{56}, \bibinfo{number}{9} (\bibinfo{year}{2018}), \bibinfo{pages}{146--152}.
\newblock


\bibitem[Groen et~al\mbox{.}(2024)]%
        {groen2024tractor}
\bibfield{author}{\bibinfo{person}{Joshua Groen} {et~al\mbox{.}}} \bibinfo{year}{2024}\natexlab{}.
\newblock \showarticletitle{TRACTOR: Traffic analysis and classification tool for open RAN}. In \bibinfo{booktitle}{\emph{ICC 2024-IEEE International Conference on Communications}}. IEEE, \bibinfo{pages}{4894--4899}.
\newblock


\bibitem[Huoh et~al\mbox{.}(2022)]%
        {huoh2022flow}
\bibfield{author}{\bibinfo{person}{Ting-Li Huoh}, \bibinfo{person}{Yan Luo}, \bibinfo{person}{Peilong Li}, {and} \bibinfo{person}{Tong Zhang}.} \bibinfo{year}{2022}\natexlab{}.
\newblock \showarticletitle{Flow-based encrypted network traffic classification with graph neural networks}.
\newblock \bibinfo{journal}{\emph{IEEE Trans. Netw. Serv. Manag.}} \bibinfo{volume}{20}, \bibinfo{number}{2} (\bibinfo{year}{2022}), \bibinfo{pages}{1224--1237}.
\newblock


\bibitem[Akbari et~al\mbox{.}(2021)]%
        {akbari2021look}
\bibfield{author}{\bibinfo{person}{Iman Akbari} {et~al\mbox{.}}} \bibinfo{year}{2021}\natexlab{}.
\newblock \showarticletitle{A look behind the curtain: Traffic classification in an increasingly encrypted web}.
\newblock \bibinfo{journal}{\emph{Proceedings of the ACM on Measurement and Analysis of Computing Systems}} \bibinfo{volume}{5}, \bibinfo{number}{1} (\bibinfo{year}{2021}), \bibinfo{pages}{1--26}.
\newblock


\bibitem[Li et~al\mbox{.}(2022)]%
        {li2022radionet}
\bibfield{author}{\bibinfo{person}{Haipeng Li} {et~al\mbox{.}}} \bibinfo{year}{2022}\natexlab{}.
\newblock \showarticletitle{RadioNet: Robust deep-learning based radio fingerprinting}. In \bibinfo{booktitle}{\emph{2022 IEEE Conference on Communications and Network Security (CNS)}}. IEEE, \bibinfo{pages}{190--198}.
\newblock


\bibitem[Kotaru et~al\mbox{.}(2015)]%
        {kotaru2015spotfi}
\bibfield{author}{\bibinfo{person}{Manikanta Kotaru} {et~al\mbox{.}}} \bibinfo{year}{2015}\natexlab{}.
\newblock \showarticletitle{Spotfi: Decimeter level localization using wifi}. In \bibinfo{booktitle}{\emph{Proceedings of the 2015 ACM conference on special interest group on data communication}}. \bibinfo{pages}{269--282}.
\newblock


\bibitem[Mitchell et~al\mbox{.}(2022)]%
        {mitchell2022deep}
\bibfield{author}{\bibinfo{person}{Frost Mitchell} {et~al\mbox{.}}} \bibinfo{year}{2022}\natexlab{}.
\newblock \showarticletitle{Deep learning-based localization in limited data regimes}. In \bibinfo{booktitle}{\emph{Proceedings of the 2022 ACM workshop on wireless security and machine learning}}. \bibinfo{pages}{15--20}.
\newblock


\bibitem[Nardin et~al\mbox{.}(2023)]%
        {nardin2023jamming}
\bibfield{author}{\bibinfo{person}{Andrea Nardin}, \bibinfo{person}{Tales Imbiriba}, {and} \bibinfo{person}{Pau Closas}.} \bibinfo{year}{2023}\natexlab{}.
\newblock \showarticletitle{Jamming source localization using augmented physics-based model}. In \bibinfo{booktitle}{\emph{ICASSP 2023-2023 IEEE International Conference on Acoustics, Speech and Signal Processing}}. IEEE, \bibinfo{pages}{1--5}.
\newblock


\bibitem[Tedeschini et~al\mbox{.}(2024)]%
        {tedeschini2024real}
\bibfield{author}{\bibinfo{person}{Bernardo~Camajori Tedeschini} {et~al\mbox{.}}} \bibinfo{year}{2024}\natexlab{}.
\newblock \showarticletitle{Real-time Bayesian neural networks for 6G cooperative positioning and tracking}.
\newblock \bibinfo{journal}{\emph{IEEE J. Sel. Areas Commun.}} \bibinfo{volume}{42}, \bibinfo{number}{9} (\bibinfo{year}{2024}), \bibinfo{pages}{2322--2338}.
\newblock


\bibitem[Sadr et~al\mbox{.}(2021)]%
        {sadr2021uncertainty}
\bibfield{author}{\bibinfo{person}{Mohammad Amin~Maleki Sadr} {et~al\mbox{.}}} \bibinfo{year}{2021}\natexlab{}.
\newblock \showarticletitle{Uncertainty estimation via Monte Carlo dropout in CNN-based mmWave MIMO localization}.
\newblock \bibinfo{journal}{\emph{IEEE Signal Processing Letters}}  \bibinfo{volume}{29} (\bibinfo{year}{2021}), \bibinfo{pages}{269--273}.
\newblock


\bibitem[Wang et~al\mbox{.}(2023)]%
        {wang2023through}
\bibfield{author}{\bibinfo{person}{Jiacheng Wang} {et~al\mbox{.}}} \bibinfo{year}{2023}\natexlab{}.
\newblock \showarticletitle{Through the wall detection and localization of autonomous mobile device in indoor scenario}.
\newblock \bibinfo{journal}{\emph{IEEE J. Sel. Areas Commun.}} \bibinfo{volume}{42}, \bibinfo{number}{1} (\bibinfo{year}{2023}), \bibinfo{pages}{161--176}.
\newblock


\bibitem[Schulz et~al\mbox{.}(2017)]%
        {schulz2017nexmon}
\bibfield{author}{\bibinfo{person}{Matthias Schulz}, \bibinfo{person}{Daniel Wegemer}, {and} \bibinfo{person}{Matthias Hollick}.} \bibinfo{year}{2017}\natexlab{}.
\newblock \showarticletitle{Nexmon: Build your own wi-fi testbeds with low-level mac and phy-access using firmware patches on off-the-shelf mobile devices}. In \bibinfo{booktitle}{\emph{Proceedings of the 11th Workshop on Wireless Network Testbeds, Experimental evaluation \& CHaracterization}}. \bibinfo{pages}{59--66}.
\newblock


\bibitem[Jiang et~al\mbox{.}(2013)]%
        {jiang2013rejecting}
\bibfield{author}{\bibinfo{person}{Zhiping Jiang}, \bibinfo{person}{Jizhong Zhao}, \bibinfo{person}{Xiang-Yang Li}, \bibinfo{person}{Jinsong Han}, {and} \bibinfo{person}{Wei Xi}.} \bibinfo{year}{2013}\natexlab{}.
\newblock \showarticletitle{Rejecting the attack: Source authentication for wi-fi management frames using csi information}. In \bibinfo{booktitle}{\emph{2013 Proceedings IEEE INFOCOM}}. IEEE, \bibinfo{pages}{2544--2552}.
\newblock


\bibitem[Zhang et~al\mbox{.}(2023)]%
        {zhang2023tag}
\bibfield{author}{\bibinfo{person}{Pinchang Zhang} {et~al\mbox{.}}} \bibinfo{year}{2023}\natexlab{}.
\newblock \showarticletitle{Tag-based PHY-layer authentication for RIS-assisted communication systems}.
\newblock \bibinfo{journal}{\emph{IEEE Trans. Dependable Secure Comput.}} \bibinfo{volume}{20}, \bibinfo{number}{6} (\bibinfo{year}{2023}), \bibinfo{pages}{4778--4792}.
\newblock


\bibitem[Hoang et~al\mbox{.}(2024)]%
        {hoang2024physical}
\bibfield{author}{\bibinfo{person}{Tiep~M Hoang} {et~al\mbox{.}}} \bibinfo{year}{2024}\natexlab{}.
\newblock \showarticletitle{Physical layer authentication and security design in the machine learning era}.
\newblock \bibinfo{journal}{\emph{IEEE Commun. Surv. Tutorials}} \bibinfo{volume}{26}, \bibinfo{number}{3} (\bibinfo{year}{2024}), \bibinfo{pages}{1830--1860}.
\newblock


\bibitem[Xie et~al\mbox{.}(2022)]%
        {xie2022physical}
\bibfield{author}{\bibinfo{person}{Ning Xie} {et~al\mbox{.}}} \bibinfo{year}{2022}\natexlab{}.
\newblock \showarticletitle{Physical layer authentication with high compatibility using an encoding approach}.
\newblock \bibinfo{journal}{\emph{IEEE Trans. Commun.}} \bibinfo{volume}{70}, \bibinfo{number}{12} (\bibinfo{year}{2022}), \bibinfo{pages}{8270--8285}.
\newblock


\bibitem[Gr{\"u}nwald(2024)]%
        {grunwald2024beyond}
\bibfield{author}{\bibinfo{person}{Peter~D Gr{\"u}nwald}.} \bibinfo{year}{2024}\natexlab{}.
\newblock \showarticletitle{Beyond Neyman--Pearson: E-values enable hypothesis testing with a data-driven alpha}.
\newblock \bibinfo{journal}{\emph{Proceedings of the National Academy of Sciences}} \bibinfo{volume}{121}, \bibinfo{number}{39} (\bibinfo{year}{2024}), \bibinfo{pages}{e2302098121}.
\newblock


\bibitem[An et~al\mbox{.}(2021)]%
        {an2021tag}
\bibfield{author}{\bibinfo{person}{Yongli An}, \bibinfo{person}{Shikang Zhang}, {and} \bibinfo{person}{Zhanlin Ji}.} \bibinfo{year}{2021}\natexlab{}.
\newblock \showarticletitle{A tag-based PHY-layer authentication scheme without key distribution}.
\newblock \bibinfo{journal}{\emph{IEEE Access}}  \bibinfo{volume}{9} (\bibinfo{year}{2021}), \bibinfo{pages}{85947--85955}.
\newblock


\bibitem[Tomasin et~al\mbox{.}(2022)]%
        {tomasin2022challenge}
\bibfield{author}{\bibinfo{person}{Stefano Tomasin} {et~al\mbox{.}}} \bibinfo{year}{2022}\natexlab{}.
\newblock \showarticletitle{Challenge-response physical layer authentication over partially controllable channels}.
\newblock \bibinfo{journal}{\emph{IEEE Communications Magazine}} \bibinfo{volume}{60}, \bibinfo{number}{12} (\bibinfo{year}{2022}), \bibinfo{pages}{138--144}.
\newblock


\bibitem[Wang et~al\mbox{.}(2025)]%
        {wang2025generative}
\bibfield{author}{\bibinfo{person}{Jiacheng Wang} {et~al\mbox{.}}} \bibinfo{year}{2025}\natexlab{}.
\newblock \showarticletitle{Generative AI based secure wireless sensing for ISAC networks}.
\newblock \bibinfo{journal}{\emph{IEEE T INF FOREN SEC.}} (\bibinfo{year}{2025}).
\newblock


\bibitem[Zhang et~al\mbox{.}(2022)]%
        {zhang2022csi}
\bibfield{author}{\bibinfo{person}{Yong Zhang} {et~al\mbox{.}}} \bibinfo{year}{2022}\natexlab{}.
\newblock \showarticletitle{CSI-based location-independent human activity recognition using feature fusion}.
\newblock \bibinfo{journal}{\emph{IEEE Transactions on Instrumentation and Measurement}}  \bibinfo{volume}{71} (\bibinfo{year}{2022}), \bibinfo{pages}{1--12}.
\newblock


\bibitem[Chen et~al\mbox{.}(2018)]%
        {chen2018wifi}
\bibfield{author}{\bibinfo{person}{Zhenghua Chen} {et~al\mbox{.}}} \bibinfo{year}{2018}\natexlab{}.
\newblock \showarticletitle{WiFi CSI based passive human activity recognition using attention based BLSTM}.
\newblock \bibinfo{journal}{\emph{IEEE Trans. Mobile Comput.}} \bibinfo{volume}{18}, \bibinfo{number}{11} (\bibinfo{year}{2018}), \bibinfo{pages}{2714--2724}.
\newblock


\bibitem[Huang et~al\mbox{.}(2020)]%
        {huang2020towards}
\bibfield{author}{\bibinfo{person}{Jinyang Huang} {et~al\mbox{.}}} \bibinfo{year}{2020}\natexlab{}.
\newblock \showarticletitle{Towards anti-interference WiFi-based activity recognition system using interference-independent phase component}. In \bibinfo{booktitle}{\emph{IEEE INFOCOM 2020-IEEE Conference on Computer Communications}}. IEEE, \bibinfo{pages}{576--585}.
\newblock


\bibitem[Lu et~al\mbox{.}(2022)]%
        {lu2022cehar}
\bibfield{author}{\bibinfo{person}{Xiao Lu}, \bibinfo{person}{Yuli Li}, \bibinfo{person}{Wei Cui}, {and} \bibinfo{person}{Haixia Wang}.} \bibinfo{year}{2022}\natexlab{}.
\newblock \showarticletitle{Cehar: Csi-based channel-exchanging human activity recognition}.
\newblock \bibinfo{journal}{\emph{IEEE Internet Things J.}} \bibinfo{volume}{10}, \bibinfo{number}{7} (\bibinfo{year}{2022}), \bibinfo{pages}{5953--5961}.
\newblock


\bibitem[Kong and Chen(2024)]%
        {kong2024csi}
\bibfield{author}{\bibinfo{person}{Ruiqi Kong} {and} \bibinfo{person}{He Chen}.} \bibinfo{year}{2024}\natexlab{}.
\newblock \showarticletitle{CSI-RFF: Leveraging micro-signals on CSI for RF fingerprinting of commodity WiFi}.
\newblock \bibinfo{journal}{\emph{IEEE T INF FOREN SEC.}}  \bibinfo{volume}{19} (\bibinfo{year}{2024}), \bibinfo{pages}{5301--5315}.
\newblock


\bibitem[Iurman et~al\mbox{.}(2021)]%
        {iurman2021towards}
\bibfield{author}{\bibinfo{person}{Justin Iurman}, \bibinfo{person}{Frank Brockners}, {and} \bibinfo{person}{Benoit Donnet}.} \bibinfo{year}{2021}\natexlab{}.
\newblock \showarticletitle{Towards cross-layer telemetry}. In \bibinfo{booktitle}{\emph{Proceedings of the 2021 Applied Networking Research Workshop}}. \bibinfo{pages}{15--21}.
\newblock


\bibitem[Ashok et~al\mbox{.}(2024)]%
        {ashok2024traceweaver}
\bibfield{author}{\bibinfo{person}{Sachin Ashok} {et~al\mbox{.}}} \bibinfo{year}{2024}\natexlab{}.
\newblock \showarticletitle{Traceweaver: Distributed request tracing for microservices without application modification}. In \bibinfo{booktitle}{\emph{Proceedings of the ACM SIGCOMM 2024 Conference}}. \bibinfo{pages}{828--842}.
\newblock


\bibitem[Toslali et~al\mbox{.}(2021)]%
        {toslali2021automating}
\bibfield{author}{\bibinfo{person}{Mert Toslali} {et~al\mbox{.}}} \bibinfo{year}{2021}\natexlab{}.
\newblock \showarticletitle{Automating instrumentation choices for performance problems in distributed applications with VAIF}. In \bibinfo{booktitle}{\emph{Proceedings of the ACM Symposium on Cloud Computing}}. \bibinfo{pages}{61--75}.
\newblock


\bibitem[Shen et~al\mbox{.}(2023)]%
        {shen2023network}
\bibfield{author}{\bibinfo{person}{Junxian Shen} {et~al\mbox{.}}} \bibinfo{year}{2023}\natexlab{}.
\newblock \showarticletitle{Network-centric distributed tracing with deepflow: Troubleshooting your microservices in zero code}. In \bibinfo{booktitle}{\emph{Proceedings of the ACM SIGCOMM 2023 Conference}}. \bibinfo{pages}{420--437}.
\newblock


\bibitem[Korany and Mostofi(2021)]%
        {korany2021counting}
\bibfield{author}{\bibinfo{person}{Belal Korany} {and} \bibinfo{person}{Yasamin Mostofi}.} \bibinfo{year}{2021}\natexlab{}.
\newblock \showarticletitle{Counting a stationary crowd using off-the-shelf wifi}. In \bibinfo{booktitle}{\emph{Proceedings of the 19th annual international conference on mobile systems, applications, and services}}. \bibinfo{pages}{202--214}.
\newblock


\bibitem[Wang et~al\mbox{.}(2024)]%
        {wang2024generative}
\bibfield{author}{\bibinfo{person}{Jiacheng Wang} {et~al\mbox{.}}} \bibinfo{year}{2024}\natexlab{}.
\newblock \showarticletitle{Generative artificial intelligence assisted wireless sensing: Human flow detection in practical communication environments}.
\newblock \bibinfo{journal}{\emph{IEEE J. Sel. Areas Commun.}} \bibinfo{volume}{42}, \bibinfo{number}{10} (\bibinfo{year}{2024}), \bibinfo{pages}{2737--2753}.
\newblock


\bibitem[Sch{\"a}fer et~al\mbox{.}(2021)]%
        {schafer2021human}
\bibfield{author}{\bibinfo{person}{J{\"o}rg Sch{\"a}fer} {et~al\mbox{.}}} \bibinfo{year}{2021}\natexlab{}.
\newblock \showarticletitle{Human activity recognition using CSI information with nexmon}.
\newblock \bibinfo{journal}{\emph{Applied Sciences}} \bibinfo{volume}{11}, \bibinfo{number}{19} (\bibinfo{year}{2021}), \bibinfo{pages}{8860}.
\newblock


\bibitem[Islam et~al\mbox{.}(2022)]%
        {islam2022deep}
\bibfield{author}{\bibinfo{person}{Mohammad~Ariful Islam} {et~al\mbox{.}}} \bibinfo{year}{2022}\natexlab{}.
\newblock \showarticletitle{A deep neural network-based communication failure prediction scheme in 5g ran}.
\newblock \bibinfo{journal}{\emph{IEEE Trans. Netw. Serv. Manag.}} \bibinfo{volume}{20}, \bibinfo{number}{2} (\bibinfo{year}{2022}), \bibinfo{pages}{1140--1152}.
\newblock


\bibitem[Mitchell et~al\mbox{.}(2023)]%
        {mitchell2023learning}
\bibfield{author}{\bibinfo{person}{Frost Mitchell} {et~al\mbox{.}}} \bibinfo{year}{2023}\natexlab{}.
\newblock \showarticletitle{Learning-based techniques for transmitter localization: A case study on model robustness}. In \bibinfo{booktitle}{\emph{2023 20th Annual IEEE International Conference on Sensing, Communication, and Networking (SECON)}}. IEEE, \bibinfo{pages}{133--141}.
\newblock


\end{thebibliography}


\end{document}